\documentclass[11pt]{article}
\usepackage{graphicx}
\usepackage{amsfonts}
\usepackage{latexsym}
\usepackage{cite}

\def    \beq    {\begin{equation}} \def \eeq    {\end{equation}}
\def    \bea    {\begin{eqnarray}} \def \eea    {\end{eqnarray}}
\def    \lf     {\left (} \def  \rt     {\right )}

    \def      \comma  {\; , \; \;} \def       \pl
{\partial} 
\def \half {1 \over 2}
\def \aad { \alpha {\dot{\alpha}}}
\def \bbd { \beta {\dot{\beta}}}
\def \bd { \dot{\beta}}
\def \ad {{\dot{\alpha}}}
\def \de {\delta_{\epsilon}}
\def \nn {\nonumber}

\begin{document}
\title{Les Houches lectures on supersymmetric gauge theories \footnote{These notes are based on lectures delivered by E. Rabinovici at the Les Houches Summer School: Session 76: Euro Summer School on Unity of Fundamental Physics: Gravity, Gauge Theory and Strings, Les Houches, France, 30 Jul - 31 Aug 2001.}}
\author{D.~S.~Berman, E.~Rabinovici \\ 
Racah Institute, Hebrew University, \\ Jerusalem, ISRAEL}

%
\maketitle
\begin{abstract}
We introduce simple and more advanced concepts that have played a key role in
the development of supersymmetric systems.
This is done by first describing various supersymmetric quantum mechanics
models. Topics covered include the basic construction of supersymmetric
field theories, the phase structure of supersymmetric systems with
and without gauge particles, superconformal theories and infrared duality in both field theory and string theory.
A discussion of the relation of conformal symmetry to a vanishing vacuum
energy (cosmological constant) is included.
\end{abstract}
\newpage
\tableofcontents

\newpage

\section{Introduction}

At least three phases of gauge theory are manifest in nature. The weak interactions are in the Higgs phase. The electromagnetic interaction is in the Coulomb phase and the colour interactions are in the confining phase. None of these phases exhibits supersymmetry explicitly. What is then the motivation to introduce and study supersymmetry? Here is a list of reasons that have motivated different people over the years.
\begin{itemize}
\item In their seminal paper Golfand and Likhtman \cite{Golfand:iw} introduced supersymmetry in order to constrain the possible forms of interaction, little did they know that the constrained interactions \cite{WessZ} will give rise to a multitude of vacua. 
\item The local version of supersymmetry contains automatically gravity \cite{Bagger:1990qh,West:tg}, the force in nature that has yet to be tamed by theoretical physics. 
\item The weak interaction scale and the Plank scale are widely separated for theoreticians this could be a problem, called the Hierarchy problem. This problem is much softer in supersymmetric theories \cite{weinberg1,weinberg2,susskind}. 
\item In various supersymmetric models the gauge couplings unify sensibly\cite{Ellis,Amaldi,Langacker}.
\item Strong coupling problems are generically intractable in field theory. In the presence of supersymmetry constraints the problems can be solved in interesting cases. In particular the idea of monopole condensation as the responsible mechanism for confinement becomes tractable. Such models serve as analytical laboratories for physical ideas.
\item Perturbative string theory seams to need supersymmetry in order to be stable. In fact, in order for string theories to make sense the number of Bosonic degrees of freedom can differ from those of the Fermionic degrees of freedom by at most those of a d=2 field theory \cite{KS}. 
\end{itemize}

The construction and analysis of supersymmetric systems requires the use of many different concepts and ideas in field theory. In section two we will introduce several of these ideas by using a simple quantum mechanical context. The ideas include: supersymmetry and its spontaneous breaking; index theorems as a tool to answer physical questions; the impact of conformal symmetries; and treating theories with no ground states. 

Section three will contain a review of the methods used to construct supersymmetric field theories and supersymmetric gauge theories. There are many excellent reviews on the subject, we refer the reader to them for a more 
complete exposition \cite{Fayet:1976cr,Bagger:1990qh,West:tg,argyres,SI,H-P,Gates,weinbook,FF}.

In section four, we review the intricate phase structure of classical and quantum supersymmetric gauge theories. It may occasionally seem that one sees only the trees and not the forest however note that many of the trees are central problems in field theory. They include the understanding of confinement, chiral symmetry breaking and the emergence of massless fermions to mention just a few. The material in this section is elaborated in similar or greater detail in \cite{SI,argyres}. This section includes also a discussion of conformal field theories and some properties in dimension greater than two including four. In particular, the vacuum energy of these theories is discussed in relation to the cosmological constant problem.

In section five we describe the phenomena of infra-red duality in supersymmetric gauge theories. This duality is described from both a field theoretical and string theoretical point of view. In the process we will discuss connections between string theory in the presence of branes and gauge theories. 

Many topics in the study of supersymmetric theories which are as worthy have not been covered for lack of time or because they were covered by other lectures. Examples of such topics are:  large N gauge theories \cite{malda}; non-supersymmetric deformations of the models described eg. \cite{ofer+}; supersymmetric matrix models \cite{banks+}; supersymmetry on the lattice \cite{rab+} as well as many others.

\section{Supersymmetric Quantum Mechanics}

The ideas in this section were introduced in \cite{Witten:nf,Witten:df}.
First we will examine the quantum mechanical realization of the supersymmetry algebra so as to introduce various ideas that will later carry over to field theory. The questions we wish to examine are: what is supersymmetry; how is SUSY broken spontaneously; and what are nonrenormalisation theorems. Along the way we will introduce some useful tools such as the Witten index. Now we begin with a one variable realization of N=1 SUSY. Later we will present a two variable realization where we can introduce the notion of a flat direction.

Quantum mechanics is a one dimensional field theory. The Hamiltonian is the only member of the Poincare group in that case. 
Thus, the basic anti-commutation relations that define the supersymmetric algebra are:
\begin{equation}
	\{ Q_i, Q_j^{+} \} =2 H \delta_{ij} \qquad \qquad i, j=1..N
	\label{susy}
\end{equation}
H is the Hamiltonian and $Q,Q^+$ are called the supercharges. N denotes the number of supersymmetries. 

A rather  general N=1 realization with n bosonic and n Fermionic degrees of freedom is given by:
\beq
Q= \sum^n_{\alpha=1} \psi^+_{\alpha} (-p_{\alpha} + i {{\pl W }\over {\pl x_\alpha}}) \comma
\eeq
 $W(x,..,x_n)$ is a general function of the n bosonic variables.
The Hamiltonian is then given by:
\beq
H={1 \over 2} \{Q,Q^+ \} = {1 \over 2} \lf \lf \sum_\alpha p_\alpha^2 + {{\pl W }\over {\pl x_\alpha}}^2 \rt 1_{L \times L} - \sum_{\alpha \beta} B_{\alpha \beta} {{\pl^2W} \over {\pl x_\alpha \pl x_\beta}} \rt
\eeq	
and B is: 
\beq
B_{\alpha \beta} = {1 \over 2} [\psi_\alpha^+, \psi_\beta]
\eeq
and the $\psi$ variables obey the following anticommutation relations.
\beq
\{ \psi^+_\alpha, \psi_\beta \} = \delta_{\alpha \beta} \comma \{ \psi_\alpha, \psi_\beta \} =0 \comma \{ \psi^+_\alpha, \psi_\beta^+ \}=0
\eeq
The dimension of the Fermionic Hilbert space is $L=2^n$. Before moving on to n=1 realization of this super algebra, we first will recall some basic facts about the Bosonic  harmonic oscillator. 
\beq 
H= {p_q^2 \over 2m} + {1 \over 2} m \omega ^2 q^2
\eeq
define 
The energy scale is extracted by defining dimensionless variables x and $p_x$.
\beq 
x=({{m \omega} \over \hbar})^{1 \over 2} q \comma \qquad p_x = (m
\omega \hbar)^{-{1 \over 2}} p_q \label{ssho}
\eeq
This gives the following commutation relations:
\beq
-i \hbar = [p_q, q] = \hbar [p_x,x] \comma
\eeq
the Hamiltonian is now given by:
\beq
H= \hbar \omega {\half} ( p_x^2 + x^2) 
\eeq
$\hbar \omega$ is the energy scale in the problem. It is useful to define creation and annihilation operators, $a^+, a$ by:
\beq
x={1\over {\sqrt{2}}} ( a+a^+) \comma p = {i \over {\sqrt{2}}} (a^{+} -a ) \label{sho}
\eeq
and so we obtain the commutation relation:
\beq 
[a, a^+] =1 \; \; .
\eeq
The number operator, N is constructed out of the creation and annihilation operators:
\beq
N = a^+ a
\eeq
The spectrum of this operator is given by the nonnegative integers.
The Hamiltonian in these variables then becomes:
\beq
H=\hbar \omega(N+ {\half})
\eeq
and the energy spacings are given by the scale $\hbar \omega$.

The energy of the ground state differs from the classical minimum energy by ${\half} \hbar \omega$. This is essentially due the uncertainty relation.

The eigenstates, $|n>$ can be recast in terms of the variable x. The ground state, $|0>$ is obtained by solving the equation:
\beq
a |0> =0\; \; . \label{va}
\eeq
Note that this equation is a first order as opposed to the second order equation that one would have to solve if one attempted to directly solve the Schroedinger equation. This is possible only for the ground state. Using equations \ref{sho},\ref{va}:
\beq
(x +i p ) \phi(x)=0
\eeq
That is:
\beq
(x+ {d \over {d x}}) \phi(x) =0
\eeq
yielding the ground state wave function:
\beq
\phi(x)={\rm{c \;  exp}}(-{{x^2 \over 2}})
\eeq
The state $|n>$ is given by:
\beq
|n> = {{ (a^+)^n} \over {\sqrt{n!}}}|0> \comma
\eeq
which can now be expressed in x-space . This completes the solution of the Bosonic harmonic oscillator.

Also note that one could have written H as:
\beq 
H= \hbar \omega ( a a^+ - {\half})
\eeq
So why not obtain a ground state energy of $- {\half} \hbar \omega$ by solving 
\beq 
a^+ |\rm{GS}> =0
\eeq
This first order equation always has a solution. It is given by:
\beq
\phi(x)= {\rm{c  \; exp}}({\half} x^2)\; .
\eeq

To find a quantum state it is not enough that it be a zero energy solution of the Schr\"{o}dinger equation but it must also be {normalizable}. (By normalizable we include {\it{plane wave}} normalisability).

The above solution is not even plane wave normalizable and hence it is not a quantum state.

Physically, how do we motivate this restriction? After all, the universe may well be finite. In such a circumstance the wave function will be exponentially confined to the edge of the universe and thus irrelevant for bulk physics. There may be situations where one will want to study the physics on the boundary. In such cases the ``non-normalizable'' states should be kept and may play an important role. From now on we will only accept plane wave normalizable states and nothing ``worse''.

Also recall that, the simple harmonic oscillator is a useful approximation for small excitations above the minimum of a generic potential.

We will now note some basic facts concerning the validity of a perturbative expansion. Consider the Hamiltonian,
\beq 
H= {p_q^2 \over 2m} + {\half} g q^n
\eeq
One may wonder whether if one can make a perturbative expansion in small or large g or small or large m. To answer this one needs to find out if one can remove the g,m dependence the same type of rescaling used for the harmonic oscillator \ref{ssho}.
Is it possible to define a new set of dimensionless canonical variables $p_x, x$ that preserve the commutation relations, such that:
\beq 
[p_q,q]=[p_x,x] \hbar
\eeq
and 
\beq
H=h(m,g) {\half} (p_x^2 + x^n)
\eeq
We will make the following ansatz:
\beq
q=f(m,g) x  \comma \qquad p_q= { 1 \over f(m,g) } p_x \comma
\eeq
giving 
\beq
2H = {{p_x^2} \over {m f^2(m,g)}} +g f(m,g)^n x^n \;\; \comma
\eeq
and so one may choose
\beq
g f(m,g)= ({1 \over m f(m,g)^2})^{1 \over {n+2}} \; \; .
\eeq
The Hamiltonian becomes:
\beq
H= g^{1-{n \over {n+2}}} m^{-{n \over {n+2}}}{\half} ( p_q^2 + q^n ) \; .
\eeq

The role of g and m is just to determine the overall energy scale. They may not serve as perturbation parameters. (This does not apply to the special case of $n=-2$, which we will return to later).
Thus, Hamiltonians of the form $p^2 + x^n$ can't be analyzed perturbatively. Further analysis shows that for Hamiltonians of the form $ p^2+x^n+x^m$ with $m < n$, perturbation theory is valid.

After reviewing the Bosonic harmonic oscillator, we review the harmonic  Fermionic Oscillator. The commutation relations become anticommutation relations.
\beq 
\{ a_F,a_F^+ \} = 1 \comma \{ a_F, a_F \}=\{ a_f^+ , a_F^+ \} =0
\eeq

The Hamiltonian (that does not have a classical analogue) is taken to be, 
\beq
H=\hbar \omega_F (a_F^+ a_F\; + \;{\rm{const.}}) \comma
\eeq
where $\omega_F$ is the Fermionic oscillator frequency and,
\beq
[a_F^+, H]=-a_F^+ \comma \qquad [a_F, H]= a_F \; \; .
\eeq
This demonstrates that $a_F^+$, $a_F$ are creation and annihilation operators.
The spectrum of $N_F$ is 0,1. This is essentially a manifestation of the Pauli exclusion principle.
The states are given by 
\beq
|0> \comma a_F^+ |0> 
\eeq

This algebra can be realized using the Pauli matrices $\{ \sigma^i \}$ as follows. Define,
\beq 
\sigma_- = {\half} (\sigma_1 - i \sigma_2 ) \comma \sigma_+ = {\half} (\sigma_1 +i \sigma_2) 
\eeq
We then identify,
\beq
a_F= \sigma_- \comma a_F^+= \sigma_+
\eeq
Hence the Hamiltonian becomes: 
\beq
H={\half}\hbar \omega_F (\sigma_3 +1) +c
\eeq
For the Bosonic case, we have
\beq 
H_B={\half} \hbar \omega_B \{ a_B, a_B^+ \} =\hbar \omega_B( N_B + {\half})
\eeq
This shifts the energy levels by  $+{\half} \hbar \omega$.
By analogy, we choose $H_F$ to be,
\beq
H_F={\half} \hbar \omega_F [a^+_F, a_F] =\hbar \omega_F( N_F - {\half})
\eeq
With this choice, the vacuum energy may cancel between the Bosons and Fermions.
We see that the total Hamiltonian becomes:
\beq
H=\hbar \omega_B(N_B+{\half})+\hbar \omega_F(N_F-{\half})
\eeq
The Fermionic and Bosonic number operators both commute with the total Hamiltonian.

Consider the case when $\omega=\omega_F=\omega_B$, here the vacuum energy precisely vanishes.
\beq
 E_n =\hbar \omega (n_B +n_F) 
\eeq
There is now a symmetry {\bf{except}} for the $E=0$ state.

Define
\beq
Q= {1 \over {\sqrt{2}}} (\sigma_1 p +\sigma_2 x)
\eeq
\beq
H= {\half} \{ Q, Q \}=Q^2
\eeq

Q commutes with H.

\bea
Q^2&=&{\half} (\sigma_1^2 p^2 + \sigma_2^2 x^2 +\sigma_1 \sigma_2 px +\sigma_2 \sigma_1xp) \\
&=& {\half} ( p^2 +x^2) 1_{2 \times 2} + i \sigma_3 (px -xp) \\ 
&=& {\half} (p^2 +x^2) 1_{2 \times 2} + {\half} \sigma_3
\eea
hence,
\beq
H= {\half} \omega ((p^2 +x^2)1_{2 \times2} + \sigma_3)
\eeq

Q anticommutes with $\sigma_3$ and commutes with H. We will label a state by its energy E and its Fermion number, $N_F=0,1$, ie. $|E,0>$ denotes a state with energy E and $N_F=0$.
 Thus,
\bea
H (Q |E,0>) &=& E (Q |E,0>) \\
N_F (Q |E,0>) &=& 1 (Q|E,0>) \\
Q(Q|E,0>) &=& E |E,0>
\eea
This implies
\bea
Q(|E,0>) &=& \sqrt{E} |E,1> \\
Q (|E,1>) &=& \sqrt{E} |E,0>
\eea
which is valid only for $E \neq 0$.

For any state $|E,N_F>$ with $E\neq 0$ there is a state with equal energy and different $N_F$ obtained by the application of Q.

The ground state $ |0,0>$ however is annihilated by Q and so is not necessarily paired.

Note, that there is a $Q_2$ such that also $H= Q_2^2$.
\beq
Q_2= {1 \over {\sqrt{2}}} (\sigma_2 p - \sigma_1 x) = {1 \over {\sqrt{2}}} \sigma_2 (p + i \sigma_3 x)
\eeq
and , $ \{ Q_1,Q_2 \}=0$.

\beq
Q_1= i \sigma_3 Q_2
\eeq
We can then define:
\beq
Q_{\pm}= {1 \over {\sqrt{2}}} (Q_1 \pm i Q_2)
\eeq
\beq
H={\half} \{ Q_+ , Q_- \} \comma \qquad \{ Q_+,Q_+ \}=0 \comma \{Q_-,Q_- \}=0
\eeq

This was a ``free'' theory, (ie. a simple harmonic oscillator).
It can be generalized to more complicated cases. We introduce new supercharges that depend on a potential W(x) as follows:

\bea
Q_1 &=& {1 \over {\sqrt{2}}} (\sigma_1 p + \sigma_2 W'(x) ) \\
Q_2 &=& {1 \over {\sqrt{2}}} (\sigma_2 p - \sigma_1 W'(x) ) \; \; .
\eea
In the previous case, with the simple harmonic oscillator, $ W(x)={\half} x^2 $. The generalized Hamiltonian is now:
\beq
H= Q_1^2 + Q_2^2 = {\half} (p^2 +W'(x)^2 ) 1 + {\half} W''(x) \sigma_3
\eeq

\beq
[H,Q_i]=0 \comma [H,\sigma_3] =0 \comma  \{Q_i, \sigma_3 \}=0
\eeq
Thus  $E \geq 0 $ and for $E > 0$ the spectrum is paired.

There is an analogue in field theory. The energy gap in the Bosonic sector ($N_F=0$) matches the energy gap in the Fermionic ($N_F=1$) sector. In field theory the energy gap between he first excited state and the ground state is the particle mass, thus the mass of a free Boson equals that of a free Fermion.

Consider the E=0 case.

In general one may have any number of zero energy states in each $N_F$ sector. For $W(x) = {\half} x^2$, $n(E=0, N_F=0)=1$ and $n(E=0,N_F=1)=0$ where n denotes the number of states of given E and $N_F$.
The full spectrum can't be solved for general potential W(x). It is necessary to solve two 2nd order equations:

\bea
\lf \matrix{ p^2 +(W')^2+W'' & 0 \cr
             0 & p^2 +(W')^2 -W'' } \rt  
\lf \matrix{\phi_1(x) \cr \phi_0(x)}  \rt = 2E \lf \matrix{\phi_1(x) \cr \phi_0(x)} \rt
\eea
The zero energy solutions on the other hand, can be found (if they exist) for any W(x) by using the following:
\beq 
H \phi =E \phi \comma H= Q^2
\eeq
So 
\beq 
H \phi =0 \Leftrightarrow Q \phi =0 \; \; .
\eeq

Proof: 
\beq 
0=<0|H|0>=<0|QQ|0>= (|| Q|0> ||) \Rightarrow Q|0> =0 
\eeq
Where we have used that Q is Hermitian.
One can now solve the first order equation to find the zero energy states,
\beq
Q \phi =0
\eeq
\beq
{1 \over {\sqrt{2}}}( \sigma_1 p+ \sigma_2 W'(x)) \phi =0
\eeq
This leads to two independent first order differential equations:
\bea
\lf -{ d \over {dx}} +W'(x) \rt \phi_1(x) &=&0 \\
\lf {d \over {dx}} +W'(x) \rt \phi_0(x)& =& 0
\eea
These can be integrated to give the following solutions:
\bea
\phi_1(x) &=& \phi_1(0) exp( -W(0) ) exp (W(x))\\
\phi_0(x) &=& \phi_0(0) exp(W(0) ) exp(-W(x))
\eea 

A zero energy solution always exists.
Is it a physically acceptable solution? 
If $W(x) \rightarrow \infty$ as $|x| \rightarrow \infty$ then $\phi_0(x)$ is a normalizable solution and one must set $\phi_1(x)=0$. This corresponds to a Bosonic ground state.
If on the other hand $W(x) \rightarrow -\infty$ as $|x| \rightarrow \infty$ then $\phi_1(x)$ is a normalizable solution and one must set $\phi_0$ to zero and obtain a Fermionic  ground state. For $W(x) \rightarrow {\rm{const.}}$ as $|x| \rightarrow \infty$ then both solutions are possible.
There is at most one unpaired solution at E=0.
For $W(x) = \sum_{n=0}^N a_n x^n $ there is an E=0 solution if N is even. There is no $E=0$ solution if N is odd. Only the leading asymptotic behavior matters. The fact that $n_{E=0}$ depends on such global information is very useful. It will appear again in a variety of contexts. 

\subsection{Symmetry and symmetry breaking}

One can break a symmetry explicitly, as it occurs in the Zeeman effect where one introduces a magnetic field in the action that does not possess all the symmetries of the original action. The symmetry may be broken spontaneously. This occurs when the ground state does not posses the symmetry of the action.
Spontaneous symmetry breaking plays an important role in the standard model and in a variety of statistical mechanical systems.

Assume S is a symmetry of the Hamiltonian,  $[H,S]=0$.  Let $|G.S.>$ denote the energy ground state of the system. If the ground state is not invariant under S, that is, if $S|G.S> \neq |G.S.>$ then it is said that the symmetry is spontaneously broken. When the symmetry is continuous, it has generators, g and the symmetry is spontaneously broken if $g|G.S.>\neq 0$. In the case of a continuous symmetry, such a symmetry breaking may result in massless particles (called Goldstone Bosons).

Returning to our model, Q is the generator of a continuous symmetry (supersymmetry). The relevant question is whether $Q |G.S.> =0$, if $Q|0>=0$ then there is a supersymmetric ground state and spontaneous supersymmetry breaking does not occur.

$Q|G.S.> \neq 0 $ iff $E_{G.S.} \neq 0$
and $Q|G.S.>=0$ iff $E_{G.S.} =0 $. For the previously considered potential, $W(x)=\sum^N_ia_i x^i$, one sees that if N is even then there is no spontaneous symmetry breaking since there are zero energy normalizable solutions and conversely if N is odd  then there is spontaneous symmetry breaking.

\begin{figure}[] 
\begin{center}
\includegraphics[width=10.truecm]{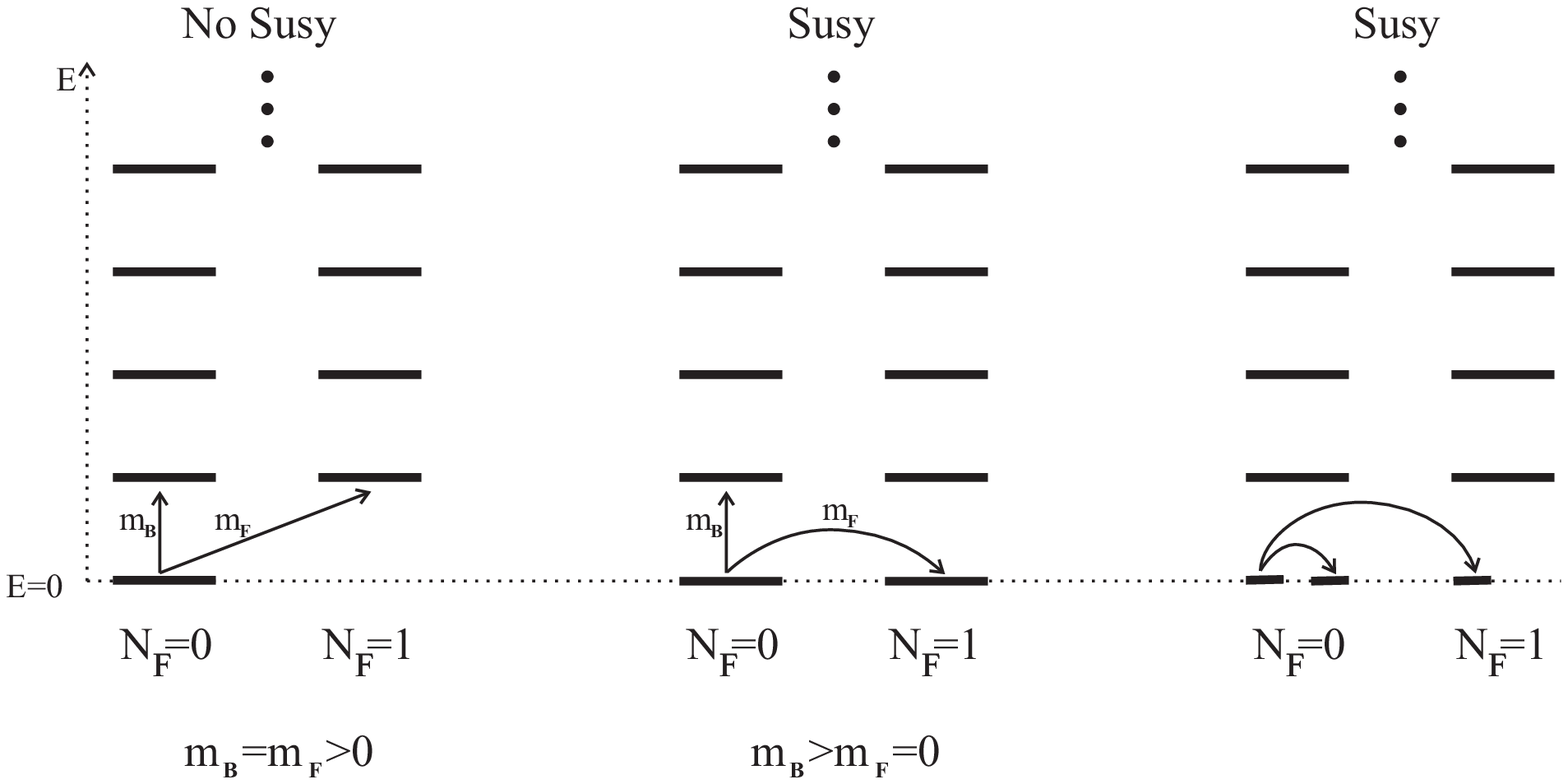} 
\caption[]{Spectra}
\label{figure}
\end{center}
\end{figure}

The consequences on the spectrum are as follows. When there is no spontaneous supersymmetry breaking Fermions and Bosons have the same mass. If supersymmetry is spontaneously broken then the ground state will not have zero energy. Previously it was shown that any state with non-zero energy must be paired with at least one other state (its supersymmetric partner). This implies there will be at least two ground states with identical energies and one must make a choice for the vacuum. For any choice of ground state, there is a zero energy gap between that ground state and the other possible ground state with Fermion number different by one. For example, taking the ground state to be the state with Fermion number zero implies that there is a Fermionic state with the same energy as the ground state thus the Fermion mass will now be zero while the Boson will remain with mass $\Delta$. After spontaneous symmetry breaking the Boson and Fermion masses are no longer equal. The zero mass Fermion is called the Goldstino. It arises from the breaking of a Fermionic symmetry just as the Goldstone Boson arises from breaking a Bosonic symmetry. The Goldstone Boson and the Goldstino have special low energy couplings.

\subsection{A  nonrenormalisation theorem}

Consider now the  system from a perturbative perspective.
The Hamiltonian is
\beq
H={\half} \lf - \hbar ^2 {{\pl^2} \over {\pl x^2}} + W'(x)^2 \rt 1 + {\half} \sigma_3 W''(x) \; \; .
\eeq
First we will do a classical analysis and then we will examine the $\hbar$ corrections to the classical result.
The classical limit is taken by $ \hbar \rightarrow 0$.
\beq
H \rightarrow {\half} (W'(x))^2 \geq 0
\eeq
For N even $W'(x)$ has at least one zero where  $W(x_0) =0$. Classically therefore the ground state has zero energy. This the same as the exact result calculated previously by solving the Schroedinger equation! So how do classical results become exact?
First let us note that in the Bosonic case, this does not work. For example, the ground state of the simple harmonic oscillator has an energy ${\half} \hbar \omega$ above the classical ground state energy.
More generally for the Bosonic Hamiltonian:
\beq
H_B={\half} (p^2 + V(x)) \comma
\eeq
when $x_0$ is a minimum of V(x) so that $ E_{cl}= {\half} V(x_0)$ we have a perturbed energy given by:
\beq
 E_{pert}= E_{cl} + {\half} \hbar V''(x_0)  \; \; . 
\eeq

In a theory with Fermions though, due to the presence of a Yukawa coupling, there is another source for $\hbar$ correction.

This is analogous to how the zero point energies of a free oscillator cancel between the Bosons and Fermions. 

Calculating the perturbative energy of the ground state for a supersymmetric theory we have:
\beq
E^{SUSY}_{pert}= {\half} ( (W'(x_0))^2 + \hbar W''(x_0) - \hbar W''(x_0) )
\eeq
Thus the Bosonic correction is canceled by the Fermionic correction and the classical result is exact.

The result is true to all orders in perturbation theory. This generalizes under some circumstances to supersymmetric field theories and is known by the name ``nonrenormalisation theorem'' \cite{West:1976wz}.
As classical results are easier to obtain, this can be made into a very powerful tool.

Consider the case $W(x)=x^3$, N is odd and the supersymmetry is broken; there is no ground state with $E=0$. However, $V_{cl}= {\half} (3x^2)^2$ and $E^{G.S.}_{cl}=0$. In this case the classical result is not exact. There are nonperturbative effects that provide corrections. 

In order to see how nonperturbative effects become relevant we will actually consider the following:
\beq
W(x)={1 \over 3}x^3 +a x     \, \qquad a<0 .
\eeq
This also has $E=0$ classical solutions and the leading term in W(x) is odd thus there are no exact quantum $E=0$ states; the potential is:
\beq
|W(x)'|^2=(x^2+a)^2
\eeq
This potential clearly has a vacuum degeneracy. We will label the vacuum states $|+> ,|->$. Perturbatively, the energies of these two states are equal. The energy eigenstates are:
\bea
|S> &=& { 1\over {\sqrt{2}}} (|+> + |->)\\
|A> &=& { 1 \over {\sqrt{2}}} (|+> - |-> ) 
\eea
$E(|A>)-E(|S>)>0$ and the ground state is the symmetric state; its energy gap is:
\beq
\Delta E= {\rm{a \; \; exp}}(-{c \over \lambda})
\eeq
$a,c$ are numerical coefficients. This is a tunneling phenomenon (essentially a nonperturbative effect) that is associated with the presence of instantons.

Note that for $a>0$ there are no zero energy solutions already classically.

\subsection{A two variable realization and flat potentials}

Consider a two  variable realization of supersymmetry. The supercharge is given by:
\beq
Q= \sum^2_{\alpha=1} \psi^+_\alpha (-p_{\alpha} + i {{\pl W }\over {\pl x_\alpha}})
\eeq
where $W(x,y)$ is a general function of the Bosonic variables x,y. One can realize $\psi_1$ and $\psi_2$ by the following matrices:
\bea
\psi_1=\lf \matrix{0&1&0&0 \cr
                   0&0&0&0 \cr
                   0&0&0&1 \cr
                   0&0&0&0 } \rt
\eea

\bea 
\psi_2= \lf \matrix{0&0&-1&0 \cr
                    0&0&0&-1 \cr
                    0&0&0&0 \cr
                    0&0&0&0 } \rt
\eea

The Hamiltonian has the following structure:
\bea
H=\lf \matrix{ H_0 &0& 0& 0 \cr
             0  & H_{11}& H_{12}&0 \cr 
             0 & H_{21} &H_{22} & 0 \cr
	     0& 0& 0& H_2 \cr } \rt
\eea

The possible states are: 
\bea
(1,1) &=& a_{F_1}^+ a_{F_2}^+ |00> \\
(0,1); (1,0) &=&  a_{F_1}^+ |00> ,\; a_{F_2}^+ |00> \\
(0,0) &=& |00>
\eea

With,

\bea
H_0 &= &{\half} [ - \Delta + (\nabla W)^2 - \Delta W] \\
H_2 &=&{\half} [ -\Delta + (\nabla W)^2 + \Delta W]
\eea

\bea
H_1 = {\half} \lf ( - \Delta + (\nabla W)^2 - \Delta W) 1 + 2 \lf \matrix{ \pl_{11} W & \pl_{12} W \cr \pl_{21} W & \pl_{22} W } \rt \rt
\eea

For any W the n=0,2 sectors can be solved exactly for E=0 just as before. 

Consider the potential $W= x(y^2+c)$ then $V_{cl}= {\half} [ (y^2 +c)^22 +4x^2 y^2]$. For $c>0$, $V_{cl}={\half} c^2 > 0 $ which leads to classical SUSY breaking \cite{Forge:kg}.

Note, $V_{cl}={\half} c^2 $ at $y=0$. This is a ``flat direction'', the potential is the same for all values of x. Flat directions imply the presence of many vacua. Such flat directions appear in abundance in supersymmetric models and lead in some contexts to supersymmetry breaking. Flat directions in purely Bosonic models are lifted by quantum fluctuations. In supersymmetric systems the Bosonic and Fermionic fluctuations cancel and the flat directions remain.

Classically the ground state is non zero which implies  that the supersymmetry is broken. What about quantum mechanical effects, can supersymmetry actually be restored? (There are cases where symmetries have been known to be restored quantum mechanically.)

For n=0,2 sectors, the answer is no; neither ground state is normalisable. What about n=1 sector? To find these states we need to solve the following  pseudo-analytic equations:
\bea
\pl_1 (e^W \phi_{1,0}) &=& \pl_2 (e^W \phi_{0,1}) \\
\pl_1(e^{-W} \phi_{0,1} )&=& -\pl_2 (e^{-W} \phi_{1,0})
\eea
Unfortunately, these cannot be solved in general. 
We will now show that from more general arguments that supersymmetry cannot be restored by quantum effects.

\beq
E^W = {{<\phi|H|\phi>} \over {<\phi|\phi>} }
\eeq
Consider perturbing the potential, 
\beq
W^\lambda = W^0 + \lambda x^i n^i
\eeq
where $n^i n^i =1$ and $W^0 = x(y^2 +c)$ is the unperturbed potential.
\beq
E^W(\lambda)= E^W(0) + {\half} \lambda^2 + \lambda \int d^2x (n \cdot \nabla W) \phi^2 
\eeq
where $\phi$ denotes the unperturbed solution. (There is no $\lambda$ dependence from the Yukawa term.)
$\pl_x W = y^2 +x >0 \comma \phi^2 \geq 0$ and $\int d^2x \phi^2 =1$,
thus, 
\beq
\int d^2x (n \cdot \nabla W) \phi^2  >0
\eeq
So we may write:
\beq
E^W(\lambda)= E^W(0) + {\half} \lambda^2 + \lambda r^2
\eeq
where r is some finite real quantity.

$E^{W(\lambda )} \geq 0$; if we now assume $E^{W(0)} =0$ then we obtain a contradiction because by taking $\lambda$ to be small and negative we obtain $E^{W(\lambda )} <0$. 

This argument is rather general and also works for the 3 variable potential 
\beq
W=a yz +bx(y^2+c)    \label{orafm}
\eeq
(this is the potential used to break supersymmetry spontaneously in scalar field theories.)

There is a short more elegant argument for non-restoration of supersymmetry that is based on an index theorem.

The ``Witten index'', \cite{Witten:df} is defined by,

\beq
\Delta = Tr(-1)^F = \sum_{E} (n^{F=even}(E) -n^{F=odd}(E))
\eeq
The trace indicates a sum over all states in the Hilbert space, F is the fermion number of the state.  $n^{F=even}(E),\; n^{F=odd}(E)$ denote the number of solutions with Fermi number even/odd with energy E.
Since the Bosons and fermions are paired at all energies greater than zero then
\beq
\Delta= n^{F=even} (E=0) - n^{F=odd} (E=0)
\eeq
If $\Delta$ is calculated for some potential W it will not change under perturbations of W (only $E_n$ will change). In particular, if $\Delta\neq 0$ for some W then there will be no spontaneous symmetry breaking under any perturbations in W.

On the other hand if $\Delta=0$ then we do not know whether SUSY may be broken or not since either 
\beq
n^{F=even} = n^{F=odd} \neq 0
\eeq or 
\beq
n^{F=even} = n^{F=odd}=0 \; \; . 
\eeq
Thus for $\Delta=0$ one needs more information.

\begin{figure}[hbtp] 
\centering
\includegraphics[width=10.truecm]{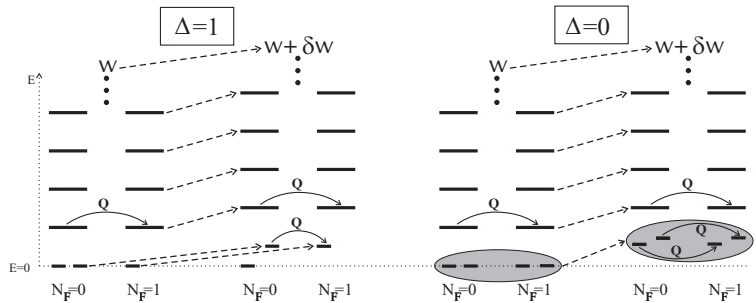} 
\caption[]{Perturbing the spectrum; $\Delta$ is unchanged}
\label{figW} 
\end{figure}

Returning to the case described above. We have calculated the classical Witten Index to be: 
\beq
\Delta_{cl} =0  \; \; .
\eeq
One considers turning on quantum corrections as a perturbation. Since the index is invariant under perturbations, one obtains:
\beq
0= \Delta_{cl}=\Delta_{quantum}=n^q_{N_F=0}(E=0)+ n^q_{N_F=2}(E=0) - n^q_{N_F=1}(E=0)  \; \; .
\eeq
We have already shown that 
\beq 
n^q_{N_F=0}(E=0)= n^q_{N_F=2}(E=0)=0
\eeq
 which in turn implies
\beq
n_{N_F=1}(E=0)=0 \comma
\eeq
thus, there is no supersymmetry restoration.

\subsection{Geometric meaning of the Witten Index}
Consider a supersymmetry sigma model with a D dimensional target space whose metric is $g_{ij}$ \cite{Witten:df},
\beq
S=\int dt g_{ij}(x){{dx^i} \over {dt}} {{dx^i} \over {dt}} +i \bar{\psi} \gamma^0 D_t \psi_i + {1 \over 4} R_{ijkl} \psi^i \psi^k \psi^j \psi^l \comma
\eeq
$D_t $ denotes a covariant derivative, $R_{ijkl}$ is the curvature.
\beq
\{ \psi_i, \psi_j \} =\{\psi^*_i, \psi_j^* \}=0 \comma \{\psi_i,\psi_j^* \} = g_{ij} \comma
\eeq
\beq
Q= i \sum^D_i= \phi_i^* p_i  \comma \qquad Q^* = - i \sum^D_i= \phi_i^* p_i
\eeq
and
\beq
p_i=- D_{x^i} \; \; .
\eeq

The Hilbert space may be graded according to the Fermion occupation number:
\bea
\matrix{
\phi_0&=&|0,..,0> &  \rm{Dim} &=&1       \cr 
\phi_1&=&\{ |0,..,1,..,0>\} & \rm{Dim} &=&D    \cr
\phi_2&=&\{ |0,..,1,..,1,..,0> \} & \rm{Dim} &=& {\half} D(D-1) \cr
\cdot & & \cdot \cr
\cdot & & \cdot \cr
\phi_D&=& |1,1,..,1> &\rm{Dim} &=& 1}
\eea
Generically the Dimension of a state with p-Fermions is given by:
\beq
\rm{Dim(p,D)}= {{D!} \over {(D-p)!p!}}
\eeq
This is identical to the dimension of p-forms in D dimensions.
Q acts by adding a Fermion hence is a map
\beq
Q: \phi_p \rightarrow \phi_{p+1}
\eeq

From simple Fermi statistics we see that, Q is nilpotent.
Hence, we have an isomorphism between p-forms with an exterior derivative, d and the space of states with Fermion occupation number p and supercharge Q.

As with any Nilpotent operator one can examine its cohomology defined by:
\beq
H^p = {{ \{ d \phi_p =0 \} }\over { \{ \phi_p =d \phi_{p-1} \} }}
\eeq

The dimension of the cohomology is denoted by the Betti number $b^p ={\rm{dim}} H^p$.
The Euler characteristic is then given by the alternating sum of the Betti numbers.
\beq
\chi=\sum_p (-1)^p b^p
\eeq

Recall that to find the E=0 states, one solved the equation $Q \phi_p =0$, hence the E=0 states are in the cohomology of Q. 
Thus one has the remarkable correspondence,
\beq
Tr(-1)^F = \chi \; .
\eeq
The Euler characteristic is a topological invariant and is independent of geometrical perturbations of the manifold. This explains why the Witten index is stable against non-singular perturbations. The Witten Index is given a very physical realization in the following example.

\subsection{Landau levels and SUSY QM}

Consider an electron moving in two dimensions in the presence of a perpendicular magnetic field.
\beq
H={\half} [ (p_x+A_x)^2 1 + (p_y +A_y)^2 1 + B_z \sigma_3 ]
\eeq
Two Bosons x,y and one Fermion. This is a less familiar realization of supersymmetry as the number of Bosonic and Fermionic oscillators differ. It is particular to the quantum mechanical system.
The supercharges are:

\bea
Q^1 &=& {1 \over {\sqrt{2}}} \lf  (p_x+A_x)\sigma_1 + (p_y+A_y)\sigma_2 \rt \\
Q^2 &=& {1 \over {\sqrt{2}}} \lf  (p_x+A_x)\sigma_2 - (p_y+A_y)\sigma_1 \rt
\eea
\beq
[Q^i,H]=0 \comma \{ Q^i, \sigma_3 \} =0
\eeq
In Lorentz gauge, $\pl_i A^i=0$ we can write $A^i = \epsilon^{ij} \pl_j a$
$B_z=- \nabla^2 a$.
Take $Q=Q^1$, Q is just like the Dirac operator.

To find E=0 states we again use $Q \phi =0$.
Multiplying by $\sigma_1$ gives:
\bea
\lf (p_x +A_x) 1 +i \sigma_3(p_y+A_y) \rt \lf \matrix{\phi_1 \cr \phi_{-1}} \rt = \lf \matrix{ 0 \cr 0} \rt \; \; .
\eea 
One can solve these equations as follows:
\beq
\phi_1 (x,y) =exp(-a) f(x+iy) \comma \phi_{-1} = exp(a) g(x-iy)
\eeq
where f and g are arbitrary functions. The normalisability of these solutions depends on the function a.

In a constant magnetic field, taking for example, $a= -{\half} y^2 B$.
Depending on the sign of B, either $\phi_1$ or $\phi_{-1}$ may be normalizable.
If $B>0$, take $\phi_1=0$ and 
\beq
\phi_{-1} =exp(-{\half} y^2 B) g(x-iy)
\eeq
From translational invariance, a convenient choice is
\beq
g(x-iy)=exp(ik(x-iy))
\eeq
There are an infinite number of E=0 states. (This is true also for any E not just $E\neq 0$.)

Let us examine this from a topological point of view.
The total magnetic flux is:
\beq
\Phi= {1 \over {2 \pi}} \int d^2 r B_z = -{1\over {2\pi}} \int_0^{2 \pi} d \theta r \pl_r a \mid_{r=\infty} \; \; .
\eeq

 $\Phi=0$ implies $a(r=\infty)=0$ and one has only plane wave normalizable states. Take a constant negative B field with $a= - { 1\over 4} r^2 B$.
The solution is:
\beq
\phi_{+1}= {c \over {r^{|\Phi|}}} r^n exp(i n \theta)
\eeq

For B constant $\Phi \rightarrow \infty$ and there are an infinite number of allowed states, ie. all n. Now assume that $\Phi$ is finite because B is confined in a solenoid, for example. In such a case, normalisability requires that n is bounded by $[\Phi -1]$. ($\Phi$ is quantized). Thus the number of E=0 states is given by the total flux number, 
\beq
n_{(E=0)}= |\Phi|  \; \; .
\eeq
The total magnetic flux is a global  quantity that is independent of local fluctuations. Again we have shown how the number of zero energy states does not depend on the local details but only upon {\it{global}} information.

\subsection{Conformal Quantum Mechanics}

Relevant material for this section may be found in \cite{deAlfaro:1976je,Akulov:uh,Fubini:1984hf}. Recall the Hamiltonian:
\beq
H={\half} (p^2 +g x^{-2}) \label{cqm}
\eeq
is special since g has a meaning in that it does more than determine an energy scale.
H is part of the following algebra:
\beq
[H,D] = iH \comma [K,D] =iK \comma [H,K] =2 i D  \; \; .
\eeq
This forms an SO(2,1) algebra where:
\beq 
D= -{1 \over 4} (xp+px)  \comma K={\half} x^2
\eeq
and H is as given above.
The Casimir is given by:
\beq
{\half} (HK +KH) -D^2 = {g \over 4} - {3  \over {16}}
\eeq
The meaning of D and K is perhaps clearer in the Lagrangian formalism:
\beq
{\cal{L}}= {\half} ( \dot{x}^2 - {g \over {x^2}} ) \comma S= \int dt {\cal{L}}
\eeq
Symmetries of the action S, and not the Lagrangian ${\cal{L}}$ alone,  are given by:
\beq
t' = { {at+b} \over {ct+d}}  \comma x'(t')= {1 \over {ct +d}} x(t) 
\eeq
\bea
A= \lf \matrix{a &b \cr c&d\cr} \rt \comma \qquad \rm{det} A = ad-bc=1
\eea
H acts as translation
\bea
A_T=\lf \matrix{1&0\cr\delta &1 } \rt \comma \qquad t'=t+\delta
\eea
D acts as dilation
\bea
A_D= \lf  \matrix{ \alpha &0 \cr 0& {1 \over \alpha}} \rt \comma \qquad t'=\alpha^2 t 
\eea
K acts as a special conformal transformation
\bea
A_K= \lf \matrix{1 & \delta \cr 0&1} \rt \comma \qquad t'= {t \over {\delta t+1}}
\eea

The spectrum of the Hamiltonian \ref{cqm} is the open set $(0,\infty)$, the spectrum is therefore continuous and bounded from below. The wave functions are given by:
\beq
\psi_E(x)= \sqrt{x} J_{\sqrt{g +{1 \over 4}}}(\sqrt{2E} x)  \; \; \quad E\neq 0 \; \; .
\eeq

We will now attempt to find the zero energy state. Take the ansatz $\phi(x)=x^\alpha$:
\beq
H= \lf - {{d^2} \over {dx^2}} + {{g} \over {x^2}} \rt x^\alpha =0 \; \; . 
\eeq
This implies
\beq
g= -\alpha(\alpha-1) 
\eeq
solving this equation gives
\beq
\alpha=-{\half} \pm {{\sqrt{1+4g}} \over 2}
\eeq
This gives two independent solutions and by completeness all the solutions.
$\alpha_+ >0$,  does not lead to a normalizable solution since the function diverges at infinity.
$\alpha_- <0$, is not normalizable either since the function diverges at the origin (a result of the scale symmetry).

Thus, there is no normalizable E=0 solution (not even plane wave normalizable)!

\begin{figure}[hbtp] 
\centering
\includegraphics[width=8.0 truecm]{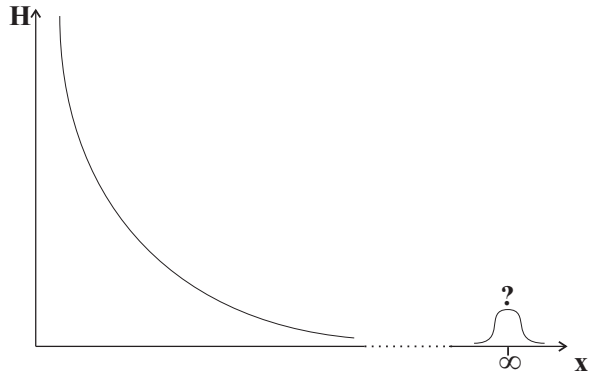} 
\caption[]{The absence of a normalisable ground state for this potential}
\label{figure} \end{figure}

Most of the analysis in field theory proceeds by identifying a ground state and the fluctuations around it.
How do we deal with a system in the absence of a ground state?

One possibility is to accept this as a fact of life. Perhaps it is possible to view this as similar to cosmological models that also lack a ground state such those with Quintessence.

Another possibility is to define a new evolution operator that does have a ground state.
\beq
G= u H +v D +w K
\eeq
 This operator has a ground state if $v^2-4uw <0$.
Any choice explicitly breaks Lorentz and scale invariance.
Take for example,
\beq
G={\half} ({1 \over a} K+ a H) \equiv R  \label{Rop}
\eeq
a has dimension of length. The eigenvalues of R are
\beq
r_n=r_0+ n \comma \qquad r_0={\half} (1 +\sqrt{g+{1\over4}})
\eeq
 
One must interpret what this means physically. Surprisingly this question arises in the context of black hole physics.

Consider a particle of mass m and charge q falling into a charged black hole. The black hole is BPS meaning that its mass, M and charge, Q are related by $M=Q$. The blackhole metric and vector potential are given by:
\beq
ds^2=-(1+{M \over r})^{-2} dt^2 +( 1+{M \over r})^2 (dr^2 +r^2 d\Omega^2) \comma \quad A_t= {r \over M}
\eeq

Now consider the near Horizon limit ie.$ r <<M$, which we will take by $M \rightarrow \infty$ keeping r fixed.
\beq
ds^2=-({r \over M})^2 dt^2 +({M \over r})^2 dr^2 + M^2 d \Omega^2
\eeq

This produces an $AdS_2 \times S^2$ geometry.

We also wish to keep fixed $M^2 (m-q)$ as we scale M. This means we must scale $(m-q) \rightarrow 0$. That is the particle itself becomes BPS in the limit.
The Hamiltonian for this in falling particle in this limit is given by our old friend:
\beq
H= {{p_r^2} \over {2m}} + { g \over {2r^2}} \comma \quad g= 8M^2(m-q) + {{4l(l+1)} \over M}
\eeq
For l=0, we have $g> 0$ and there is no ground state. This is associated with the coordinate singularity at the Horizon. The change in evolution operator is now associated with a change of time coordinate. One for which the world line of a static particle passes through the black hole horizon instead of remaining in the exterior of the space time.

\subsection{Superconformal quantum mechanics}

The bosonic conformal mechanical system had no ground state.
The absence of a $E=0$ ground state in the supersymmetric context leads to the breaking of supersymmetry. This breaking has a different flavor from that which was discussed earlier. Next the supersymmetric version of conformal quantum mechanics is examined to see if supersymmetry is indeed broken.
For this the superpotential is chosen to be,
\beq
W(x)={\half} g\; \log x^2 \comma
\eeq
yielding a Hamiltonian:
\beq
H= {\half} \lf \lf p^2 + \lf {{dw} \over {dx}}\rt^2 \rt 1- \sigma_3 {{d^2W} \over {dx^2}} \rt  \; \; . 
\eeq

Representing $\psi$ by$ {\half} \sigma_-$  and $\psi^*$by ${\half} \sigma_+$ gives the supercharges:
\beq
Q=\psi^+(-ip + {{dW } \over {dx}}) \comma Q^+=\psi (ip + {{dw }\over {dx}})
\eeq
One now has a larger algebra, the superconformal algebra,
\bea
\{ Q,Q^+ \} &=&2H  \comma \{ Q, S^+ \} =g-B +2iD \\
\{ S,S^+ \} &=&2K   \comma \{ Q^+,S \}=g-B-2iD
\eea
A realization is:
\beq
B=\sigma_3 \comma S=\psi^+ x \comma S^+ = \psi x \; \; .
\eeq

The zero energy solutions are
\beq
exp(\pm W(x) ) = x^{\pm g} \comma
\eeq
neither solution is normalizable.

H factorizes:
\bea
2H= \lf \matrix{ p^2 +{{g(g+1)} \over x^2}  & 0 \cr 0 & p^2 +{{ g(g-1)} \over x^2} } \rt
\eea
and we may solve for the full spectrum:
\beq 
\psi_E(x)= x^{\half} J_{\sqrt{\nu} }(x\sqrt{2E} )\comma \qquad E \neq 0 \comma
\eeq
Where $\nu=g(g-1)+{1\over 4}$ for $N_F=0$ and $\nu=g(g+1)+{1 \over 4}$ for $N_F=1$. Again the spectrum is continuous though there is no normalizable zero energy state.
Again we must interpret the absence of a normalizable ground state.
One can again define a new operator with a normalizable ground state. By inspection the operator \ref{Rop} can be used provided one makes the following identifications:
\bea
N_F&=&1 \qquad g_B =g_{susy}(g_{susy}+1)\comma \\
N_F&=&0 \qquad g_B=g_{susy}(g_{susy}-1)
\eea
Thus the spectrum differs between the $N_F=1$ and $N_F=0$ sectors and so supersymmetry would be broken. One needs to define a whole new set of operators:
\bea
M=Q-S \qquad M^+=Q^+-S^+ \\
N=Q^+ +S^+ \qquad N^+=Q+S^+
\eea
which produces the algebra:
\bea
{1 \over 4} \{ M, M^+ \}&=& R + {\half} B - {\half}  g \equiv T_1 \\
{1 \over 4} \{ N, N^+ \}&=& R + {\half} B + {\half}  g \equiv T_2 \\
{1 \over 4} \{ M, N \} &=& L_- \qquad  {1 \over 4} \{ M^+,N^+ \} = L_+ \\
L_\pm &=& -{ \half} (H- K \mp 2 i D)
\eea
$T_1, T_2, H$ have a doublet spectra. ``Ground states'' are given by:
\beq
T_1|0>=0 \; ; \quad T_2 |0>=0 \; ; \quad  H|0>=0 \; \; .
\eeq
A physical context arises when one considers a supersymmetric particle falling into a black hole \cite{Claus:1998ts,Kallosh:1999mi}. This is the supersymmetric analogue of the situation already discussed.

\section{Review of Supersymmetric Models}
\subsection{Kinematics}

For a more detailed discussion of the material presented in this section see \cite{Bagger:1990qh}. The possible particle content of supersymmetric (SUSY) theories is determined by the SUSY algebra, its N=1 version is:
\beq
\{ Q_\alpha, \bar{Q}_{\dot{\alpha}} \} = 2 \sigma_{\alpha \dot{\beta}}^\mu P_\mu \comma 
\{ Q_\alpha, Q_\beta \} = \{ {\bar{Q}}_{\dot{\alpha}} , {\bar{Q}}_{\dot{\beta}} \} = 0  \; \; .
\eeq
\beq
[P_\mu,Q_\alpha] = [ P_\mu, {\bar{Q}}_{\dot{\alpha}} ] =[P_{\mu},P_{\nu}]=0
\eeq
This algebra can be generalized to include a higher number of supersymmetries $N=2,4$ by:
\beq
\{ Q^i_\alpha, \bar{Q^j}_{\dot{\alpha}} \} = 2 \delta^{ij} \sigma_{\alpha \dot{\beta}}^\mu P_\mu + \delta_{\alpha \dot{\beta} } U_{ij} + (\gamma_5 )_{\alpha \dot{\beta}} V_{ij} \comma
\eeq
U and V are the central charges ie. they commute with all other charges (they are antisymmetric in ij). They are associated with BPS states such as monopoles. We will discuss in this section the d=4 realisations with, $\mu, \nu=0,1,2,3$ the space-time indices. In four dimensions we have two component Weyl Fermions. Those with $\alpha$ or $\beta$ indices transform under the $(0,{\half})$ representation of the Lorentz group; and those with dotted indices,$ {\dot{\alpha}}$ or ${\dot{\beta}}$ transform under the $({\half},0)$ representation.

Consider first the massless representations of N=1 supersymmetry. 
The simplest is the the chiral multiplet: 
\beq
(-{\half} ,0,0,{\half}) \qquad (\phi,\psi)  \qquad (2,2)  
\eeq
In the above table, first are written the helicities; then the associated component fields, $\phi$ denotes a complex scalar and $\psi$ a Weyl Fermion; and finally are the number of physical degrees of freedom carried by the Bosons and Fermions.
The vector multiplet has:
\beq
(-1,-{\half}, {\half},1) \qquad (\lambda_\alpha , A_\mu)  \qquad (2,2)  
\eeq
$\lambda $ is a Weyl Fermion and $A_{\mu}$ is a vector field.
With N=2 supersymmetry, there is a massless vector multiplet:
\bea 
  \lf -1 \matrix{-{\half}&0&{\half} \cr -{\half} &0& {\half} } 1 \rt \qquad (\phi,\psi) + (\lambda_\alpha,A_\mu)  \qquad (4,4)
\eea
and a massless hypermultiplet which is given by:
\bea
 \lf \matrix{ -{\half} \cr -{\half} } \comma \matrix{0 \cr 0 \cr 0 \cr 0 } \comma \matrix{{\half} \cr {\half}} \rt \qquad (\phi_1,\psi_1) +(\phi_2,\psi_2) \qquad (4,4) \; \; .
\eea

For Massive multiplets, in $N=1$, there is again the chiral multiplet which is the same as the massless multiplet but with now massive fields. The massive vector multiplet becomes: 
\bea
\lf -1 \matrix{-{\half}&0&{\half} \cr -{\half} &0& {\half} } 1 \rt \qquad
(h,\psi_\alpha,\lambda_\alpha, A_{\mu}) \qquad
(4,4)
\eea
Where h is a real scalar field. The massive vector multiplet has a different field content than the massless vector multiplet because a massive vector field has an additional physical degree of freedom. One sees that the massive vector multiplet is composed out of a massless chiral plus massless vector multiplet. This can occur dynamically; massive vector multiplets may appear by a supersymmetric analogue of the Higgs mechanism.
With N=4 supersymmetry, the massless vector multiplet is:
\bea
\lf -1 \comma \matrix{ -{\half} \cr -{\half} \cr-{\half} \cr-{\half} } \comma \matrix{0 \cr 0\cr 0 \cr0 \cr 0 \cr 0} \comma \matrix{{\half} \cr {\half} \cr{\half} \cr{\half} } \comma 1 \rt \quad (\lambda^a,\phi^I,A_\mu) \qquad (8,8)
\eea
where $I=1..6  ,\; a=1..4$.

\subsection{Superspace and Chiral fields}

Spacetime can be extended to include Grassmann spinor coordinates, ${\bar{\theta}}_{\ad} ,\theta_\alpha$.  Superfields are then functions of this superspace.
Constructing a Lagrangian out of superfields provides a useful way to construct explicitly supersymmetric Lagrangians.
Recall the integration formulas for Grassmann variables:
\beq
\int d \theta_{\alpha} \theta_{\alpha} = {{\pl} \over { \pl \theta_{\alpha} }} =1 \comma \int d \theta_{\alpha} =0 
\eeq
The following identity will be of use:
\beq
\int d^2 \theta d^2 {\bar{\theta}} {\cal{L}} = \int d^2 \theta { {\pl^2 {\cal{L}}} \over {\pl {\bar{\theta}}_1 {\bar{\theta}}_2 }}
\eeq
The supercharges can be realized in superspace by generators of supertranslations:
\beq
Q_{\alpha}= {\pl \over {\pl \theta_{\alpha}}} - i \sigma^{\mu}_{\aad}  {\bar{\theta}}^{\ad} \pl_{\mu}  \comma \quad {\bar{Q}}_{\ad} = -{\pl \over {\pl {\bar{\theta}}_{\ad}}} + i \theta^{\alpha} \sigma^{\mu}_{\aad}  \pl_{\mu} \; .
\eeq
It will also be useful to define a supercovariant derivative:
\beq
D_{\alpha}= {\pl \over {\pl \theta_{\alpha}}} + i \sigma^{\mu}_{\aad}  {\bar{\theta}}^{\ad} \pl_{\mu}  \comma \quad {\bar{D}}_{\ad} = -{\pl \over {\pl {\bar{\theta}}_{\ad}}} - i \theta^{\alpha} \sigma^{\mu}_{\aad}  \pl_{\mu} \; .
\eeq
A superfield $\Phi$ is called  ``chiral'' if:
\beq
{\bar{D}}_{\ad} \Phi =0 \; .
\eeq
One introduces the variable,
\beq
y^\mu = x^{\mu} + i \theta \sigma ^{\mu} {\bar{\theta}}
\eeq
which produces the following expansion for a chiral field,
\beq
\Phi(y) = A(y) +\sqrt{2} \theta \psi(y) + \theta \theta F(y)
\eeq
The Taylor expansion terminates because of the anticommuting property of the Grassmann coordinates.
As a function of x it may be written as follows: 
\bea
\Phi(x) &=& A(x) + i \theta \sigma^\mu {\bar{\theta}} \pl_\mu A(x) + {1 \over 4} \theta \theta {\bar{\theta}} {\bar{\theta}} \Box A(x) + \sqrt{2} \theta \psi(x) \\ &-& {i \over {\sqrt{2}}} \theta \theta \pl_\mu \psi(x) \sigma^{\mu} {\bar{\theta}} + \theta \theta F(x)
\eea
The key point is that 
\beq
{\cal{L}}= \int d^2 \theta  \Phi(x)
\eeq
is a invariant under supersymmetric transformations (up to a total derivative).

After the integration some terms will disappear from the expansion of $\Phi(x)$ leaving only:
\beq
\Phi(x)= A(x) + \sqrt{2} \theta \psi(x) + \theta \theta F(x)
\eeq
A(x) will be associate with a complex Boson; $\psi(x)$ will be associated with a Weyl Fermion and F(x) acts as an auxiliary field that contributes no physical degrees of freedom. These are called the component fields of the superfield.
The product of two chiral fields also produces a chiral field. Therefore, any polynomial , $W(\Phi)$ can be used to construct a supersymmetry invariant as 
\beq
{\cal{L}}= \int d^2 \theta W(\Phi) = F_{W(\Phi)}
\eeq
is a supersymmetry invariant. This is used to provide a potential for the chiral field.
The kinetic terms are described by:
\beq
\int d^2 \theta d^2 {\bar{\theta}} {\bar{\Phi}}_i \Phi_j = {\bar{\Phi}}_i \Phi_j \mid_{\theta \theta {\bar{\theta}} {\bar{\theta}}}
\eeq
After expanding and extracting the $\theta \theta {\bar{\theta}} {\bar{\theta}}$ term is (up to total derivatives):
\beq
F^*_iF_f -|\pl_\mu A|^2 + {i\over 2} \pl_\mu {\bar{\psi}} {\bar{\sigma}}^{\mu} \psi 
\eeq
One thus composes the following Lagrangian:
\bea
{\cal{L}}&=& {\bar{\Phi}}_i \Phi_i  \mid _{\theta \theta {\bar{\theta}} {\bar{\theta}}} + [\lambda_i \Phi_i + {\half} m_{ij} \Phi_i \Phi_j +{1 \over 3} g_{ijk}\Phi_i \Phi_j \Phi_k ]_{\theta \theta}   \\  
&=& i \pl {\bar{\psi}}_i {\bar{\sigma}} \psi_i + A^*_i \Box A_i + F_i^* F_i + \nn  \lambda _i F_i + m_{ij} ( A_i F_j - {\half} \psi_i \psi_j ) \\ &+& g_{ijk} (A_i A_j F_k - \psi_i \psi_j A_k) + h.c.
\eea
One must now eliminate the auxiliary fields $F_i , F_i^*$.
The equation of motion for $F^*_k$ is as follows:
\beq
F_k+\lambda_k^* +m_{ij} A^*_i + g_{ijk}^* A_i^* A_j^*
\eeq
This gives:
\bea
{\cal{L}}&=& i \pl {\bar{\psi}}_i {\bar{\sigma}} \psi_i + A^*_i \Box A_i - {\half} m_{ij}  \psi_i \psi_j  -{\half} m_{ij}^* \psi^*_i \psi^*_j \nn \\ &-& g_{ijk}  \psi_i \psi_j A_k -g^*_{ijk} {\bar{\psi}}_i {\bar{\psi}}_j
A^*_k -F^*_i F_i
\eea
where the last term is a potential for $A, A^*$; these are known as the F terms, $V_F(A^*,A)$. (Note, $V_F \geq 0$). At the ground state this must vanish ie. $V_F(A^*,A)=0$. This in turn implies that $F_i=0 $ for the ground state. Although this is a classical analysis so far, in fact it is true to all orders in perturbation theory as there exists a non renormalization theorem for the effective potential.

\subsection{K\"ahler Potentials}

To describe the supersymmetric Lagrangian for scalar fields spanning a more complicated manifold it is convenient to introduce the following supersymmetry invariant:
\beq
\int d^4 \theta K(\Phi, {\bar{\Phi}}) \label{ka}  \; \; .
\eeq
$K(\Phi,{\bar{\Phi}})$ is called the K\"ahler potential.
One may add any function of $\Phi$ or ${\bar{\Phi}}$ to the integrand since these terms will vanish after integration.
For the usual kinetic terms, K is taken to be given by $K = \Phi {\bar{\Phi}}$ which produces the $-\delta_{ij} \pl_{\mu} A^{*i} \pl^{\mu} A^j$ kinetic terms for the scalars. For the case of a sigma model with a target space whose metric is $g_{ij}$; this metric is related to the K\"ahler potential by:
\beq
g_{ij}= { {\pl^2 K}  \over {\pl {\bar{\Phi}}_i \pl \Phi_j}}  \label{km}  \; \; .
\eeq
The above supersymmetry invariant \ref{ka} which previously gave the usual kinetic terms in the action, produces for general K the action of a supersymmetric sigma model, with the target space metric given by equation \ref{km}.

{\subsection{F-terms}}

In this section we examine the vanishing of the potential generated by the F terms. The issues we are interested in are whether supersymmetry is spontaneously broken; is there a non renormalization theorem; and are there other internal symmetries broken.
\beq
V_F=0 \Leftrightarrow F_i=0 \qquad \forall i \comma
\eeq
are n (complex) equations with n (complex) unknowns. Generically, they have a solution.
Take the example of the one component WZ model, where
\beq
F_1 =-\lambda -m A +g A^2 \; \; .
\eeq
This has a solution. There is no supersymmetry breaking classically. 
The Witten index ${\rm{Tr}}(-1)^F=2$. This implies the classical result is exact. Note, that for $\lambda=0$,
\beq
V=A^*A|(gA-m)|^2
\eeq
and hence there will be a choice of vacuum:
either $<A>={m \over g}$ or $<A>=0$.

The claim is that the form of the effective interacting superpotential $W_{eff}$ will be the same as the classical superpotential $W_{cl}$. 
Take $W_{cl}$ to have the form:
\beq
W= {\half} m \Phi^2 + {1 \over 3} \lambda \Phi^3 \; \; .
\eeq
To show that the form of W remains invariant, the following ingredients are used: the holomorphic dependence of $W(\Phi,m,\lambda)$, that is W is independent of $m^*, \lambda^*, \Phi^*$; and
the global symmetries present in the theory \cite{Seiberg:1993vc}.

\subsection{Global Symmetries}

R-symmetry is a global U(1) symmetry that does not commute with the supersymmetry. The action of the R-symmetry on a superfield $\Phi$ with R-character n as follows. 
\bea
R \Phi(\theta,x)&=& {\rm{exp}}(2in\alpha) \Phi( {\rm{exp}}(-i \alpha \theta),x)\\
R {\bar{\Phi}}({\bar{\theta}},x)&=& {\rm{exp}}(-2in\alpha) \Phi( {\rm{exp}}(i \alpha {\bar{\theta}}),x)
\eea
Since the R-charge does not commute with the supersymmetry, the component fields of the chiral field have different R-charges. For a superfield $\Phi$ with R-character n, the R-charges of the component fields may be read off as follows: 
\beq
R({\rm{lowest \; \; component\; \;of}}\; \Phi)=R(A)\equiv n \comma R(\psi)=n-1 \comma \quad R(F)=n-2
\eeq
The R-charge of the Grassmann variables is given by:
\beq
 R(\theta_\alpha) =1 \comma R(d \theta_\alpha) =-1
\eeq
with, barred variables having opposite R charge. The kinetic term ${\bar{\Phi}} \Phi$ is an R invariant. (${\bar{\theta}}
{\bar{\theta}} \theta \theta$ is an invariant.)
For the potential term,  
\beq
\int d^2 \theta W
\eeq
to have zero R charge requires that $R(W)=2$.
For the mass term $W={\half} m \Phi^2$, 
\beq
W= m\psi \psi + m^2 |A|^2 \comma
\eeq
to have vanishing R-charge requires
\beq
R(\Phi) = R(A)=1 \comma R(\psi)=0  \label{rchg}
\eeq
Adding the cubic term: 
\beq
W_3={\lambda \over 3} \Phi^3
 \eeq
produces
\beq
V=|\lambda|^2 |A|^4 + \lambda A \psi \psi \; \; .
\eeq
This term is not R-invariant with the R-charges given by \ref{rchg}. To restore R-invariance requires  $\lambda$ is assigned an R-charge of -1. This can be viewed as simply a book keeping device or more physically one can view the coupling as the vacuum expectation value of some field. The expectation value inherits the quantum numbers of the field. This is how one treats for example the mass parameters of Fermions in the standard model. There is also one other global U(1) symmetry that commutes with the supersymmetry. All component fields are charged the same with respect to this U(1) symmetry. Demanding that the terms in the action maintain this symmetry requires an assignment of U(1) charges to $\lambda$, and m. 

The charges are summarized in the following table:
\bea
\matrix{    & U(1) & U(1)_R \cr
        \Phi & 1 & 1 \cr
	m & -2 & 0 \cr
	\lambda & -3 & -1 \cr
	W &0 & 2 \cr }
\eea

These symmetries are next used to prove the nonrenormalisation theorem.

\subsection{The effective potential}

That the effective potential be invariant under the global U(1) implies that
\beq
W_{eff}(\Phi,m,\lambda)= g_1(m\Phi^2, \lambda\Phi^3) \comma
\eeq
combining this with the invariance under the U(1) R-symmetry implies
\beq
W_{eff}= m \Phi^2 g ({{\lambda \Phi} \over m})  \; \; .
\eeq
Consider expanding in $|\lambda|<<1$,
\beq
m \Phi^2 g ({{\lambda \Phi} \over m}) = {\half} m \Phi^2 + {1 \over 3}\Phi^3 + \sum_{n=2}^\infty a_n {{\lambda^n \Phi^{n+2} }\over m ^{n-1}}
\eeq

As $\lambda \rightarrow 0$ one must recover the classical potential. This requires that $n \geq 0$. Consider the limit $m \rightarrow 0$, for the function to be free of singularities implies that $n < 1$. There can therefore be no corrections to the form of $W_{eff}$ \cite{Seiberg:1993vc}. In particular, if he classical superpotential has unbroken supersymmetry so does the full theory.  (On the other hand, the Kahler potential is renormalized).

\subsection{Supersymmetry breaking}

Let us examine now how supersymmetry may be spontaneously broken. 
The following anecdote may be of some pedagogical value \cite{story}. It turns out that
a short time after supersymmetry was introduced arguments were published
which claimed to prove that supersymmetry cannot be broken spontaneouly. Supersymmetry resisted breaking attmepts for both theories of scalars and gauge
theories. One could be surprised that the breaking was first achieved in
the gauge systems. This was done by Fayet and Illiopoulos. The presence in
the collaboration of a student who paid little respect to the general
counter arguements made the discovery possible. Fayet went on to discover
the breaking mechanism also in supersymmetric scalar theories as did
O'Raighfeartaigh.

The Fayet-O'Raifeartaigh potential \cite{O'Raifeartaigh:pr,Fayet:1975ki} is the field theory analogue of the potential given by equation \ref{orafm} for supersymmetric quantum mechanics. In order to break supersymmetry a minimum of three chiral fields are needed:

\beq
{\cal{L}}_{Potential}= \lambda \Phi_0 + m \Phi_1 \Phi_2 + g \Phi_0 \Phi_1\Phi_1 + {\rm{h.c.}}
\eeq

Minimizing the potential leads to the following equations:
\bea
0&=& \lambda + g \Phi_1 \Phi_1 \\
0&=& m \Phi_2 +2 g \Phi_0 \Phi_1 \\
0&=& m \Phi_1
\eea
These cannot be consistently solved so there cannot be a zero energy ground state and supersymmetry must be spontaneously broken.
To find the ground state one must write out the full Lagrangian including the kinetic terms in component fields and then minimize. Doing so one discovers that in the ground state $A_1=A_2=0$ and $A_0$ is arbitrary. The arbitrariness of $A_0$ is a flat direction in the potential, like in the quantum mechanical example, \ref{orafm}. Computing the masses by examining the quadratic terms for component fields gives the following spectrum: the six real scalars have masses: $ 0,0,m^2m^2,m^2\pm2g\lambda$; and the Fermions have masses: $0,2m$. The zero mass Fermion is the Goldstino.
We turn next to supersymmetric theories that are gauge invariant.

\subsection{Supersymmetric gauge theories}

A vector superfield contains spin 1 and spin ${\half}$ component fields. It obeys a reality condition $V= {\bar{V}}$. 
\bea
V&=&B+ \theta \chi {\bar{\theta}} {\bar{\chi}} + + \theta^2 C+ {\bar{\theta}}^2 {\bar{C}} - \theta \sigma^{\mu} {\bar{\theta}}  A_\mu \\
&+& i \theta^2 {\bar{\theta}} ({\bar{\lambda}} +{\half} {\bar{\sigma}}^\mu \pl_\mu \chi) -i {\bar{\theta}}^2 \theta (\lambda -{\half} \sigma^\mu \pl_\mu {\bar{\chi}})\\ &+&{\half} \theta^2 {\bar{\theta}}^2 (D^2 +\pl^2B)
\eea
B,D, $A_\mu$ are real and C is complex.
There will be a local U(1) symmetry with gauge parameter, $\Lambda$ an arbitrary chiral field:
\beq
V \rightarrow V + i(\Lambda - {\bar{\Lambda}})  \label{symv}
\eeq
B, $\chi$, and C are gauge artifacts and can be gauged away. The symmetry is actually $U(1)_{\bf{C}}$ as opposed to the usual $U(1)_{\bf{R}}$ because although the vector field transforms with a real gauge parameter, the other fields transform with gauge parameters that depend on the imaginary part of $\Lambda$.

It is possible to construct a chiral superfield, $W_{\alpha}$, from V as follows
\beq
W_\alpha = -{1 \over 4} {\bar{D}} {\bar{D}} D_{\alpha} V \comma {\bar{D}}_{\dot{\beta}} W_{\alpha} =0
\eeq
One may choose a gauge (called the Wess Zumino gauge) in which B, C and $\chi$=0 and then expand in terms of component fields,
\bea
V(y)&=&-\theta \sigma^{\mu} {\bar{\theta}} A_{\mu} + i \theta^2 {\bar{\theta}} {\bar{\lambda}} -i {\bar{\theta}} ^2 \theta \lambda  + {\half} \theta^2 {\bar{\theta}}^2 D  \\
W_\alpha(y) &=& -i \lambda_\alpha +(\delta_\alpha^\beta D -{i \over 2} (\sigma^{\mu}{\bar{\sigma}}^{\nu})^\beta_\alpha F_{\mu \nu} ) \theta_\beta + (\sigma^{\mu} \pl_\mu {\bar{\lambda}})_\alpha \theta^2
\eea
Where $A_{\mu}$ is the vector field, $F_{\mu \nu}$ its field strength, $\lambda$ is the spin ${\half}$ field and D is an auxiliary scalar field. Under the symmetry \ref{symv}, the component fields transform under a now $U(1)_{\bf{R}}$ symmetry as:
\beq
A_{\mu} \rightarrow A_{\mu} - i \pl_{\mu}  (B- B^*) \comma \lambda \rightarrow \lambda \comma D \rightarrow D
\eeq
Note, W is gauge invariant.
The following supersymmetric gauge invariant Lagrangian is then constructed:
\beq
{\cal{L}}=\int d^2 \theta ({{-i \tau \over 16 \pi}}) W^{\alpha} W_{\alpha} + h.c.
\eeq
where the coupling constant $\tau$ is now complex,
\beq
\tau= { \theta \over {2 \pi}} + i { {4\pi} \over g^2}
\eeq
Expanding this in component fields produces,
\beq
{\cal{L}}= {-1 \over {4 g^2}} F_{\mu \nu } F^{\mu \nu} + {1 \over {2 g^2}}D^2-{i \over {g^2}}\lambda \sigma D {\bar{\lambda}} + {\theta \over {32 \pi^2}} (*F)^{\mu \nu} F_{\mu \nu} \; \; . \label{kinv}
\eeq
D is clearly a non propagating field. The $theta$ term couples to the instanton number density (this vanishes for abelian fields in a non-compact space). A monopole in the presence of such a coupling will get electric charge through the Witten effect.
The supersymmetries acting on the component fields are, (up to total derivatives):
\bea
\de A &=& \sqrt{2} \epsilon \psi \\
\de \psi & =& i \sqrt{2} \sigma^\mu {\bar{\epsilon}} \pl_\mu A + \sqrt{2} \epsilon F \\
\de F &= &i \sqrt{2} {\bar{\epsilon}} {\bar{\sigma}}^\mu \pl_\mu \psi \\
\de F_{\mu \nu}&=& i( \epsilon \sigma_{\mu} \pl_\nu {\bar{\lambda}} +{\bar{\epsilon}} {\bar{\sigma}}_\mu  \pl_\nu \lambda) - ( \mu \leftrightarrow \nu) \\
\de \lambda &=& i \epsilon D + \sigma^{\mu \nu} \epsilon F_{\mu \nu} \\
\de D &=& {\bar{\epsilon}} {\bar{\sigma}}^\mu \pl_\mu \lambda - \epsilon \sigma^\mu \pl_\mu {\bar{\lambda}} \; \; .
\eea

One may also add to the action a term linear in the vector field V,  known as a Fayet-Iliopoulos term \cite{Fayet:jb}:
\beq
2K \int d^2 \theta d^2 {\bar{\theta}} V = KD= \int d \theta^\alpha W_\alpha + h.c.
\eeq
Its role will be discussed later.
The U(1)  gauge fields couple to charged chiral matter through the following term
\beq
{\cal{L}}=\sum_i \int d^2 \theta d^2 {\bar{\theta}} {\bar{\Phi}}_i \exp (q_iV) \Phi_i  \label{int}
\eeq
Under the gauge transformation
\beq
V\rightarrow V+ i (\Lambda - {\bar{\Lambda}}) \comma \Phi_i \rightarrow \exp (-i q_i \Lambda) \Phi_i
\eeq
Since there are chiral Fermions there is the the possibility for chiral anomalies. In order that the theory is free from chiral anomalies one requires:
\beq
\sum q_i = \sum q_i^3 =0 \; \; .
\eeq
Writing out the term \ref{int} in components produces:
\bea
{\cal{L}}&=& F^* F - |\pl_\mu \phi + {{i q}\over 2} A_\mu \phi|^2 -i {\bar{\phi}} {\bar{\sigma}} (\pl_\mu + {{i q}\over 2} q A_{\mu}) \psi \\ &-& {{iq} \over {\sqrt{2}}} ( \phi{\bar{\lambda}}{\bar{\psi}} - {\bar{\phi}}\lambda \psi ) +{\half} q D {\bar{\phi}} \phi \; \; .
\eea
There are two auxiliary fields, the D  and F fields.

Adding the kinetic term \ref{kinv} for the vector field and a potential, ${\tilde{W}}(\Phi)$ for the matter, gives the total Lagrangian,
\beq
{\cal{L}}=\int d^2 \theta \lf W^\alpha W_\alpha + \int d^2 {\bar{\theta}} {\bar{\Phi}}^i \exp (q_i V) \Phi_i + {\tilde{W}}(\Phi) \rt
\eeq
this produces the following potential, 
\beq
V= \sum_i|{{ \pl{\tilde{W}}} \over {\pl \phi^i}}|^2 + {q^2 \over 4} ((2K + \sum|\phi_i|^2)^2
\eeq
The first term is the F-term and the second is the D-term. Both these terms need to vanish for unbroken supersymmetry.

Some remarks about this potential:

Generically the F-terms should vanish since there are n equations for n unknowns. If $<\phi_i>$ =0, that is if the U(1) is not spontaneously broken then supersymmetry is broken if and only if $K_{F.I.}\neq 0$. When K=0 and the F-terms have a solution then so will the D-term and there will be no supersymmetry breaking. 

These ideas are demonstrated by the following example. Consider fields $\Phi_1,\Phi_2$ with opposite U(1) charges and Lagrangian given by:
\bea
{\cal{L}}&=& { 1\over 4} (W^\alpha W_\alpha +h.c.) + {\bar{\Phi}}_1 \exp (eV) \Phi_1 \\ &+& {\bar{\Phi}}_2 \exp (-eV) \Phi_2 + m\Phi_1 \Phi_2 + h.c. + 2KV
\eea
This leads to the potential:
\beq
V={\half}D^2 + F_1 F_1^* + F_2 F_2^*
\eeq
where 
\bea
D+K+{e \over 2} (A_1^* A_1- A_2^* A_2)=0  \\
F_1+mA_2*=0 \\
F_2+mA_1^*=0
\eea
Leading to the following expression for the potential:
\beq
V={\half} K^2 +(m^2 +{\half} eK ) A_1^* A_1 +(m^2-{\half}eK) A^*_2 A_2 +{1 \over 8} e^2 (A_1^* A_1 -A^*_2A_2)^2
\eeq

Consider the case, $m^2 > {\half} e K$. The scalars have mass, $\sqrt{m^2 + {\half} eK} $ and $\sqrt{m^2 -{\half} eK}$. The vector field has zero mass. Two Fermions have mass m and one Fermion is massless. Since the vector field remains massless then the U(1) symmetry remains unbroken. For $K \neq 0$, supersymmetry is broken as the Bosons and Fermions have different masses. (For K=0 though the symmetry is restored.) The massless Fermion (the Photino) is now a Goldstino. Note that a {\it{trace}} of the supersymmetry remains as ${\rm{TrM}}_B^2={\rm{TrM}}_F^2$ even after the breaking of supersymmetry.

Next, consider the case, $m^2 < {\half} eK$; at the minimum, $A_1=0,A_2=v$ where $v^2 \equiv 4 {{{\half} e K - m^2} \over {e^2}}$. The potential expanded around this minimum becomes, with $A \equiv A_1$ and ${\tilde{A}} \equiv A_2 -v$:
\bea
V&=&{{2m^2} \over e^2} (eK -m^2) + {\half} ( {\half}e^2 v^2) A_{\mu} A^{\mu}\\
&+&2m^2 A^*A = {\half} ({\half}e^2 v^2) ({1\over {\sqrt{2}}} ({\tilde{A}}+{\tilde{A}}^*))^2\\
&+& \sqrt{m^2+{\half} e^2 v^2} (\psi {\tilde{\psi}} + {\bar{\psi}} {\bar{\tilde{\psi}}} ) + 0 \times \lambda {\bar{\lambda}}
\eea
The first term implies that supersymmetry is broken for $m>0$.
The photon is now massive, $m^2_{\gamma}={\half} e^2 v^2$ implying that the U(1) symmetry is broken as well.  The Higgs field, ${1\over {\sqrt{2}}} ({\tilde{A}}+{\tilde{A}}^*))^2$ has the same mass as the photon. Two Fermions have non-zero mass and there is one massless Fermion, the Goldstino. 

In the above example there is both supersymmetry breaking and U(1) symmetry breaking except when $m=0$ in which case the supersymmetry remains unbroken.

Next consider a more generic example where there is U(1) breaking but no supersymmetry breaking, $\Phi$ is neutral under the U(1) while $\Phi_+$ has charge
+1 and $\Phi_- $ has charge -1. The potential is given by:
\beq
{\cal{L}}={\half} m \Phi^2 + \mu \Phi_+ \Phi_- + \lambda \Phi \Phi_+ \Phi_- +h.c.
\eeq

There are two branches of solutions to the vacuum equations (a denotes the vacuum expectation value of A etc.):
\beq
a_+ a_- =0\comma a= -{\lambda \over m}
\eeq
which leads to no U(1) breaking and
\beq
a_+ a_- = -{1 \over 8} ( \lambda - {{m \mu} \over g}) \comma a= -{ \mu \over g}
\eeq
which breaks the U(1) symmetry.
Note, the presence of a flat direction:
\beq
a_+ \rightarrow e^\alpha a_+ \comma a_- \rightarrow e^{-\alpha a_-}
\eeq
leaves $a_-a_+$ fixed and the vacuum equations are still satisfied.

Typical generic supersymmetry breaking leads to the breaking of R-symmetry. Since there is a broken global symmetry this would lead to the presence of Goldstone Boson corresponding to the broken $U(1)_R$. Inverting this argument leads to the conclusion that supersymmetric breaking in nature cannot be generic since we do not observe such a particle.

So far we have only dealt with U(1) vector fields. One can also consider non-abelian groups. The fields are in an adjoint representation of the group, $A_{\mu}^a,\lambda^a,D^a$, the index a is the group index, a = 1..dim(group) and $D^a=\sum_i {\bar{\phi}}_i T^a_{R(\Phi_i)} \Phi_i$. 
Only if there is a U(1) factor will the Fayet-Iliopoulos term be non-vanishing.

We wish to examine the properties of potentials with flat directions.
Take the simple example of an abelian theory with no Fayet-Iliopoulos terms and no F-terms. This gives a U(1) theory with oppositely charged fields $\Phi_+,\Phi_-$. The vanishing of the D-term implies:
\beq
D=|\Phi_+|^2- |\Phi_-|^2=0
\eeq
Four real fields obeying the constraint:
\beq
\Phi_+=\exp (i\alpha) \Phi_-
\eeq
implies a three dimensional configuration space. However there is still the U(1) gauge symmetry that one must mod out by. This leaves a two dimensional moduli space. A convenient gauge fixing is,
\beq
\Phi_+=\Phi_- \; \; .
\eeq
There remains a $Z_2$ residual gauge symmetry however:
\beq
\Phi_+ \rightarrow -\Phi_+ \; \; .
\eeq
After moding out by this action, the classical moduli space is given by the orbifold, $ C \over Z_2$. There is therefore a fixed point at $\Phi_+=0$, which will be a singularity. What is the physical interpretation of this singularity in moduli space?

Let us study the space in terms of gauge invariant variables:
\beq
M\equiv \Phi_+ \Phi_-
\eeq
Using $\Phi_- =\Phi_+$, the Kahler potential becomes:
\beq
K={\bar{\Phi}}_+ \Phi_+ + {\bar{\Phi}}_- \Phi_- = 2 {\bar{\Phi}}_+ \Phi_+=+2\sqrt{M{\bar{M}}}
\eeq
The metric will then be:
\beq
ds^2= {\half} { {dM d{\bar{M}}} \over {\sqrt{M {\bar{M}}}}} \; \; .
\eeq
There is a singularity in moduli space when M=0. By expanding the D-term, ones sees that M is a parameter that determines the mass of the matter fields. Thus the singularity in moduli space is a signature of a particle becoming massless. Note, there is no nonrenormalisation theorem for the Kahler potential.

When K=0, (as is the case above) the moduli space is determined in the following way. In the absence of an F-term then there is always a solution to the D-term equations. The moduli space is the space of all fields $\Phi_i$, such that $D^a=0$, modulo $G_{\bf{R}}$ gauge transformations . This is equivalent to the space of all constant $\Phi$ fields modulo the complex $G_{\bf{C}}$ gauge transformations.
When K=0 but in the presence of an F-term, then provided there are solutions to the F-terms equations, the D-terms automatically vanish. The moduli space is then given by the space of fields that solve the F-term equations modulo complex $G_{\bf{C}}$ gauge transformations. 
Moreover the moduli space, $\Phi_i \over G_{\bf{C}}$ is spanned by a basis of the independent gauge singlets (such as $m=\phi_+ \phi_-$).

\section{Phases of gauge theories}
First we will gain our intuition from D=4 lattice gauge theories, ie. $Z_N$ valued gauge fields with coupling g \cite{Banks:1977cc,Horn:fy,Elitzur:1979uv,Ukawa:yv,Elitzur:2000ps,Cardy:qy}

The effective ``temperature'' of the system will be given by, $T={{Ng^2} \over {2 \pi}}$.

For a given theory there is a lattice of electric and magnetically charged operators where the electric charge is denoted by n and the magnetic charge by m. An operator with charges (n,m) is {\it{perturbative}} ie. it is an irrelevant operator and weakly coupled to system, so long as the free energy, $F>0$, that is,
\beq
n^2 T + {m^2 \over T} > {C \over N} \comma
\eeq
however when the free energy is negative for the operator (n,m), it condenses indicating the presence of a relevant operator and hence an infra red instability, this occur when, 
\beq
n^2 T + {m^2 \over T} < {C \over N} \; \; .
\eeq
Keeping  N,C fixed and vary T. How does the theory change?

There are three phases depending which operators condense.
At small T, there is electric condensation which implies that there is electric charge screening, magnetic charges are confined, and the log of the Wilson loop is proportional to the length of the perimeter of the loop. (This is called the Higgs phase).

At high T, magnetic condensation occurs, this is the dual of electric condensation. Magnetic charges are screened, electric charges are confined and the log of the Wilson loop is proportional to the area. (This is called the confinement phase.) For intermediate values of T it is possible that there is neither screening of charges nor confinement, this is the Coulomb phase.

\begin{figure}[hbtp] 
\centering
\includegraphics[width=10.truecm]{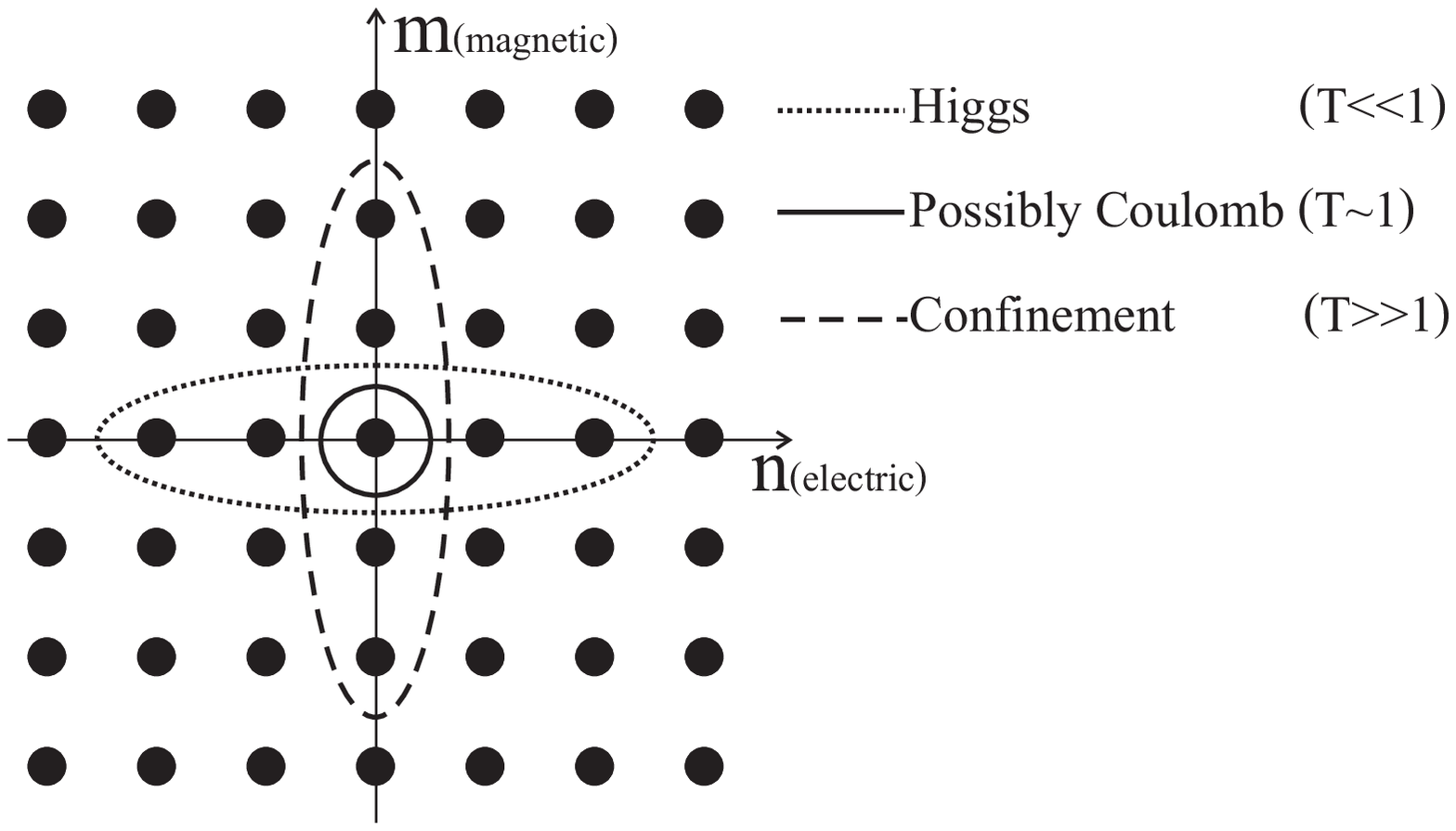} 
\caption[]{The different possible phases}
\label{figure} \end{figure}

In the presence of a $theta$ parameter, an electric charge picks up a magnetic charge and becomes dyonic \cite{WE}.
\beq
n' = n + {\theta \over {2 \pi}} m
\eeq

This lead to a tilted lattice of dyonic charges and one may condense dyons with charges $(n_0,m_0)$. This leads to what is called oblique confinement with the charges commensurate with $(n_0,m_0)$ being screened and all other charges being confined \cite{Cardy:qy}.

How does this relate to QCD? There are ideas that confinement in QCD occurs due the condensation of QCD monopoles \cite{Nielsen:cs,Mandelstam:1974pi,'tHooft:1977hy}. It is difficult to study this phenomenon directly. The Dirac monopole in a U(1) gauge theory is a singular object; however by embedding the monopole in a spontaneously non-abelian theory with an additional scalar field one may smooth out the core of the monopole and remove the singularity. One may proceed analogously, by enriching QCD; adding scalars and making the the theory supersymmetric one can calculate the condensation of monopoles in a four dimensional gauge theory. This has been achieved for gauge theories with N=1,2 supersymmetries. There are many new methods that have been utilized and the phase structures of these theories have been well investigated \cite{SI,Seiberg:1994aj,Seiberg:1994rs}. Novel properties of these theories have been discovered such as new types of conformal field theories and new sorts of infra-red duality. To this we turn next.

\begin{figure}[hbtp] 
\centering
\includegraphics[width=8.truecm]{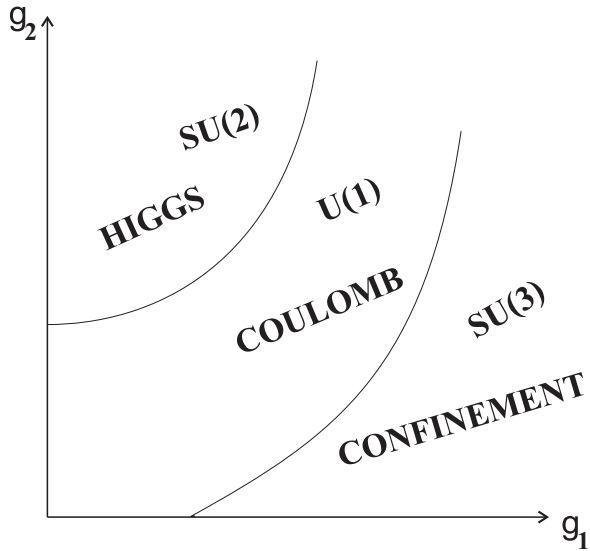} 
\caption[]{Possible phases of gauge theories ($g_1$ and $g_2$ are some relevant /marginal couplings)}
\label{figure} \end{figure}

\section{Supersymmetric gauge theories/ super QCD}

The goal will be to examine theories that are simple supersymmetric extensions of QCD \cite{SI}. Consider the case of an N=1 vector multiplet with gauge group $SU(N_C)$, and $N_F$ chiral multiplets in the fundamental representation of $SU(N_C)$ and $N_F$ chiral multiplets in the antifundamental representation.
The Lagrangian is:
\bea
{\cal{L}}&=&  \int (-i \tau) {\rm{Tr}} W^\alpha W_\alpha d^2 \theta  + h.c. \nn \\
&+& Q^+_F \exp (-2V) Q_F +{\tilde{Q}} \exp (2V) Q_F^+ |_{{\bar{\theta}}{\bar{\theta}}\theta \theta} + m_F {\tilde{Q}}_F Q |_{\theta \theta}
\eea
where the coupling is:
\beq
\tau = {\theta \over {2 \pi }} +i {{4 \pi }\over {g^2}}  \; \; \label{tau} .
\eeq

Apart from the local $SU(N_C)$ gauge symmetry, the fields are charged under the following global symmetries.

\bea
\matrix{ 
  &SU(N_F)_L &\times& SU(N_F)_R &\times &U(1)_V& \times&U(1)_A&\times&U(1)_{R_C} \cr
Q^i_a&N_F&&1&&1&&1&&1 \cr
{\tilde{Q}}_i^a& 1 &&{\bar{N}}_F &&-1&&1&&1 \cr
W_\alpha & 1&&1&&0&&0&&1 \cr } 
\eea
Note, when $N_C=2$, because $2 \sim {\bar{2}}$, the global flavor symmetry is enhanced to $SO(2 N_F)_L \times SO(N_F)_R$.

 There is an anomaly of the $U(1)_A \times U(1)_R$ symmetry. A single  $U(1)$ symmetry survives the anomaly. This is denoted as $U(1)_R$ and is a full quantum symmetry. The adjoint Fermion contributes $2N_C \times R(\lambda)$ to the anomaly. The Chiral Fermions contribute, $2N_F \times R_F$. $R(\lambda)=1$, while $R_F$ is now chosen so that the total anomaly vanishes,
\beq
2N_C+2 R_F N_F =0  \; \; .
\eeq
This leads to 
\beq
R_F= - {{N_C} \over {N_F}}
\eeq
The Bosons in the chiral multiplet have an R charge one greater than the Fermions in the multiplet. Thus, 
\beq
R_B= 1- { {N_C} \over {N_F}}={{ N_F- N_C} \over {N_F}} \; \; .
\eeq
The non-anomalous  R-charge, is given by :
\beq
R = R_C - {{N_C} \over {N_F}} Q_{A} \comma
\eeq
where $R_C$ is the classical R-charge and A is the classical $U(1)_A$ charge.
This leads to the following non-anomalous global charges:
\bea
\matrix{ 
  &SU(N_F)_L &\times& SU(N_F)_R &\times &U(1)_V& \times&U(1)_R \cr
Q^i_a&N_F&&1&&1&& {{N_F -N_C} \over {N_F}}  \cr
{\tilde{Q}}_i^a& 1 &&{\bar{N}}_F &&-1&&{{N_F -N_C} \over {N_F}}   \cr
W_\alpha & 1&&1&&0&&1 \cr
} 
\eea
One is now ready to identify the classical moduli space.

\subsection{The classical moduli space}

The classical moduli space is given by solving the D-term and F-term equations:
\bea
D^a &=& Q_F^+ T^a Q_F - {\tilde{Q_F}} T^a {\tilde{Q_F}}^+ \\
{\bar{F}}_{Q_F} &=& -m_F {\tilde{Q}} \\
{\bar{F}}_{{\tilde{Q}}_F} &=& -m_F {{Q}}
\eea

For $N_F=0$ or for $N_F \neq0$ and  $m_F\neq 0$, there is no moduli space. Note, the vacuum structure is an infra-red property of the system hence having $m_F\neq0$ is equivalent to setting $N_F=0$ in the infrared. 

Consider the quantum moduli space of the case where $N_F=0$.
The Witten index, ${\rm{Tr}}(-1)^F=N_C$ ie. the rank of the group +1.
This indicates there is no supersymmetry breaking.
There are $2N_C$ Fermionic zero modes (from the vector multiplet). These Fermionic zero modes break through instanton effects the original $U(1)_R$ down to $Z_{2N_C}$. Further breaking occurs because the gluino two point function acquires a vacuum expectation value which breaks the symmetry down to $Z_2$. This leaves $N_C$ vacua. The gluino condensate is:
\beq
<\lambda \lambda> ={\rm{exp}}( {{2 \pi ik} \over {N_C}}) \Lambda^3_{N_C}
\eeq
where $\Lambda_{N_C}$ is the dynamically generated scale of the gauge theory and $k=1,..,N_C-1$ label the vacua. Chiral symmetry breaking produces a mass gap. Note, because chiral symmetry is discrete there are no Goldstone Bosons. Further details of quantum moduli spaces will be discussed later.

Consider the case where $m_F=0$ and  $0 < N_F < N_C$. The classical moduli space is determined by the following solutions to the D-term equations:
\bea
Q={\tilde{Q}}= \lf \matrix{ a_1 &0    &      &  &     & 0..0 \cr
                            0   & a_2 &      &  &     & 0..0 \cr
			        &     & \cdot &  &     & 0..0 \cr
			        &     &      &\cdot &    & 0..0 \cr
			        &     &      &     &a_{N_F} & 0..0 \cr } \rt_{N_F \times N_C}		
\eea

Where the row indicates the the flavor and the column indicates the colour. There are $N_F$ diagonal non-zero real entries, $a_i$. (To validate this classical analysis the vacuum expectation values must be much larger than any dynamically generated scale, ie. $a_i>>\Lambda$. The gauge symmetry is partially broken:
\beq
SU(N_C) \rightarrow SU(N_C -N_F) \; .
\eeq
This is for generic values of $a_i$. By setting some subset of $a_i$ to zero one may break to a subgroup of $SU(N_C)$ that is larger than $SU(N_C-N_F)$.
Also, if $N_F=N_C-1$ then the gauge group is complete broken. This is called the Higgs phase.

The number of massless vector Bosons becomes
\beq
N_C^2 - ((N_C-N_F)^2-1)= 2N_C N_F -N_F^2 \comma
\eeq
the number of massless scalar fields becomes,
\beq
2N_C N_F -(2N_C N_F -N_F^2)= N_F^2 \; .
\eeq
The matrix 
\beq
M_{{\tilde{i}} j}\equiv {\tilde{Q}}_{\tilde{i}} Q_j
\eeq
forms a gauge invariant basis. The Kahler potential is then,
\beq
K= 2 {\rm{Tr}} \sqrt{(M{\bar{M}})}
\eeq
and as before when singularities appear ie. $detM =0 $ this signals the presence of enhanced symmetries.

When $N_F \geq N_C$, one has the following classical moduli space.

\bea
Q= \lf \matrix{ a_1 &0    &      &  &     \cr
                0   & a_2 &      &  &    \cr
                    &     & \cdot &  &     \cr
		    &     &      &\cdot&     \cr
	            &     &      &    &a_{N_C}   \cr
                0   &	0 & \cdot \cdot& \cdot \cdot& 0 \cr
                \cdot&\cdot & \cdot \cdot& \cdot \cdot& 0 \cr
                0   &	0 & \cdot \cdot& \cdot \cdot& 0 \cr } \rt_{N_F \times N_C}
\comma \quad
{\tilde{ Q}}= \lf \matrix{ {\tilde{a}}_1 &0    &      &  &     \cr
                            0   & {\tilde{a}}_2 &      &  &    \cr
                                &     & \cdot &  &     \cr
		                &     &      &\cdot&     \cr
	                        &     &      &    & {\tilde{a}}_{N_C}  \cr
                            0   &   0 & \cdot \cdot& \cdot \cdot& 0 \cr
                            \cdot&\cdot & \cdot \cdot& \cdot \cdot& 0 \cr
                            0   &   0 & \cdot \cdot& \cdot \cdot& 0 \cr } \rt_{N_F \times N_C}
\eea
with the constraint that 
\beq
|a_i|^2 -|{\tilde{a}}_i|^2=\rho  \; .
\eeq
Generically the $SU(N_C)$ symmetry is completely broken. However, when $a_i={\tilde{a}}_i=0$ then a subgroup of the $SU(N_C)$ can remain. 

We will now consider some special cases, first the classical moduli space for $N_F=N_C$.  The dimension of the moduli space is given by:
\beq
2N_C^2-(N_C^2-1)=N_C^2+1=N_F^2+1
\eeq
There are $N_F^2$ degrees of freedom from $M_{{\tilde{i}}j}$ and naively one would have two further degrees of freedom from:
\beq
B= \epsilon_{i_1 \cdots i_{N_C}} Q^{i_1}_{j_1} \cdots Q^{i_{N_C}}_{j_{N_F}} \comma {\tilde{B}}= \epsilon_{i_1 \cdots i_{N_C}} {\tilde{Q}}^{i_1}_{j_1} \cdots {\tilde{Q}}^{i_{N_C}}_{j_{N_F}}  \; \; .
\eeq
There is however a classical constraint:
\beq
{\rm{det}} M -B {\tilde{B}} =0
\eeq
which means M, B and ${\tilde{B}}$ are classically dependent. This leaves only $N_F^2+1 $ independent moduli.
 
Generically, as well as the gauge symmetry being completely broken the global flavor symmetry is also broken. There is a singular point in the moduli space where $M=0=B={\tilde{B}}$.

Next, consider the case, $N_f=N_C+1$, again there are $N_F^2 $ moduli from $M_{i{\tilde{j}}}$. There are also, $2(N_C+1)$ degrees of freedom given by:
\beq
B_i= \epsilon_{i i_1 \cdots i_{N_C}} Q^{i_1}_{j_1} \cdots Q^{i_{N_C}}_{j_{N_F}}
\comma {\tilde{B}}_{\tilde{i}}= \epsilon_{i_1 \cdots i_{N_C}} {\tilde{Q}}^{i_1}_{j_1} \cdots {\tilde{Q}}^{i_{N_C}}_{j_{N_F}} \; \; .
\eeq
However there are again the classical constraints:
\bea
{\rm{det}}M &-& M_{i {\tilde{j}}} B^i B^{\tilde{j}}=0 \\
M_{{\tilde{j}} i} B^i& =& M_{i {{\tilde{j}}}} B^{\tilde{j}} =0
\eea
giving again an $N_F^2+1$ dimension moduli space. (The moduli space is not smooth). There is a generic breaking of gauge symmetry.

\subsection{Quantum moduli spaces}
One is required to examine on a case by case basis the role that quantum effects play in determining the exact moduli space. Quantum effects both perturbative and nonperturbative can lift moduli. In what follows we examine the quantum moduli space for the separate cases:
$1 \leq N_F \leq N_C -1\comma \; N_F=N_C \comma \; N_F=N_C+1 \comma \; N_C+1 <  N_F \leq {{3N_C} \over 2} \comma \; {{3N_C} \over 2} < N_F < 3N_C \comma N_F=3N_C \comma  N_F>3N_C$.

\subsection{Quantum moduli space for $0<N_F < N_C$}

Classically, the dimension of the moduli space is $N_F^2$ from $Q, {\tilde{Q}}$. The following table summarizes the charges under the various groups \cite{Amati,ads,taylor}.
\bea
\matrix{ &SU(N_C) & 
  SU(N_F)_L & SU(N_F)_R & U(1)_V  & U(1)_A & U(1)_{R_{cl}} & U(1)_R \cr
Q^i_a&N_C&N_F&1&1&1&1& {{N_F -N_C} \over {N_F}}  \cr
{\tilde{Q}}_i^a&{\bar{N_C}}& 1 &{\bar{N}}_F &-1&1&1&{{N_F -N_C} \over {N_F}}   \cr
\Lambda ^{3 N_C -N_F} & 1&1&1&0& 2N_F &2 N_C&0       \cr
M & 1&N_F&{\bar{N}}_F&0&2&2&2-{{2N_C} \over N_F} \cr
{\rm{det}}M & 1&1&1&0&2N_F&2N_F&2(N_F-N_C) } 
\eea
$\Lambda $, the dynamically generated QCD scale is assigned charges as m and g were before. The power $3N_C -N_F$ is the coefficient in the one loop beta function. There is no Coulomb phase so $W_{\alpha}$ does not appear.

The symmetries imply, the superpotential, W, has the following form:
\beq
W = (\Lambda^{3N_C-N_F})^a \; ({\rm{det}}M)^b \; c
\eeq
a, b are to be determined. c is a numerical coefficient, If c does not vanish, the classical moduli space gets completely lifted by these nonperturbative effects.

\begin{figure}[hbtp] 
\centering
\includegraphics[width=6.truecm]{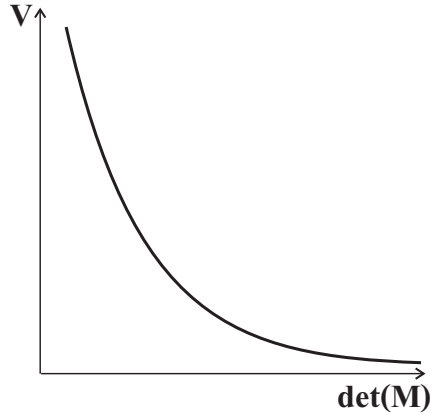} 
\caption[]{The potential for $1 < N_F < N_C$, it has no ground state}
\label{figure} \end{figure}

One examines the charges of W under the various symmetries.
Automatically, the charges of W for the flavor symmetries, $SU(N_F)_L \times SU(N_F)_R$ and the $U(1)_V$ vanish. 

If one requires the  $U(1)_A$ charge to vanish then this implies $a=-b$. Requiring the $U(1)_R$ charge to vanish implies that $b={1 \over {N_F-N_C}}$.
These restrictions fix:
\beq
W=c \lf { {\Lambda ^{3N_C-N_F}} \over { {\rm{det}} M}} \rt ^{1\over {N_C-N_F}}
\eeq
For non vanishing c, all the moduli are now lifted and there is no ground state.

What is the value of c? This is a difficult to calculate directly unless there is complete Brouting. For $N_F=N_C-1$ there is complete symmetry breaking and one can turn to weak coupling. From instanton calculations one calculates that $c \neq 0$ and the prepotential for the matter fields is
\beq
W \sim \lf { {\Lambda^{2N_C+1}} \over {{\rm{det}}M}}\rt  \; \; .
\eeq
One may now go to $N_F < N_C-1$ by adding masses and integrating out the heavy degrees of freedom.
This produces,  
\beq
<M^i{}_j>_{min}= (m^{-1})^i{}_j(\Lambda^{3N_C-N_F}{\rm{det}}m )^{ 1\over{N_C}}
\eeq

\begin{figure}[hbtp] 
\centering
\includegraphics[width=6.truecm]{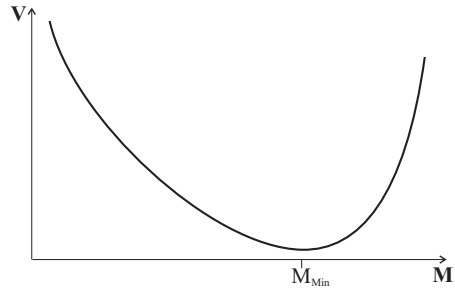} 
\caption[]{The potential with finite masses has a ground state. $M_{min}\rightarrow \infty$ as $ m \rightarrow 0$.}
\label{figure} \end{figure}

\subsection{Integrating in}

This method involves the addition of very massive fields to known effective actions and extrapolating to the case where the additional degrees of freedom are massless, see \cite{I,IS1,IS2,ISS,ILS,efgr,efgr2,efgr-proc,efgr3}. It is rather surprising that anything useful can be learned by this flow in the opposite direction to the usual infra red. We will show that under certain circumstances it is possible to derive in a rather straight forward way the potential for light fields. We thus give some of the flavour of this possibility. It gives results for the phase structure in many cases. We will also discuss when these conditions are met. We begin by reviewing the conventional method of integrating out; heavy degrees of freedom are integrated out to obtain an effective potential for the light degrees of freedom. 

Consider a theory containing gauge invariant macroscopic light fields of the following nature:

Fields X: built out of $d_A$ degrees of freedom

Fields M: built out of $u_i$ degrees of freedom

Fields Z: built out of both $d_a$ and $u_i$.

The $d_a$ dofs will be kept light throughout all the discussion.The $u_i$ and with them the macroscopic fields M and Z will be considered as heavy in part of the discussion. Assume that one is given the effective potential $W_u(X,M,Z,\Lambda_u)$ describing all the light macroscopic degrees of freedom. ($\Lambda_u$ is the dynamically generated scale of the theory. The effective potential is the Legendre transform
\beq
W_u(\Phi)=( {\tilde{W}}(g_i) - \sum g_i \Phi_i)_{<g_i>}
\eeq
where
\beq
{{\pl {\tilde{W}}(g_i)} \over {\pl g_i}} = < \Phi_i> \; .
\eeq
Consider next making the microscopic ``up'' fields massive and integrating them out retaining only the light degrees of freedom and the couplings ${\tilde{m}}, \lambda$ to the macroscopic degrees of freedom containing the removed fields: M and Z. One thus obtains:
\beq
{\tilde{W}}_d(X,{\tilde{m}},\lambda,\Lambda_u)= (W_u(X,M,Z,\Lambda_u) = {\tilde{m}} M + \lambda Z )_{<M>,<Z>}
\eeq
For the case ${\tilde{m}} \rightarrow \infty$ one may tune the scale $\Lambda_u$ so as to replace an appropriate combination of $\lambda,{\tilde{m}}$ and $\Lambda_u$ by $\Lambda_d$ the scale of the theory of the remaining light degrees of freedom. One ends with $W_d(X,\Lambda_d)$. It is convenient for general ${\tilde{m}}\neq 0$ to write,
\beq
{\tilde{W}}_d(X,{\tilde{m}},\lambda,\Lambda_u)= W_d(X,\Lambda_d) + W_I(X,{\tilde{m}},\lambda,\Lambda_d)
\eeq
where $W_d(X,\Lambda_d) $ is the exact result for infinitely heavy $U_i$, and to partition again,
\beq
W_I=W_{{\rm{tree}},d} + W_{\Delta}
\eeq
where
\beq
W_{{\rm{tree}},d} = (W_{{\rm{tree}}})_{<u_i>}
\eeq
has by definition no scale dependence. $W_{\Delta}$ should obeys the constraints:
\beq
W_{\Delta} \rightarrow 0 \comma \quad {\rm{when}} \quad \Lambda_u \rightarrow 0 \comma \; or\; {\tilde{m}} \rightarrow \infty \; .
\eeq
Now one reverses the direction and integrates in. Namely, given an exact form for $W_d(X,\Lambda_d)$ one can obtain:
\beq
W_u(X,M,Z,\Lambda_u)= (W_d(X,\Lambda_d) + W_{\Delta} + W_{{\rm{tree}},d} - W_{{\rm{tree}}} )_{<{\tilde{m}}>,<\lambda>} \; \; \; .
\eeq

All the essential complexities of the flow lie in the term $W_{\Delta}$. Imagine however that this term would vanish. In this case the calculations become much simpler. For the cases corresponding to the colour group $SU(2)$ this indeed turns out to be the case. This can be seen in the following manner.

One starts with $SU(2)$ without adjoint fields ($N_a=0$), and quark flavors $Q_i \comma \; i=1,..N_F \; \; (N_F\leq 4)$ for the down theory. one brings down infinitely heavy fields in the adjoint representation, that is , one resuscitates the full up theory, which contains $N_a$ adjoint fields $\Phi_\alpha$ with the superpotential:
\beq
W_{\rm{tree}} = mX + {\tilde{m}M} + \lambda Z \comma
\eeq
where $M=\Phi \Phi$ and $Z= Q \Phi Q$. For convenience one writes,
\beq
W_{\Delta}(,{\tilde{m}},\lambda, \Lambda) = W_{{\rm{tree}},d}f(t)
\eeq
where $X= WW$ and t being any possible singlet of $SU(2N_F) \times U(1)_Q \times U(1)_\Phi \times U(1)_R$; $\Phi$ is the adjoint field we add. The quantum numbers of all relevant fields and parameters are given by:
\bea
\matrix{&U(1)_Q & U(1)_\Phi &U(1)_R \cr
X &2&0&0 \cr
\lambda&-2&-1&2 \cr
\Lambda^{b_1} &2N_F&4N_a&4-4N_a-2N_F \cr
{\tilde{m}} &0&-2&2&  \cr
W_{\Delta} &0&0&2&  }
\eea
where $b_1 = 6-2N_a -N_F$. Writing t as,
\beq
t \sim (\Lambda^{b_1})^a {\tilde{m}}^b X^c \lambda^d  \; \; .
\eeq
It is a singlet provided that
\beq
b= (2N_a + 2 - N_F) a \comma \quad c=(N_F-4) a \comma \quad d=(2N_F-4)a \; .
\eeq
Recall the constraints
\beq
W_{\Delta} \rightarrow 0 \quad {\rm{for}} \quad {\tilde{m}} \rightarrow \infty \quad and \quad W_{\Delta} \rightarrow 0 \quad {\rm{for}}\quad \Lambda \rightarrow 0
\eeq
as well as the fact that in the Higgs phase one can decompose W as:
\beq
W(\Lambda^{b_1}) = \sum_{n=1}^{\infty} a_n (\Lambda^{b_1})^n \; \; .
\eeq
One shows that 
\beq
W_{\rm{tree},d} \sim {1 \over {\tilde{m}}}
\eeq
which implies that for all values of $N_f$ the vanishing of $W_{\Delta} $. We see this explicitly in the following cases.

For $N_F=0 $ and $N_a=1$
\beq
b=4a
\eeq
leading to
\beq
W_{\Delta}= \sum_{a=1}^{\infty} r_a ( \Lambda^{b_1} )^a {\tilde{m}}^{a-1}  \; \; \; .
\eeq
The constraints imply that $r-a$ indeed vanishes for all a's.

For $N_a+1, \; N_F=1,2,3,\; b_1=4-N_F$,
\beq
W_{\Delta} = {{r({\tilde{m}}\Lambda)^{b_1} }\over {{\tilde{m}}}} + ...
\eeq
implying that $r=0$ as well.

Starting from known results for $N_a=N_F=0$ one can now obtain the effective potential for all relevant values of $N_a$ and $N_F$. The equations of motion of these potentials can be arranged in such a manner as to coincide with the singularity equations of the appropriate elliptic curves derived for systems with N=2 supersymmetry and SU(2) gauge group. (The role of these elliptic curves in N=2 theories will be described later.)

\subsection{Quantum moduli space for $N_F \geq N_C$}
There is a surviving moduli space. In the presence of a mass matrix, $m_{ij}$ for matter one obtains,
\beq
<M^i{}_j>=(m^{-1})^i{}_j (\Lambda^{3N_C-N_F} {\rm{det}}m)^{1 \over {N_C}}
\eeq
Previously, for the case of $N_F < N_C$, it turned out that $m \rightarrow 0$ implied ${{<M^i{}_j>} \rightarrow \infty}$ thus explicitly lifting the classical moduli space. For $N_F \geq N_C$ it is possible to have $m \rightarrow 0$ while keeping $<M^i{}_j>$ fixed. 

\subsection{$N_F = N_C$}
Quantum effects alter the classical constraint to:
\beq
{\rm{det}}M - B {\tilde{B}} = \Lambda^{2N_C}
\eeq
This has the effect of resolving the singularity in moduli space. The absence of a singularity means there will not be additional massless particles.
In this case,
\beq
M_i{}^j= (m^{-1})_i{}^j ( {\rm{det}}m )^{1 \over {N_C}} \Lambda^2
\eeq
This implies,
\beq
{\rm{det}}M=\Lambda^{2N_C}
\eeq
since m cancels, thus this also holds in the limit $m \rightarrow 0$.
On the other hand, $<B>=<{\tilde{B}}>=0$ if ${\rm{det}}m \neq 0$ because all fields carrying B number are integrated out. Therefore ${\rm{det}}M \neq B{\tilde{B}} $ through quantum effects.

Note, $R(M_{IJ})=0$ for $N_F =N_C$. Writing out an expansion that obeys the R-charge conservation: 
\beq
{\rm{det}}M - B {\tilde{B}} = \Lambda^{2N_C} \lf 1 + \sum_{ij} c_{ij}{{ (B {\tilde{B}} )^i \Lambda^{2N_C}{}^j } \over { ({\rm{det}}M)^{i+j}}} \rt
\eeq
then by demanding that there be no singularities at small $<M>$ or at large $<B>$ implies that all $c_{ij}$ must vanish and the hence ${\rm{det}}M - B {\tilde{B}} = \Lambda^{2N_C}$ obeys a nonrenormalisation theorem. 
Note,
\beq
R(M)=0 \Rightarrow W=0 \; .
\eeq
There are allowed {\it{soft}} perturbations, mass terms given by:
\beq
W={\rm{tr}}(mM) + bB + {\tilde{b}} {\tilde{B}}
\eeq
One check is to integrate out to give the case $N_F=N_C-1$ yielding, 
\beq
W={{m \Lambda^{2N_C}} \over {{\rm{det}}M}} = {{{\tilde{\Lambda}}^{2N_C+1}} \over { {\rm{det}}M }}
\eeq

What is the physics of this theory, is it in a Higgs/confinement phase?
For large, $M/B/{\tilde{B}}$ one is sitting in the Higgs regime; however, for small $M/B/{\tilde{B}}$ one is in the confining regime. Note that M cannot be taken smaller than $\Lambda$.

Global symmetries need to be broken in order to satisfy the modified constraint equation.

Consider some examples:
With the following expectation value,
\beq
<M^i{}_j>=\delta^i{}_j \Lambda^2 \comma \quad <B {\tilde{B}}>=0
\eeq
the global symmetries are broken to:
\beq
SU(N_F)_V \times U_B(1) \times U_R(1)
\eeq
and there is chiral symmetry breaking.
When,
\beq
<M^i{}_j>=0  \comma \quad <B {\tilde{B}}>\neq 0
\eeq
then the group is broken to:
\beq
SU(N_F)_L \times SU(N_F)_R \times U_R(1)
\eeq
which has chiral symmetry and also has confinement. This is an interesting situation because there is a dogma that as soon as a system has a bound state there will be chiral symmetry breaking \cite{casher}.

In both cases the 't Hooft anomaly conditions \cite{'tHooft:wj} are satisfied. These will be discussed later.

\subsection{$N_F = N_C+1$}
The moduli space remains unchanged. The classical and quantum moduli spaces are the same and hence the singularity when $M=B={\tilde{B}}=0$ remains. This is not a theory of massless gluons but a theory of massless mesons and baryons.
When, $M,B,{\tilde{B}}\neq0$ then one is in a Higgs/confining phase. At the singular point when, $M=B={\tilde{B}}=0$ there is no global symmetry breaking but there is ``confinement'' with light baryons.

There is a suggestion that in this situation, $M, B, {\tilde{B}} $ become dynamically independent. The analogy is from the nonlinear sigma model, where because of strong infra red fluctuations there are n independent fields even though there is a classical constraint.
The effective potential is:
\beq
W_{eff} = {1 \over {{\Lambda}^{2N_C-1}}} (M^i{}_j B_i {\tilde{B}}^j - {\rm{det}}M )
\eeq
the classical limit is taken by: 
\beq
\Lambda \rightarrow 0
\eeq
which in turn imposes the classical constraint.
Again the system obeys the 't Hooft anomaly matching conditions.

\begin{figure}[hbtp] 
\centering
\includegraphics[width=4.truecm]{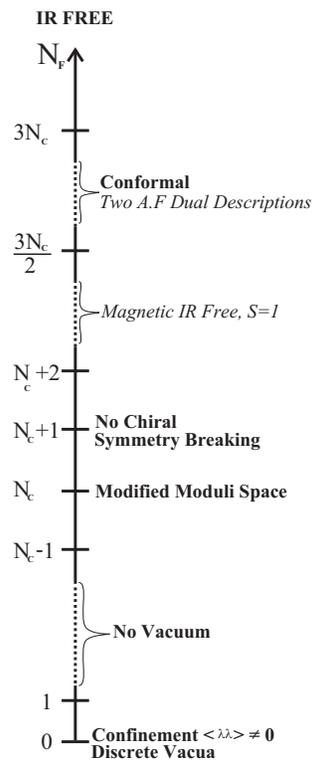} 
\caption[]{The phases of super QCD}
\label{figure} \end{figure}

\subsection{Higgs and confinement phases}

This section is discussed in \cite{Banks:1979fi}. While one is discussing the confinement phase in supersymmetric gauge theories one should recall that for gauge theories such as $SU(N)$ Yang Mills, with matter in a nontrivial representation of the center of the group, which is $Z_N$ for $SU(N)$, the difference between the Higgs and confinement phases is purely quantitative. There is no phase boundary. This contrasts the situation of pure QCD or super QCD where all the particles are in the adjoint representation which is trivial under the center. In such a case there is a phase transition and there is a qualitative difference between the phases. So what about the standard model, $SU(2) \times U(1)$. Is it in a Higgs or confinement phase? Below we present the spectrum in the two pictures.
In the Englert picture:
\bea
s=0 &&\; I={\half} \; \qquad \phi \rightarrow \phi_{real} \\
s={\half}&& \; I={\half} \; \qquad (l)_L \; (q)_L \;  \\
s={\half} &&\; I=0 \; \qquad (l)_R \; (q)_R
\eea
In the confinement picture, 
\beq
(l)_R, (q)_R 
\eeq
are $SU(2)_L$ singlets. Along with,
\bea
s={\half} &&\; \; \; \phi^+\psi_i \comma \quad \epsilon_{ij} \psi_i \phi_j \\
s=0 &&\; \; \; \phi^+_i \phi_i  \\
s=1 &&\; \; \; \phi_i D_{\mu} \phi_j \epsilon_{ij} \comma \quad \phi_i^+ D_{\mu} \phi^+_j \epsilon_{ij} \comma \quad \phi^+_i D_{\mu} \phi_i
\eea
One may choose a gauge,
\bea
\phi(x)= \Omega(x) \lf \matrix{ \rho(x) \cr 0 } \rt
\eea
then 
\beq
B_{\mu}= \Omega^+ D_{\mu} \Omega
\eeq
leading to the Lagrangian,
\beq
{\cal{L}}= {\rm{tr}} F_{\mu \nu}(B) F^{\mu \nu}(B)  + \pl_{\mu} \rho \pl^{\mu} \rho + \rho^2 (B^+_\mu B^\mu)_{||} +V(\rho^2)
\eeq
Unitary gauge is $\Omega=1$.
The Higgs picture also contains the operators:
\beq
\psi_1 = {{\phi_i^+ \psi_i} \over {|\phi|}} \comma \; \psi_2 ={{ \phi_i \psi_j \epsilon_{ij}} \over { |\phi|}} \comma \; {\tilde{W}}_{\mu}^+ = {{\phi^+_i} \over {|\phi|}} D_\mu  {{\phi^+_j} \over {|\phi|}} \epsilon_{ij} \comma \; {\tilde{W}}^0 = {{\phi^+_i} \over {|\phi|}} D_\mu  {{\phi^+_i} \over {|\phi|}}
\eeq

Like confinement but with the scale determined by: $|\phi|$. At finite temperature the two phases are qualitatively indistinguishable.

Examine the charges of the fields with respect to the unbroken $U(1)$.
In the Higgs picture,  
\beq
Q(\psi)= {\half} e \quad Q(\phi)={\half}e \quad  Q(W^0)=0
\eeq
However, the confined objects have integral charge.
The appropraite conserved charge is actually:
\beq
Q' = Q + T_3
\eeq
Then,
\beq
Q'(e)=1 \quad Q'(W^0)=0 \quad Q'(v)=0 \quad Q'(W^\pm) = \pm 1 \quad Q'(\rho)=0
\eeq
and
\bea
&& Q'(\psi \phi)=1 \quad Q'(\phi^+ D_\mu \phi)=0 \quad Q'( \phi+ \psi)=0 \\
&& Q'(\phi^+ D_\mu \phi^+) =1 \quad Q'(\phi^+ \phi ) =0 \quad Q'(\phi D_\mu \phi )=-1
\eea
This matches the charges of operators in the confined picture.

Is it possible to have strong ``weak'' interactions at high energies? This is theoretically possible but it is not the course chosen by nature for the standard model. This is seen empirically by the absence of radial excitations of the Z particles \cite{Abbott}.

Note, this will not work for the Georgi-Glashow model where $\phi$ is a triplet in $SO(3)$.

\subsection{Infra-red Duality}

Two systems are called infra-red dual if, when observed at longer and
longer length scales, they become more and more similar.

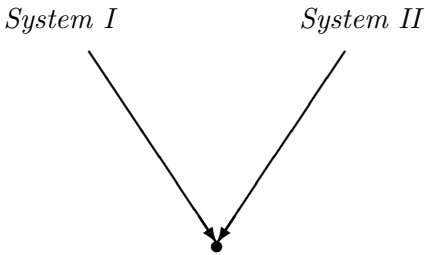
\begin{figure}[htp]
\begin{center}
\begin{picture}(130,100)
\setlength{\unitlength}{1mm}
\thicklines
\put(24,-2) {$\bullet$}
\put(42,25){\vector(-2,-3){17}}
\put(8,25){\vector(2,-3){17}}
\put(4,29){\makebox(0,0){\em System I}}
\put(44,29){\makebox(0,0){\em System II}}
\end{picture}
\end{center}
\caption{Two systems with a different ultra-violet behavior
flowing to the same infra-red fixed point.}
\label{infraredduality1}
\end{figure}

Seiberg has observed and has given very strong arguments that the
following set of N=1 supersymmetric gauge theories are pairwise
infra-red dual \cite{S2}.

$$
\begin{array} {ccrccc}
\underline{System} & &  & \underline{Dual \; System} & &\\
\\
Gauge \; Group & \#\mbox{flavors}  &               & Gauge \, Group   & \# flavor &
\#\mbox{singlets} \\
\\
SU(N_C)     & N_F        &               & SU(N_F-N_C)   & N_F    & N_F^2   \\
SO(N_C)     & N_F        &               & SO(N_F-N_C+4) & N_F    & N_F^2    \\
Sp(N_C)     & 2N_F        &               & Sp(N_F-N_C-2) & 2N_F    & N_F^2    \\
\end{array}      
$$

For a given number of colors, $N_C$, the number of flavors, $N_F$, for which the infrared
duality holds is always large enough so as to make the entries in the table meaningful.
Note that the rank of the dual pairs is usually different.
Lets explain why this result is so powerful.
In general, it has been known for quite a long time that two systems
which differ by irrelevant operator have the same infra-red behavior.
We have no indication whatsoever, that this is the case with Seiberg's
duality, where groups with different number of colors are infra-red dual.
Nevertheless, the common wisdom in hadronic physics has already identified
very important cases of infra-red duality.
For example, QCD, whose gauge group is $SU(N_C)$ and whose flavor group is 
$SU(N_F) \times SU(N_F) \times U(1)$, is expected to be infra-red dual to
a theory of massless pions which are all color singlets.
The pions, being the spin-0 Goldstone Bosons of the spontaneously broken
chiral symmetry, are actually infra-red free in four dimensions.
We have thus relearned that free spin-0 massless particles can actually be the
infra-red ashes of a strongly-interacting theory, QCD whose ultraviolet behavior 
is described by  other  particles.
By using supersymmetry, one can realize  a situation where free massless
spin-$\frac{1}{2}$ particles are also the infra-red resolution of another
theory.
Seiberg's duality allows for the first time to ascribe a similar role to massless
infra-red free spin-1 particles.
Massless spin-1 particles play a very special role in our understanding of the 
basic interactions.
This comes about in the following way: Consider, for example, the N=1 supersymmetric
model with $N_C$ colors and $N_F$ flavors.
It is infra-red dual to a theory with $N_F - N_C$ colors and $N_F$ flavors and
$N_F^2$ color singlets.
For a given $N_C$, if the number of flavors is in the interval 
$N_C +1 <  N_F < \frac{3N_C}{2}$, the original theory is strongly coupled in the
infra-red, while the dual theory has such a large number of flavors that it becomes  infra-red
free. Thus the infra-red behavior of the strongly-coupled system is described by infrared free
spin-1 massless fields (as well as its superpartners),
that is, Seiberg's work has shown that infrared free massless spin-1 particles(for example
photons in a SUSY system) could
be, under certain circumstances, just the infra-red limit of a much more complicated ultraviolet
theory.
Seiberg's duality has passed a large number of consistency checks under many
circumstances.

The infrared duality relates two disconnected systems.
From the point of view of string theory
the two systems are embedded in a larger space of models, such that a continuous
trajectory relates them. We will describe the ingredients of such an embedding \cite{EGK,EGKRS} later. In order to be able to appreciate how that is derived, we will need to learn to 
use some tools of string theory.

\begin{figure}[htp]
\begin{center}
\begin{picture}(130,180)
\setlength{\unitlength}{1mm}
\thicklines
\put(4,60){$\bullet$}
\put(34,60){$\bullet$}
\put(5,55){\makebox(0,0){I}}
\put(35,55){\makebox(0,0){II}}
\put(68,60){\makebox(0,0){$Z_2$}}

\put(-10,5){\line(0,4){40}}
\put(-10,5){\line(10,0){65}}
\put(55,5){\line(0,4){40}}
\put(-10,45){\line(10,0){65}}
\put(4,30){$\bullet$}
\put(34,30){$\bullet$}
\put(20,30){\oval(30,10)[t]}
\put(5,25){\makebox(0,0){I}}
\put(35,25){\makebox(0,0){II}}
\put(68,30){\makebox(0,0){continous}}

\end{picture}
\end{center}
\caption{In string theory, a continuous path in parameter space
relates a pair of two disjoint infrared-dual field theories.}
\label{infraredduality2}
\end{figure}
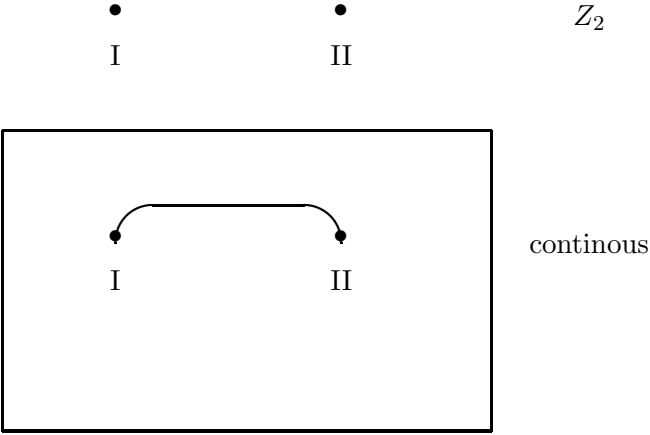

First we will describe some more details of the Seiberg infra red duality in field theory. Consider the example of N=1 supersymmetric Yang-Mills theory with gauge group $SU(N_C)$ and $N_F \comma {\bar{N}}_F$ fundamental, anti-fundamental matter. The charges of the matter fields are given by the table below:
\bea
\matrix{  & SU_L(N_F)&SU_R(N_F)&U_B(1)&U_R(1) \cr
        Q &  N_F     &   1     &  1   &1-{{N_C} \over{N_F}} \cr
{\tilde{Q}} &  1     & {\bar{N}}_F &-1&1-{{N_C} \over{N_F}} \cr }
\eea
The infra-red dual is N=1 supersymmetric Yang-Mills with gauge group $S(N_F-N_C)$ and $N_F \comma {\bar{N}}_F$ fundamental and antifundamental matter and $N_F^2$ gauge singlets. The charges are given by:
\bea
\matrix{  & SU_L(N_F)&SU_R(N_F)&U_B(1)&U_R(1) \cr
       q &  {\bar{N}}_F &1&  {{N_C} \over {N_F -N_C}} &1-{{N_C} \over{N_F}} \cr
{\tilde{q}} &  1     & {{N}}_F &-{{N_C} \over {N_F -N_C}} &1-{{N_C} \over{N_F}} \cr 
M& N_F&{\bar{N}}_F &0& 2 {{N_F-N_C} \over {N_F}} \cr }
\eea
One must also add an interaction term in the dual theory described by:
\beq
W= {1 \over \mu} M^i {}_{\tilde{j}} q_i q^{\tilde{j}}  \label{sint}
\eeq

\begin{itemize}
\item The dual theories have different gauge groups. If one regards a gauge symmetry as a redundancy in the description of the theory then this is not important. What does matter is that the two dual theories share the same global symmetries.
\item  Note, it is not possible to build a meson out of $q, {\tilde{q}}$ that has the same R-charge as a meson built from $Q,{\tilde{Q}}$. The M field in the dual theory does have the same charges as a meson built from  $Q,{\tilde{Q}}$.
\item The Baryons built from $Q,{\tilde{Q}}$ have the same charges as those built from  $q, {\tilde{q}}$. For the case $N_F=N_C+1$ then the Baryon of the $SU(N_C)$ theory becomes the q in the dual theory (which is a singlet in this case). This looks like there is a solitonic dual for the quarks in this case. 
\item If M is fundamental there should be an associated $U_M(1)$ charge which does not appear in the original $SU(N_C)$ theory. 
\item Where are the $q, {\tilde{q}}$ mesons in the original $SU(N_C)$ theory?
\item the resolution to the previous two points is provided by the interaction term, \ref{sint}. This term breaks the $U_M(1)$ symmetry and provides a mass to the $q, {\tilde{q}}$ mesons, which implies one may ignore them in the infra-red.
\end{itemize}

For the case, ${{3 N_C} \over 2} < N_F < 3N_C$, and for ${3 \over 2}(N_F-N_C)< {\tilde{N}}_F < 3(N_F-N_C)$. The operator, $M q {\tilde{q}}$ has dimension:
\beq
D(Mq {\tilde{q}})= 1 + {{3N_C} \over {N_F}} <3
\eeq
and so it is a relevant operator.
In both dual pictures there is an Infra red fixed point; both are asymptotically free and in the center of moduli space  the theories will be a conformal.

The checks of the duality are as follows:
\begin{itemize}
\item They have the same global symmetries.
\item They obey the 't Hooft anomaly matching conditions.
\item It is a $Z_2$ operation.
\item There are the same number of flat directions.
\item There is the same reaction to a mass deformation. Adding a mass in one theory is like an Englerting in the other.
\item There is a construction of the duality by embedding the field theory in string theory. This will be the subject of the next section.
\end{itemize}

The 't Hooft anomaly matching conditions are determined as follows. One takes the global symmetries in the theory and then make them local symmetries. One then calculates their anomalies. Both dual theories must share the same anomalies.
In the above example there are anomalies for:
\bea
&&SU(N_F)^3\comma SU(N_F)^2 U_R(1) \comma SU(N_F)^2 U_B(1) \comma 
 U_R(1) \comma \\ && U_R(1)^3 \comma  U_B(1)^2 U_R(1) \comma U_B(1)^3 \comma U_B(1)^2 U_R(1)
\eea
All these anomalies match between the dual theories. There was no a priori reason for them to do so.

Let us examine some of the consequences of this duality. For the case $ {{3N_C} \over 2} < N_F < 3N_C$ the two dual theories are both asymptotically free. It is symmetric around $N_F=2N_C$. Perhaps one can more learn about this system since it is a fixed point under duality. At the origin of moduli space one may have obtained a new conformal theory- this will be discussed later. For $N_C+2 \leq N_F \leq 3 {{N_C} \over 2}$, the theory is an infra red free gauge theory plus free singlets. This is the first example of a weakly interacting theory with spin one particles that in the infra red one may view as bound states of the dual theory. The panorama of these structures is given in figure 8.

Let us now enrich the structure of the theory by adding $N_a$ particles in the adjoint representation. At first we will have no matter in the fundamental representation and scalar multiplets which are adjoint valued. The potential for the scalars, $\phi_i$ is given by:
\beq
V= ([ \phi,\phi ])^2  \; .
\eeq
This potential obviously has a flat direction for diagonal $\phi$. The gauge invariant macroscopic moduli would be ${\rm{Tr}} \phi^k$. Consider the non generic example of $N_C=2$ and $N_a=1$, the supersymmetry is now increased to N=2. There is a single complex modulus, ${\rm{Tr}}\phi^2$. Classically, $SU(2)$ is broken to $U(1)$ for ${\rm{Tr}}\phi^2 \neq 0$. One would expect a singularity at 
${\rm{Tr}}\phi^2=0$. The exact quantum potential vanishes in this case \cite{Seiberg:1994aj,Seiberg:1994rs}.

\begin{figure}[hbtp] 
\centering
\includegraphics[width=10.truecm]{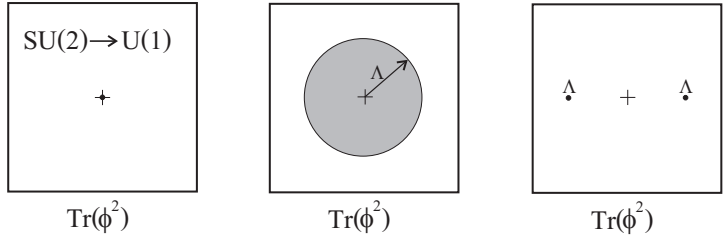} 
\caption[]{The classical, naive quantum and exact quantum moduli spaces}
\label{figmod} \end{figure}

Naively, one would expect the following. When ${\rm{Tr}}\phi^2$ is of order $\Lambda$ or smaller, one would expect that the strong infra red fluctuations would wash away the expectation value for ${\rm{Tr}}\phi^2$ and the theory would be confining. The surprising thing is that when SU(2) breaks down to U(1), because of the very strong constraints that supersymmetry imposes on the system, there are only two special points in moduli space and even there the theory is only on the verge of confinement. Everywhere else the theory is in the Coulomb phase. At the special points in the moduli space, new particles will become massless. This is illustrated in the figure \ref{figmod}.

We will now examine the effective theory at a generic point in moduli space where the theory is broken down to U(1).
The Lagrangian is given by,
\beq
{\cal{L}}= \int d^2 \theta Im (\tau_{eff} ({\rm{tr}}\phi^2,g,\Lambda) W_\alpha W^\alpha)
\eeq
The $\tau_{eff}$ is the effective complex coupling which is a function of the modulus, ${\rm{tr}}\phi^2$, the original couplings and the scale, $\Lambda$. This theory has an $SL(2,Z)$ duality symmetry. The generators of the $SL(2,Z)$ act on $\tau$, defined by \ref{tau}, as follows:
\beq
\tau \rightarrow -{1 \over \tau} \comma \; \tau \rightarrow \tau+1
\eeq
This is a generalization of the usual U(1) duality that occurs with electromagnetism to the case of a complex coupling.
Recall the usual electromagnetic duality for Maxwell theory in the presence of charged matter is:
\beq
E\rightarrow B \comma B \rightarrow -E \comma e \rightarrow m \comma m \rightarrow -e \; \; .
\eeq
This generalizes to a U(1) symmetry by defining:
\beq
E + i B \comma \; e+i m \; \; .
\eeq
The duality symmetry now acts by:
\beq
E+iB \rightarrow \exp(i\alpha) (E+iB) \comma e+im \rightarrow \exp(i\alpha) (e +im) \; \; .
\eeq

Previously for the $SU(2)$ case the moduli were given by $u={\rm{Tr}}\phi^2$ for $SU(N_C)$ the moduli are given by $u_k={\rm{Tr}}\phi^k \comma  k=2,..,N_C$. Again the classical moduli space is singular at times, there is no perturbative or nonperturbative corrections.

How does one find $\tau$ as a function of the $u$? There is a great deal of literature on the subject here we will just sketch the ideas \cite{Seiberg:1994aj,Seiberg:1994rs,Bilal:1995hc}. 

The following complex equation, 
\beq
y^2=ax^3+bx^2 +cx+d
\eeq
determines a torus. The complex structure of the torus, $\tau_{torus}$ will be identified with the complex coupling $\tau_{eff}$. a,b,c,d are holomorphic functions of the moduli, couplings and scale and so will implicitly determine $\tau_{torus}$.

When $y(x)$ and $y'(x)$ vanish for the same value of x then $\tau$ is singular. Therefore,
\beq
\tau_{eff}=i \infty \comma g^2_{eff} =0
\eeq
and the effective coupling vanishes. This reflects the presence of massless charged objects. This occurs for definite values of u in the moduli space. These new massless particles are monopoles or dyons. The theory is on the verge of confinement. For N=2 supersymmetry that is the best one can do. The monopoles are massless but they have not condensed \cite{I,IS1,IS2,efgr,efgr2,efgr-proc,efgr3}. For condensation to occur the monopoles should become tachyonic indicating  an instability that produces a condensation. One can push this to confinement by adding a mass term: ${\tilde{m}} {\rm{Tr}} \phi^2$, or generally for $SU(N_C)$ the term:
\beq
\delta W=g_k u_k \; .
\eeq
The effective prepotential is now:
\beq
W=M(u_k) q {\tilde{q}} + g_k u_k
\eeq
then
\beq
{{\pl W} \over {\pl u_k}}=0 \; \comma {{\pl W } \over {\pl (q {\tilde{q}})}}=0
\Rightarrow M(<u_k>)=0 \comma \pl_{u_k} M(<u_k>) <q {\tilde{q}}> =-g_k
\eeq
Since generically,
\beq
\pl_{u_k} M(<u_k>) \neq 0
\eeq
then there will be condensation.

We now describe how the complex elliptic curve arises using more physical terms. This is achieved here by using the integrating in method discussed earlier.
Consider the case, $N_C=2$ with arbitrary $N_F$ and $N_a$.
The fields that are the moduli in the system are:
\bea
X_{IJ}& =&\epsilon_{ab} Q^a_i Q^b_j \\
M_{\alpha \beta}&=& \epsilon_{aa'} \epsilon_{bb'} \phi_\alpha^{ab} \phi^{a'b'}_{\beta} \\
Z_{ij}&=& \epsilon_{aa'} \epsilon_{bb'} Q_i^a \phi^{a'b'}_\alpha Q^b_j
\eea
where Q are fundamental and $\phi$ are adjoint fields. $\alpha, \beta =1..,N_a \comma i,j= 1..,2N_F \comma a,b=1,2$.
We define the quantity,
\beq
\Gamma_{\alpha \beta}(M,X,Z) = M_{\alpha \beta} + {\rm{Tr}}_{2N_F} (Z_\alpha X^{-1} Z_{\beta} X^{-1})
\eeq
which we will use to write the prepotential as follows,
\bea
W_{N_F,N_a}(M,X,Z)&=&(b_1-4) \lf \Lambda^{-b_1}PfX(\det (\Gamma_{\alpha \beta}))^2 \rt ^{1 \over {4 -b_1}} \\&+&{\rm{Tr}}_{N_a} {\tilde{m}} M {\half} {\rm{Tr}}_{2N_F} mX \\ &+& { 1 \over \sqrt{2}} {\rm{Tr}}_{2N_F} \lambda^\alpha Z_\alpha
\eea
This respects the necessary symmetries and can be checked semiclassically.
Take the case $N_a=1 \comma N_F=2$.
The equations of motion from minimizing the superpotential are:
\beq
{{\pl W_{2,1}} \over { \pl M }}= {{\pl W_{2,1}} \over { \pl X}} = {{\pl W_{2,1}} \over \pl Z}=0
\eeq
which imply:
\bea
{\tilde{m}}&=& 2 \Lambda^{-1} (PfX)^{\half}  \\
m&=& R^{-1} (X^{-1} -8 \Gamma^{-1} X^{-1} (ZX^{-1})^2)\\
{ 1\over {\sqrt{2}}} \lambda &=& 4 R^{-1} \Gamma^{-1} X^{-1} Z X^{-1} \\
\eea
where 
\beq
R^{-1} \equiv \Lambda^{-} (PfX)^{\half} \Gamma  \comma X\equiv {\half} \Gamma
\eeq
The following equations are then obeyed:
\bea
X^3 -MX^2 +b X- {1 \over 128} (c-8M){\tilde{c}} =0  \\
X^2 -2MX +b=0
\eea
Taking y and y' to vanish we can compare with the elliptic cure,
\beq
y^2=x^3+ax^2+bx+c
\eeq
One can therefore identify the parameters as:
\bea
a=-M&&b=-{{\alpha} \over 4} + {{ \Lambda}^2 \over 4} Pf m \\
c={ \alpha \over 8}(2M +{\rm{Tr}}(\mu^2)) && \alpha \equiv {{\Lambda^4} \over 16} \comma \mu \equiv \lambda^{-1} m
\eea
Identifying the modular parameter of the torus from the elliptic equation involves standard techniques in algebraic geometry. This modular parameter will then be the effective coupling of the theory.  

Some comments:
\begin{itemize}
\item Some points in moduli space when $2+N_F=4$, are degenerate vacua which are possibly non-local with respect to each other. These are Argyres Douglas points \cite{AD}.
\item As you move in moduli space monopoles turn smoothly into dyons and electric charge. This is an indication of the Higgs/confinement complementarity.
\item These techniques may be extended to obtain curves for other more complicated groups.
\item One can also attempt to preserve the crucial analytic structure of the theory while breaking the supersymmetry so that one may learn about theories with fewer (or no supersymmetry). This is known as soft supersymmetry breaking \cite{luis}. 
\end{itemize}

We now rexamine some special properties of the region ${3 \over 2} N_C < N_F < 3N_C$.

\subsection{Superconformal Invariance in d=4}
For the case of ${3 \over 2} N_C < N_F < 3N_C$, at the center of moduli space when all expectation values vanish, it is claimed that the theory is described by a non-trivial conformal field theory \cite{banksz}. There are several motivations for reaching this conclusion. Examine for example, the exact $\beta$ function:
\beq
\beta(g) =-{{g^3} \over{16 \pi^2}} {{3N_C -N_F +N_F \gamma(g^2)} \over {1- {{g^2} \over {8 \pi^2}} N_C}}
\eeq
where $\gamma(g^2)$ is the anomalous dimension given perturbatively by:
\beq
\gamma(g^2) = - {{g^2} \over {8 \pi^2}} {{N_C^2-1} \over {N_C}} + O(g^4) \; .
\eeq
If one now considers a limit where $N_F,N_C$ are both taken to be large but their ratio is kept fixed then the fixed point of the $\beta$ function,  $\beta(g^*)=0$ , is at
\beq
g^{*2}= {1 \over N_C} {{ 8 \pi^2} \over {3}} \epsilon \comma
\eeq
where $\epsilon= 3-{{N_F} \over {N_C}} << 1$. Since the coupling at the fixed point is proportional to $\epsilon<<1$, both the existence of a fixed point and the perturbative evaluation for $\gamma$ is justified.
The anomalous dimension at the fixed point is:
\beq
\gamma(g^*)=1 - {{3N_C} \over {N_F}}
\eeq
The dimension of the $Q {\tilde{Q}}$ is
\beq
D(Q {\tilde{Q}})=2 +\gamma= 3 {{N_F-N_C} \over {N_F}} \label{dqq}
\eeq
This observatoin will lead to interesting conclusions.

In d=2 the conformal group is infinite dimensional and so provides very powerful constraints on the theory. In d=4 the conformal group is finite dimensional and so the conformal symmetry does not constrain the theory in the same way \cite{Mack,Dobrev}. Nevertheless there are still interesting properties of d=4 conformal theories arising from the conformal invariance. One can prove that the scaling dimension of a field is bounded by its R-charge as follows:
\beq
D \geq {3 \over 2} |R|     \label{chisat}
\eeq
where D is the scaling dimension and R is the R-charge. The bound is saturated for (anti-)chiral fields.
Consider the operator product:
\beq
O_1(x^1) O_2(x^2)  = \sum_i O^i_{12} f^i(x^1-x^2) \label{ope}
\eeq
Generally, the dimension of the product of the operator appearing on the righthand side of equation \ref{ope} is not a sum of the dimensions of the two operators appearing on the left hand side of equation \ref{ope}. The product contains a superposition of operators with different dimensions.
For chiral fields the situation is simpler. Since the R-charge is additive and chiral fields saturate the bound \ref{chisat}
\beq
R(O^i_{12}) = R (O_1(x^1)) + R(O_2(x^2)) \Rightarrow D(O_1 O_2) =D(O_1)+D(O_2) \label{ope2}
\eeq
and $f^i(x^1-x^2) $ are thus all constants. By \ref{ope2} $O^i_{12}$ is also a chiral operator. The closure property of the chiral fields under the operator product expansion leads to the name ``chiral ring''.

There are reasons to expect that at the fixed point, the infra-red non-anomalous R-charge equals the non-anomalous R-charge of the ultra violet. At the fixed point of the dimension of $Q {\tilde{Q}}$ is given by, see equation (\ref{dqq}):
\beq
D(Q{\tilde{Q}})= {3 \over 2} R(Q {\tilde{Q}}) = 3 {{N_F-N_C} \over N_F}
\eeq
The dimension of the Baryon and anti-Baryon are:
\beq
D(B)=D({\tilde{B}})= {{3N_C(N_F-N_C)} \over {2N_F}}
\eeq
For unitary representations (of spin=0) fields, where I is the identity operator and $O$ is an operator $\neq I$,
\beq
D(I)=0 \comma D(O)\geq 1
\eeq
For a free field the bound is saturated,
\beq
D(O)=1 \; .
\eeq
When $N_F= {3\over 2} N_C $, the operator $Q {\tilde{Q}}$ becomes free, since $D(Q {\tilde{Q}})=1$. For $N_C+1 < N_F < {3 \over 2} N_C $ it appears that $D(Q {\tilde{Q}})<1$ which is forbidden. This is an indication that one is using the wrong degrees of freedom and a dual description is required. This will be elaborated later.

 Non-trivial superconformal N=2 theories in d=4 occur for N=2, SU(3) theories without matter or SU(2) theories with matter \cite{Argyres:1995xn}. The key point is that a non-trivial conformal theory with vector fields contains both massless electric and magnetic excitations (these are mutually non-local).

The definition of a primary state is that it is annihilated by the generator of special conformal transformations, K. The descendents are obtained by acting on primary states with momentum operators. The Lorentz group decomposes into $SU(2)_L\oplus SU(2)_R$ with charges $j,{\tilde{j}}$ respectively. An operator with non-zero spin will be carry $j, {\tilde{j}}$ charge as well as $D$ the dimension.

Unitary chiral primaries obey: $j {\tilde{j}}=0$. The dimension of a chiral filed is:
\beq
D(O) \geq j +{\tilde{j}} +1
\eeq
as compared with a non-chiral operator:
\beq
D(O) \geq j +{\tilde{j}} +2  \; \; .
\eeq
For a free field:
\beq
D(O)= j + {\tilde{j}}+1 \; .
\eeq
This generalizes the previous results for scalar chiral fields.
$F_{\mu \nu}$  decomposes into self-dual and anti self-dual parts that form irreducible representations of $SU(2)_L\oplus SU(2)_R$. 
\beq
F=F^+ + F^- \comma F^\pm = F \pm *F
\eeq
One can show that there are states associated with the conserved currents:
\beq
J_{\mu}^\pm = \pl^\nu F^\pm_{\mu \nu} \; .
\eeq
These states are the decendents of $F^\pm$, obtained by applying the momentum operator P on the chiral field, $F^\pm$. The norm is then calculated using the following: the commutator of the d=4 conformal algebra,
\beq
[P^{\aad}, K^{\bbd}] = {i \over 2} M^{\alpha \beta} \epsilon^{\ad \bd} +{\bar{ M}}^{\ad \bd} \epsilon^{\alpha \beta} + D \epsilon^{\alpha \beta} \epsilon^{\ad \bd} \comma
\eeq
where the undotted indices are $SU(2)_L$ indices, dotted indices are $SU(2)_R$ indices and  ${i \over 2}M^{12}=J^3$; the Hermitian conjugate relation,
\beq
(P^{\aad})^+ = - \epsilon^{\alpha \beta} \epsilon^{\ad \bd} K^{\beta \bd} \; \; ;
\eeq
and that K annihilates primary states. After some algebra this produces,
\beq
|\,|J^\pm>\, |=2(D-2)
\eeq
For D=2, these are null states and $F^{\pm}$ is free. If F is not free and $D>2$, then
\beq
|\, |J^\pm>\, | > 0
\eeq
and therefore both,
\beq
J^e\equiv J^+ +J^- \comma J^m \equiv J^+-J^-
\eeq
are non-vanishing \cite{Argyres:1995xn}.

At the point in moduli space (discussed previously) where there are both electric and magnetic charges, called the Argyres Douglas point \cite{AD}, this condition is met and it is possible to have a non-trivial conformal field theory with spin one particles.

Other conformal theories occur for N=1 supersymmetry when $N_a=3$ and the  couplings are appropriately tuned; this is actually a deformed N=4 theory. For N=2 supersymmetry, the theory is conformal  when $N_F=2N_C$ and the one loop $\beta$ function vanishes; this is known to be an exact result \cite{Leigh:1995ep}.

One can deform $N=4$ theories with marginal operators such that the global symmetries are broken but the theory remains conformal. Hence, consider,
\beq
{\cal{L}}={\cal{L}}_0 + \sum_i g_i O^i
\eeq
where ${O^i}$ are the set of operators with dimension 4.

Consider N=1 with $N_a=3$ and an interaction for the adjoint fields $X_1,X_2,X_3$:
\beq
L_{int}= h X_1 X_2 X_3 \;
\eeq
When $h=g$, g being the gauge coupling, the theory has full N=4 supersymmetry.

For a general $h \phi_1..\phi_n$ perturbation one has 
\beq
\beta_h = h(\mu) (-d_w + \sum_{k=1}^n (d(\phi_k) + {\half} \gamma({\phi_k})))
\eeq
where $d_w$ is the engineering dimension of the composite perturbing term, $d(\phi_k)$ is the engineering dimension of the field, $\phi_k$ and $\gamma(\phi_k)$ is the associated anomalous dimension of the field $\phi_k$. By symmetry all fields have the same anomalous dimension and so $\gamma(\phi_k)$ is independent of k and so is denoted as simply $\gamma(\phi)$. 
In the case at hand 
\beq
d_w=3 \comma \quad \sum_{k=1}^3 d(\phi_k)=3 \comma \quad \sum_k {\half} \gamma(\phi_k) = {3 \over 2} \gamma
\eeq
and therefore
\beq
\beta_h= h(\mu) {3 \over 2} \gamma(\phi)
\eeq
and
\bea 
\beta_g &=& - f(g(\mu)) ( (3C_2(G) - \sum_k T(R_k)) + \sum_k T(R_k)\gamma(\phi_k)) \\
&=& - f(g(\mu)) 3 T(R)\gamma(\phi)
\eea
which is non vanishing. $C_2(G) $ is the second Casimir of the group G and $T(R_k)$ is associated with the representation of $\phi$.

\begin{figure}[hbtp] 
\centering
\includegraphics[width=6.truecm]{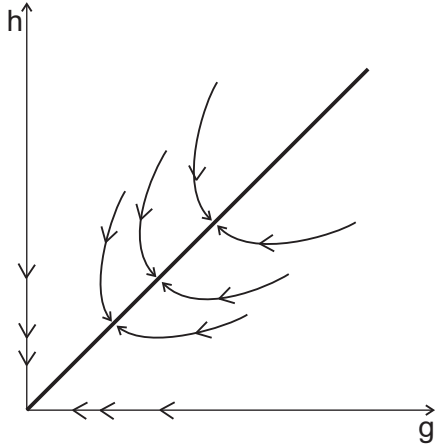} 
\caption[]{RG flow to the fixed line}
\label{figure} \end{figure}

Therefore, both $\beta_h$ and $\beta_g$ vanish if $\gamma $ vanishes. This means there is a fixed line, rather than a fixed point, where g=h and the supersymmetry is enhanced to N=4. (This is different from the more generic situation where the $\beta$ functions are not related and there are isolated fixed points.) This fixed line is infra-red stable. Also the relations between relevant deformations follow a similar pattern \cite{Einhorn:wp}.

At the fixed line, one can ask if the $N=4 $ theory has any discrete symmetries that relate theories with different moduli. The answer is yes \cite {kol}. There is a great deal of evidence that N=4 theories possess an SL(2,Z) duality \cite{olive} that identifies theories with coupling $\tau$ given by equation (\ref{tau}):
\beq
\tau \rightarrow { {a \tau +b } \over {c \tau +d}}  \comma \quad ad-bc =1 \comma a,b,c,d \in Z  \;.
\eeq

To conclude, supersymmetric gauge theories have a very rich phase structure and many outstanding dynamical issues can be discussed reliably in the supersymmetric arena that are hard to address elsewhere.

\section{Comments on vacuum energies in scale invariant theories}

The puzzle to be addressed in this section is that of the cosmological constant problem \cite{zel,Weinberg:1988cp}. The problem was originally stated as follows. Given that there is an effective scale below which the physics is known then (such as QED or the standard model) integrating out the more energetic degrees of freedom above this scale leads to a vacuum energy that is proportional to the cut off.  For either choice of the above physics, the vacuum energy calculated in such a manner however would give a cosmological constant many orders of magnitude above the observed value. 
One should observe some important caveats in this argument \cite{Amit:ri,Rabinovici:tf} and also \cite{fub,Einhorn:wp}. When one writes down a low energy effective action for a spontaneously broken theory  one should respect all the symmetries that appear in the original action. If the original theory is scale invariant then the effective action of the spontaneously broken theory should reflect this symmetry. The consequence of scale invariance is a zero vacuum energy whether or not scale invariance is spontaneously broken.

As an example consdier $N=4$ super Yang-Mills in four dimensions. The potential is given by :
\beq
V= [\phi,\phi]^2
\eeq
thus the theory has flat directions. This result is exact to all orders in perturbation theory and non-perturbatively. Giving an expectation value to a field will break the scale invariance spontaneously as well as generically breaking the gauge symmetry down to $U(1)^r$ where r is the rank of the unbroken gauge group.
An analysis of the spectrum shows that a gauge singlet particle emerges, which is the dilaton (the Goldstone Boson associated with spontaneously broken scale invariance). The vacuum energy remains zero.

Thus provided there is a translationally invariant ground state,  global supersymmetry can't be spontaneously broken whether or not scale invariance is spontaneoulsy broken. The presence of a ground state is crucial; recall the example discussed in quantum mechanics where the action  scale invariance and supersymmetry were spontaneously broken but the system had no ground state.

The next example is the O(N) model in three dimensions, described by the Lagrangian,
\beq
{\cal{L}}=\pl_{\mu} {\vec{\phi}} \cdot \pl^{\mu} {\vec{\phi}} - g_6 (|\vec{\phi}|^2)^3
\eeq
where the fields $\vec{\phi}$ are in the vector representation of O(N). In the limit, $N \rightarrow \infty$,
\beq
\beta_{g_6}=0
\eeq
(${1 \over N}$ corrections break the conformality). $g_6$ is a modulus.
An effective potential can be written down for the O(N) invariant field,
\beq
\sigma = |\vec{\phi}|^2
\eeq
the effective potential is:
\beq
V(\sigma)= f(g) |\sigma|^3  \; \; .
\eeq
One has the following possibilities, summarized in the  table below:
\bea
\matrix{ & <\sigma> & {\rm{S.B.}} & {\rm{masses}}& V \cr
f(g) >0  &   0     &      {\rm{No}} & 0  &0 \cr
f(g)=0   &  <\sigma>=0& {\rm{No}} &0&0 \cr 
f(g)=0   & <\sigma>\neq 0& {\rm{Yes}} & <\sigma>,0 &0}
\eea
where S.B. indicates spontaneous symmetry breaking of scale invariance and V is the vacuum energy. (For $f(g) <0$ the theory is unstable). Note, the vacuum energy  always vanishes.
To summarize, in the situation where $<\sigma> \neq0$ and the scale invariance is spontaneously broken, one could write down the effective theory for energy scales below $<\sigma>$ and integrate out the physics above that scale. The vacuum energy remains zero however and not proportional to $<\sigma>^3$ as is the naive expectation.

The final example will be the $O(N) \times O(N)$ model with two fields in the vector representation of O(N), with Lagrangian:
\bea
{\cal{L}}&=&\pl_{\mu} {\vec{\phi_1}} \cdot \pl^{\mu} {\vec{\phi_1}} 
+\pl_{\mu} {\vec{\phi_2}} \cdot \pl^{\mu} {\vec{\phi_2}} -  \lambda_{6,0} (|\vec{\phi_1}|^2)^3 \\
&-& \lambda_{4,2} (|\vec{\phi_1}|^2)^2 (|\vec{\phi_2}|^2)
- \lambda_{2,4} (|\vec{\phi_1}|^2) (|\vec{\phi_2}|^2)^2
- \lambda_{0,6} (|\vec{\phi_2}|^2)^3
\eea
Again the $\beta$ functions vanish in the strict $N \rightarrow \infty $ limit.
There are now two possible scales, one associated to the break down of a global symmetry and another with the break down of scale invariance.
The possibilities are summarized by the table below:
\bea
\matrix{ & O(N)&O(N)& {\rm{scale}}& {\rm{massless}}&{\rm{massive}} &V \cr
        & +&+&+&{\rm{all}}&{\rm{none}}&0 \cr
        &-&+&-&(N-1)\pi's, D&N,\sigma&0 \cr
	&+&-&-&(N-1)\pi's, D & N,\sigma&0 \cr
	&-&-&-& 2(N-1)\pi's,D & \sigma& 0 \cr }
\eea
Again, in all cases the vacuum energy vanishes.
Assume a hierarchy of scales where the scale invariance is broken at a  scale much above the scale at which the O(N) symmetries are broken. One would have argued that one would have had a low energy effective Lagrangian for the pions and dilaton with a vacuum energy given by the scale at which the global symmetry is broken. This is not true, the vacuum energy remains zero.

It was  proposed that the underlying physics of nature is scale invariant and the scale invariance is only removed by spontaneous symmetry breaking \cite{Amit:ri,Rabinovici:tf} and also \cite{fub,Einhorn:wp}. This would explain the vanishing of the cosmological constant. (The data in 2001 seems to indicate the presence of a small cosmological constant). A key problem with this scenario however is that we do not observe a dilaton in nature that would be expected if scale invariance is spontaneously broken. The hope is that this dilaton may somehow be given a mass with a further independent (Higgs like?) mechanism.

The total energy of the universe can be augmented even in a scale invariant theory. Such theories may have many superselection sectors (such as magnetic monopoles in N=4 super Yang-Mills). In each sector the total energy will differ from zero by the energy of these particles. If there are many of them they may mimic a very small cosmological constant \cite{rabidea}.

\section{Supersymmetric gauge theories and string theory}
We now view the supersymmetric gauge theories from a different point of view. We obtain them as the low energy limit of various string backgrounds. Many properties of gauge theories are obtained in this fashion \cite{GK,erice}. In the following discussion we will have as our goal to:

\begin{itemize}

\item Construct:
\begin{itemize}
\item a D=4 dimensional effective theory
\item with N=1 supersymmetry (SUSY)
\item with $U(N_c)$ gauge symmetry
\item with $SU(N_F) \times SU(N_F) \times U(1)$ global symmetry
\end{itemize}

\item Identify dualities in a pedestrian way.

\end{itemize}

Our tools for this project will be... comic strips \cite{HW}.
The unabridged original novel, from which this `comics illustrated' is derived
is yet to be written.
\noindent
The scaffolding for this construction will be extended objects called branes.

\subsection{Branes in String Theory}
Branes are extended object solutions which emerge non perturbatively in string theory
in a very similar way that solitons emerge in field theory.
Magnetic monopoles and vortices are examples of solitonic configurations in gauge
theories.
What are they good for?
First, there existence answers a yearning to search for more than meets the eye (or the
equations).
A yearning, which seems to be engraved in at least part of our community.
It, of course, also answers positively the important question of the
existence of a magnetic monopole, as well as that of other interesting topological
excitations.
These excitations are often very heavy and have little direct impact on
the low-energy dynamics.
However, there are circumstances, in which they
can condense and thus take over the  control and drive  the infra-red dynamics.
In the context we will discuss here, the branes will basically serve as
regulators for some field theories, playing the same role as string sizes
and lattice cut-offs.
In lattice gauge theories, such regulators have granted a very rapid
access to identifying non-perturbative features, such as confinement, in strong-coupling
approximations.
Similar consequences will occur here.
In string theory, at this stage, the regulator will also learn how to behave from the 
very theory it regulates.

\subsection{Branes in IIA and IIB String Theories}
For some years it has been known that solitons exist in the low-energy
effective theory of superstring theory/supergravity \cite{Duff}.
To appreciate that, we first review the spectrum of massless particles
in various string theories \cite{GSW}.
In the closed bosonic string theory, the spectrum consists of a graviton,
an anti-symmetric tensor, and a dilaton, denoted by $G_{\mu\nu}$, $B_{\mu\nu}$,
$\Phi$, $\mu=1,2,\ldots,24$, respectively.
The massless spectrum of the open bosonic string consists of photons $A_{\mu}$.
The bosonic sector of type-IIA theory consists of two sectors.
The first is called NS-NS sector and consists of the same spectrum as that
of the bosonic closed string theory, namely $G_{\mu\nu}$, $B_{\mu\nu}$,
$\Phi$.
The other sector is called the RR sector, consisting of one-form and three-form
vector potentials, denoted by $A_{\mu}$ and $A_{\mu\nu}$.
To each of them is associated an `electric field', carrying two and four indices,
respectively.
In type-IIB the NS-NS sector also consists of $G_{\mu\nu}$, $B_{\mu\nu}$, $\Phi$.
The RR sector consists of vector potentials which are 0-forms, 2-forms, and
4-forms.
In addition, duality relations in ten dimensions between the electric field
$E_{\mu_1,\ldots,\mu_{p+1}}$, derived from the p-form $A$, and the electric field
$E_{\mu_1,\ldots,\mu_{8-p-1}}$, derived from the (6-p)-form $A$ lead to further 
vector potentials, a 5-form in type-IIA and 4-forms and 6-forms in RR sector
of type-IIB.
Gravitons and the other particles in the NS-NS sector have a perturbative
string realization.
What about the objects in the RR sector?
First we note, that in the NS-NS sector, there exist solitons, discovered in
supergravity, whose tension $T$ is: $T_{NS-NS} = \frac{1}{g_s^2 l_s^6}$,
where $g_s$ is the string coupling and $l_s$ is the string scale.
These solitons spread over five space and one time dimensions.
The exact background corresponding to this configuration is not yet known
as the dilaton background seem to contain singularities.
In the RR sector, there are solitons as well \cite{Pol}.
The p+1 dimensional solitons have a tension which is: $T_{RR} = \frac{1}{g_s l_s^{p+1}}$.
These are called D-branes. 
The D stands for Dirichlet.
It turns out that open strings, obeying appropriate Dirichlet boundary conditions,
may end on these branes.
The solitonic sector of closed string theory contains configurations of open
strings.
One may inquire as to the effective theory on the six-dimensional world volume 
of the NS-NS soliton, the NS5 brane.
One may also inquire what are the supersymmetry properties of that theory.
It is easier to answer the second question.
The supersymmetry algebra:
\begin{eqnarray}
\{ Q_{\alpha}^i,\bar{Q}_{\beta \, j} \}
&=& 2 \sigma_{\alpha \beta}^{\mu} P_{\mu} \delta^{i}_{\;j}  \nn \\
\{ Q_{\alpha}^i,Q_{\beta}^j \}
&=& \{\bar{Q}_{\alpha \, i} , \bar{Q}_{\beta \, j} \} = 0 
\end{eqnarray}
contains on its right-hand side the generators of translation.
A soliton, existing in ten dimensions, whose extension is in p+1 space-time dimensions
breaks translational invariance in 9-p directions.
Thus it is not clear a priori if any of the supersymmetric generators,
residing in the left-hand side of the algebra, will survive intact.
It turns out that, in flat space-time, the system has 32 SUSY charges.
In the presence of the NS5 
configuration sixteen (half) of the SUSY charges survive.
They are of the form:
\beq
\epsilon_L Q_L + \epsilon_R Q_R  \label{survivingcharge}
\eeq
where $\epsilon_L$ and $\epsilon_R$ obey the constraints:
\begin{eqnarray}
\epsilon_L &=& \Gamma^0 \ldots \Gamma^5 \epsilon_L  \nn \\
\epsilon_R &=& \pm \Gamma^0 \ldots \Gamma^5 \epsilon_R  \; , \label{survivingns}
\end{eqnarray}
where the sign $\pm$ corresponds to type-IIA and type-IIB, respectively,
and the $\Gamma$-matrices are respective ten-dimensional Dirac $\Gamma$-matrices.

The answer to the first question is more complicated.
The massless sector on the NS5 in the type-IIA theory is a chiral system
consisting of a self-dual anti-symmetric vector potential, five scalars,
and their supersymmetric partners.
The massless sector on the NS5 in the type-IIB theory is a non-chiral
system consisting of spin-1 particles, spin-0 particles, and their
supersymmetric partners.

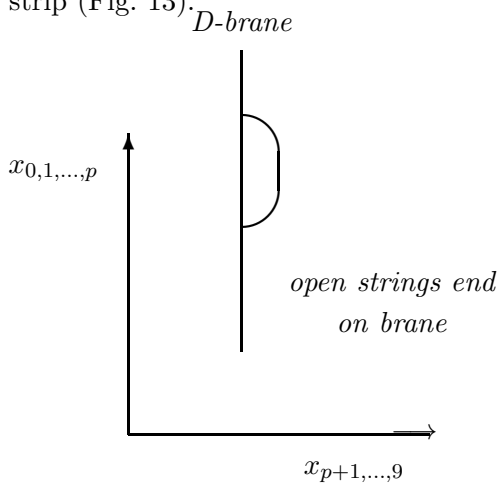
\begin{figure}[htp]
\begin{center}
\begin{picture}(130,160)
\setlength{\unitlength}{1mm}
\thicklines
\put(20,16){\line(0,0){40}}
\put(20,40){\oval(10,15)[r]}
\put(20,60){\makebox(0,0){\em D-brane}}
\put(40,25){\makebox(0,0){\em open strings end}}
\put(40,20){\makebox(0,0){\em on brane}}
\put(5,5){\vector(0,4){40}}
\put(5,5){\line(10,0){40}}
\put(43,5){\makebox(0,0){$\longrightarrow$}}
\put(-5,40){\makebox(0,0){$x_{0,1,\ldots,p}$}}
\put(35,0){\makebox(0,0){$x_{p+1,\ldots,9}$}}
\end{picture}
\end{center}
\caption{A Dp brane with an open string ending on it.}
\label{Dpbrane1}
\end{figure}

Let us return now to the RR sector.
The D-brane solitons in this sector are denoted Dp, where p denotes the 
number of {\em spatial} dimensions of the branes world volume (the spatial
volume in which the brane extends).
This is our first comic strip (Fig.~\ref{Dpbrane1}).

The coordinates $x_0,x_1,\ldots,x_p$ are unconstrained and span the 
brane's space-time p+1 dimensional world volume.
The other coordinates $x_{p+1},..,x_9$ are fixed.
The translational non-invariance of the Dp-brane reduces the number
of SUSY generators by half for any value of p.
For Dp-branes $\epsilon_R$ and $\epsilon_L$ in the surviving supersymmetric
charge (Eq.~\ref{survivingcharge}) obey the constraint
\begin{equation}
\epsilon_L = \Gamma^0 \ldots \Gamma^p \epsilon_R. \label{survivingd}
\end{equation}
Thus, for any p, there are sixteen surviving supercharges. 
Note that, applying naively Newton's law, one can estimate
the effective gravitational coupling $G_N ``M''$.
$G_N$, Newton's constant, is proportional to $g_s^2$.
``$M$'' is proportional to $\frac{1}{g_s}$ for Dp branes and to 
$\frac{1}{g_s^2}$ for the NS5 brane.
Thus $G_N ``M''$ vanishes at weak coupling for Dp branes,
unaltering the large distance geometry.

\subsection{The effective field theory on branes}
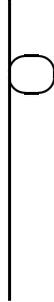
\begin{figure}[htp]
\begin{center}
\begin{picture}(130,140)
\setlength{\unitlength}{1mm}
\thicklines
\put(20,0){\line(0,10){40}}
\put(23,30){\oval(6,5)}
\end{picture}
\end{center}
\caption{A massless state propagating on a Dp brane.}
\label{Dpbrane2}
\end{figure}

The effective theory on the Dp-brane can be identified in more detail
in this case.
It can be shown to contain, in addition to the sixteen conserved
supercharges, 9-p massless scalars (corresponding to the Goldstone
Bosons resulting from the spontaneous breaking of some of the translational invariances),
spin-1 massless particles, and their spin-$\frac{1}{2}$ superpartners.
The effective theory is invariant under local U(1) gauge transformations (Fig.~\ref{Dpbrane2}).

\begin{figure}[htp]
\begin{center}
\begin{picture}(130,160)
\setlength{\unitlength}{1mm}
\thicklines
\put(-10,0){\line(0,10){50}}
\put(-13,40){\oval(6,5)}
\put(10,0){\line(0,10){50}}
\put(13,30){\oval(6,5)}
\put(30,0){\line(0,10){50}}
\put(27,14){\oval(6,5)}
\put(50,0){\line(0,10){50}}
\put(53,22){\oval(6,5)}
\put(-10,5){\line(10,0){18}}
\put(10,5){\oval(3,3)[b]}
\put(12,5){\line(10,0){16}}
\put(30,5){\oval(4,4)[b]}
\put(32,5){\line(10,0){18}}
\put(-10,21){\line(10,0){18}}
\put(10,21){\oval(4,4)[b]}
\put(12,21){\line(10,0){18}}
\put(10,38){\line(10,0){18}}
\put(30,38){\oval(4,4)[t]}
\put(32,38){\line(10,0){18}}
\put(30,18){\line(10,0){20}}
\put(-10,35){\line(10,0){20}}
\put(10,46){\line(10,0){20}}
\put(65,30){\makebox(0,0){$U(1)^4$}}
\end{picture}
\end{center}
\caption{The effective field theory describing this D4 brane configuration
is a $U(4)$ gauge theory, spontaneously broken down to $U(1)^4$.
Four massless states (corresponding to open strings ending on the same
brane) and six massive states (corresponding to open strings ending on 
different branes) are shown.}
\label{U1tothe4}
\end{figure}
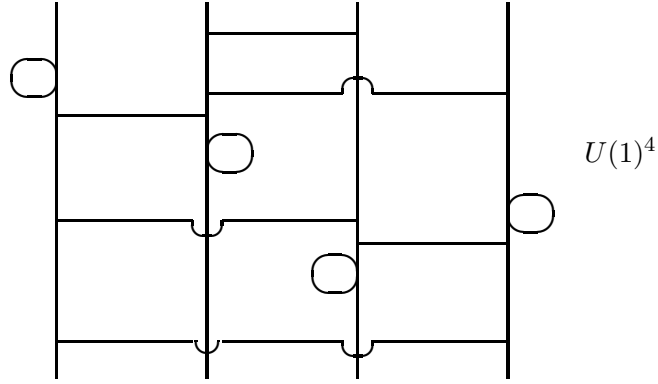

Similarly, one can construct a configuration containing $N_C$ parallel Dp-branes.
This theory still contains sixteen conserved supercharges, and the gauge symmetry has 
increased to $U(1)^{N_C}$.
The next comic strip describes this configuration for $N_C$=4 (Fig.~\ref{U1tothe4}).

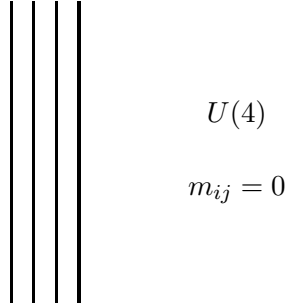
\begin{figure}[htp]
\begin{center}
\begin{picture}(130,100)
\setlength{\unitlength}{1mm}
\thicklines
\put (20,0){\line(0,10){40}}
\put (23,0){\line(0,10){40}}
\put (26,0){\line(0,10){40}}
\put (29,0){\line(0,10){40}}
\put(50,25){\makebox(0,0){$U(4)$}}
\put(50,15){\makebox(0,0){$m_{ij}=0$}}
\end{picture}
\end{center}
\caption{Four parallel D-branes, piled on top of each other.
The sixteen massless states formed are not shown in the figure.}
\label{U4}
\end{figure}

It turns out that the masses of the particles can be directly associated with
the minimal distance between the end points of the strings stretched between
different branes.
For example, a string stretched between the $i$-th and $j$-th brane represents
a particle which has a mass
$m_{ij} = \frac{1}{l_s^2} | \vec{x}_i - \vec{x}_j |$, where $\vec{x}_i$ represents
the value of the coordinates at which the brane is set. 
A parameter in field theory, the mass of a particle has a very simple geometrical
meaning.
Imagine now bringing the parallel branes together (Fig.~\ref{U4}).
This, according to the above relation, will lead to the emergence of massless
particles.
In fact, one can show that these massless particles can enhance the gauge symmetry
all the way from $U(1)^4$ to $U(4)$.

The minimal gauge symmetry for $N_C$ Dp-branes is $U(1)^{N_C}$.
The maximal gauge symmetry is $U(N_C)$.
This has a correspondence in SUSY field theory.
The Lagrangian describing the bosonic sector of a supersymmetric gauge
theory with sixteen supercharges is of the form:
\begin{eqnarray}
{\cal L} = \frac{1}{g^2_{YM, p+1}} \left( \mbox{Tr} F^2_{\mu\nu}
 + \frac{1}{l^4_s} D_{\mu} X^I D^{\mu} X_I \right)
 + \frac{1}{l_s^8 g^2_{YM, p+1}} \mbox{Tr} [X^I,X^J]^2  \; ,
\label{lagrangian}
\end{eqnarray}
where
\begin{eqnarray}
D_{\mu} X^I &=& \partial_{\mu} X^I - i [ A_{\mu}, X^I ]   \nn \\ 
\end{eqnarray}
$X^I$ is a scalar field in the adjoint representation of the gauge
group  and $g^2_{YM, p+1}$ is the effective Yang-Mills coupling in
p+1 dimensions,
it is proportional to $g_s$.
The system has indeed the same particle content and gauge symmetries as can
be inferred from figure~\ref{U1tothe4}.
Moreover, the Higgs mechanism in the presence of scalars in the adjoint
representation is known to conserve the rank of the group.
Indeed, only such residual gauge symmetry groups that preserve the rank
are allowed according to the comic strip.
The form of the potential term appearing in Eq.~\ref{lagrangian} is generic
for supersymmetric theories. 
The classical potential is flat and allows for an infinite set of vacua,
parameterized by those expectation values of the scalar fields $X^I$ for
which the potential term vanishes.
For example, in a N=1 supersymmetric gauge theory, containing two multiplets
with opposite electric charges, the potential is given by:
\begin{equation}
V=(q^{\dagger}q-\tilde{q}^{\dagger}\tilde{q})^2 \; .
\end{equation}

It is a property of supersymmetric theories that the flat potential is
retained to all orders in perturbation theory.
For some theories with a small number of SUSY charges, this result
is modified non-perturbatively \cite{SI,Amati}.
For theories with sixteen SUSY charges, the potential remains flat also
non-perturbatively.
The vacua of the system consist of those expectation values for which
$<q>$=$<\tilde{q}>~\neq~0$.
In each of these vacua, the system will have massless scalar excitations.
They are denoted moduli and their number is called the dimension of
moduli space.

Returning to the configuration of $N_C$ Dp-branes,
we have seen that U(1) symmetries remain unbroken and
therefore the theory is said to be in the Coulomb phase.
The Coulomb phase is that phase of gauge theory for which the force between both electric and
magnetic charges is a Coulomb force.
The number of massless particles is at most $N_C$ (in complex notation).
This is again apparent from the geometry of the open strings ending on
the brane configuration.
The expectation values of the Higgs fields in the adjoint representation
can be shown to have themselves a very transparent geometrical meaning:
\begin{equation}
\vec{x}_i = < \vec{X}_{ii} > \; \; \; (i=1,\ldots,N_C) \; .
\end{equation}
$\vec{x}_i$ on the left-hand side of the equation denotes the location
of the $i$-th brane, $\vec{X}_{ii}$ on the right-hand side of the equation
denotes the component of the Higgs field in the $i$-th element of the Cartan
subalgebra of $U(N_C)$.
In this context, the mass formula
\begin{equation}
m_{ij} = \frac{1}{l_s^2} | \vec{x}_i - \vec{x}_j |
\end{equation}
is just the usual mass obtained by the Higgs
mechanism. To keep fixed the mass of these ``W'' particles in the limit of the decoupling
of the string states ($l_s \rightarrow 0$),
the separations between the branes should vanish themselves in that limit,
that is, these separation should be sub-stringy. 
In this limit, one cannot resolve the $N_C$ different world volumes,
so the theory is perceived as a $U(1)^{N_C}$ gauge theory on a single p+1 dimensional
world volume.
We are by now well on our way to obtain that brane configurations will help  accomplish
 our goals, namely the brane configuration leading to an effective D=4, N=1
supersymmetrical U($N_C$) gauge theory containing in addition matter fields.

\subsection{Effective D=4 dimensional systems with N=2 supersymmetry}
\label{subseclabel}
To obtain an effective D=4 description one can either set up a single D3-brane
configuration in type IIB string theory or build a more complicated configuration in type IIA
string theory. It turns out that for our goals the latter is more useful. One constructs
a brane configuration which has a world volume of the form $M^{3,1} \times I_{[\Delta x_6]}$
where $M^{3,1}$ is the four-dimensional Minkowski space-time and $I_{[\Delta x_6]}$
is an interval of length $\Delta x_6$. This is realized by the following configuration:

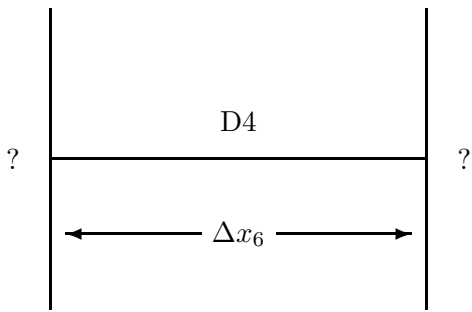
\begin{figure}[htp]
\begin{center}
\begin{picture}(150,140)
\setlength{\unitlength}{1mm}
\thicklines
\put (10,0){\line(0,10){40}}
\put (60,0){\line(0,10){40}}
\put (10,20){\line(10,0){50}}
\put (35,25){\makebox(0,0){D4}}
\put (30,10){\vector(-1,0){18}}
\put (40,10){\vector(1,0){18}}
\put (35,10){\makebox(0,0){$\Delta x_6$}}
\put(65,20){\makebox(0,0){?}}
\put(5,20){\makebox(0,0){?}}
\end{picture}
\end{center}
\caption{The effective field theory on the D4 brane will be four-dimensional for
energies much smaller than $\frac{1}{\Delta x_6}$.
The content of the field theory will depend on the branes $``?''$ on which the
D4 brane ends.}
\label{FigII.3}
\end{figure}

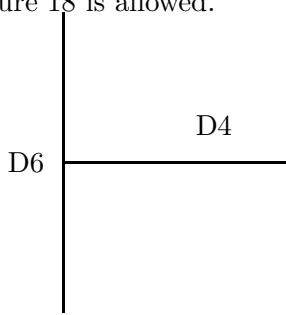
\begin{figure}[htp]
\begin{center}
\begin{picture}(150,100)
\setlength{\unitlength}{1mm}
\thicklines
\put (10,0){\line(0,10){40}}
\put (10,20){\line(10,0){30}}
\put(30,25){\makebox(0,0){D4}}
\put(5,20){\makebox(0,0){D6}}
\end{picture}
\end{center}
\caption{A ``vertex'' in which a D4 brane ends on a D6 brane.}
\label{vertexD6}
\end{figure}

In this configuration a D4-brane (whose world volume is 5 dimensional and of the type $M^{3,1}
\times I_{[\Delta x_6]}$) the two branes between which the D4-brane is suspended would be chosen
such that the effective field theory has $U(N_C)$ local gauge symmetry and N=1 supersymmetry.
The candidates for ``heavier'' such branes would be either NS5-branes or D6-branes. For either choice, the
effective field theory, that is the field theory at energy scales much smaller than $1/\Delta
x_6$, is effectively four dimensional.
Before analyzing the various effective theories resulting from the different choices of
the  branes on which the D4-brane ends
 we discuss which are the allowed `vertices', that is on which branes
are the D4-branes actually allowed to end. In the first example we will show that a D4 is allowed
to end on a D6-brane, that is the vertex appearing  in figure~\ref{vertexD6} is allowed.

A fundamental string (F1) can by definition end on any D$p$-brane, in particular in
type IIB string theory it can end on a D3-brane. Performing what is called an S-duality
transformation validates that also a D1-brane may end on a D3-brane in type IIB theory.
The world-volume of the D3-brane is chosen to extend in the $x_0, x_7, x_8, x_9$ directions 
and that of the D1-brane extends in  $x_0, x_6$. 
Establishing that the D1 may end on a D3-brane, we pause now to briefly discuss several 
types of useful discrete symmetries in string theory, called S- and T-dualities.

S-duality is a symmetry which is familiar already in some field theories. For example, in an 
N=4 supersymmetric gauge theory in D=4 dimensions the gauge coupling constant $g$ is a real
parameter. The field theory is finite, the gauge coupling does not run under the
renormalization group and thus different values of $g$ correspond to different theories.
There is evidence that the theory with coupling $g$ is isomorphic to the theory with
coupling $1/g$. Type IIB string theory has similar properties with the string coupling
playing the role of the gauge coupling. This non-perturbative symmetry, called S-duality, has a
generalization involving also the value of the $\theta$ parameter in field theory and an
additional  corresponding field in string theory. 
In its implementation in field theory electric and magnetic excitations were interchanged,
similarly in string theory different types of branes are interchanged under S-duality.
An F1 is interchanged with a D1, a D3 is left invariant and a D5 is interchanged with an
NS5. We have used some of these properties in the derivation above.

\begin{figure}[htp]
\begin{center}
\begin{picture}(130,300)
\setlength{\unitlength}{1mm}
\thicklines

\put(-30,65){\line(0,1){30}}
\put(-20,75){\line(0,1){30}}
\put(-30,65){\line(1,1){10}}
\put(-30,95){\line(1,1){10}}
\put(-38,80){\makebox(0,0){$R_i$}}
\put(-34,75){\vector(0,1){10}}

\put(-15,80){\vector(1,0){17}}
\put(-7,84){\makebox(0,0){T}}
\put(-2,81){\vector(0,1){7}}

\put(20,65){\line(0,1){30}}
\put(30,75){\line(0,1){30}}
\put(20,65){\line(1,1){10}}
\put(20,95){\line(1,1){10}}
\put(37,80){\makebox(0,0){$R_i$}}
\put(34,75){\vector(0,1){10}}

\put(45,68){\makebox(0,0){N $\longrightarrow$ D}}

\put(68,80){\makebox(0,0){`Longitudinal Duality'}}

\put(-25,60){\makebox(0,0){Dp}}
\put(-15,60){\vector(1,0){17}}
\put(25,60){\makebox(0,0){D(p-1)}}

\put(-30,10){\line(0,1){30}}
\put(-20,20){\line(0,1){30}}
\put(-30,10){\line(1,1){10}}
\put(-30,40){\line(1,1){10}}
\put(-35,29){\makebox(0,0){$R_j$}}
\put(-35,25){\makebox(0,0){$\longleftarrow$}}

\put(-15,25){\vector(1,0){17}}
\put(-7,29){\makebox(0,0){T}}
\put(-2,27){\makebox(0,0){$\longrightarrow$}}

\put(20,10){\line(0,1){30}}
\put(30,20){\line(0,1){30}}
\put(20,10){\line(1,1){10}}
\put(20,40){\line(1,1){10}}
\put(15,29){\makebox(0,0){$R_j$}}
\put(15,25){\makebox(0,0){$\longleftarrow$}}

\put(45,13){\makebox(0,0){D $\longrightarrow$ N}}

\put(68,25){\makebox(0,0){`Transverse Duality'}}

\put(-25,5){\makebox(0,0){Dp}}
\put(-15,5){\vector(1,0){17}}
\put(25,5){\makebox(0,0){D(p+1)}}

\end{picture}
\end{center}
\caption{T-duality acting on Dp branes. D and N denote Dirichlet
and Neumann boundary conditions in the compact directions, respectively.}
\label{TDualityDBrane}
\end{figure}
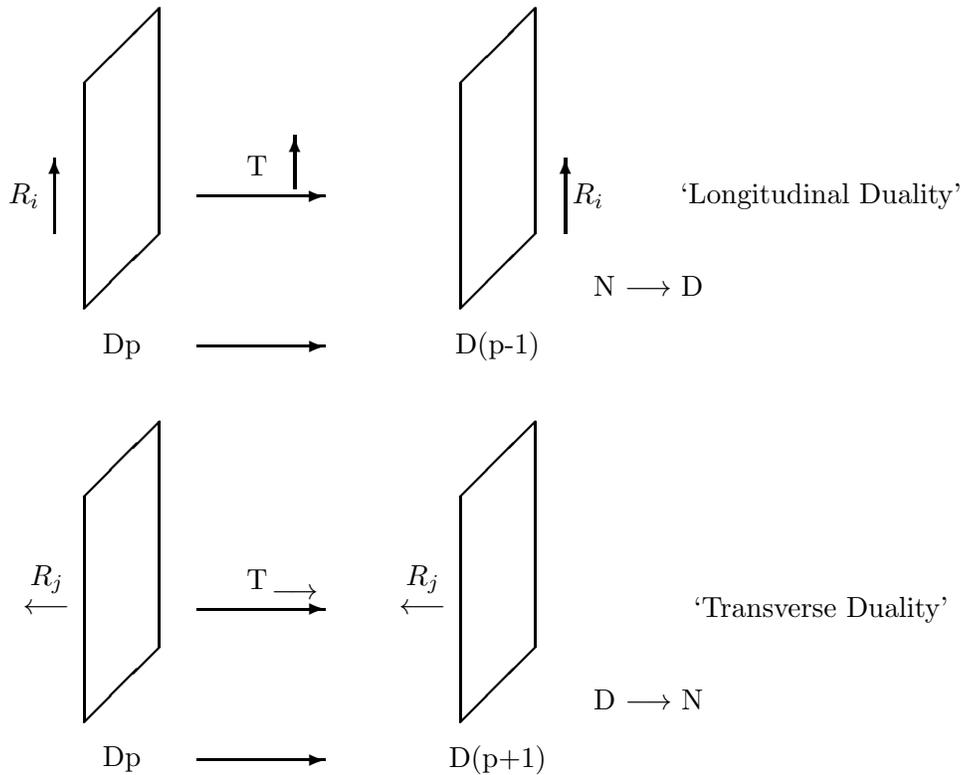

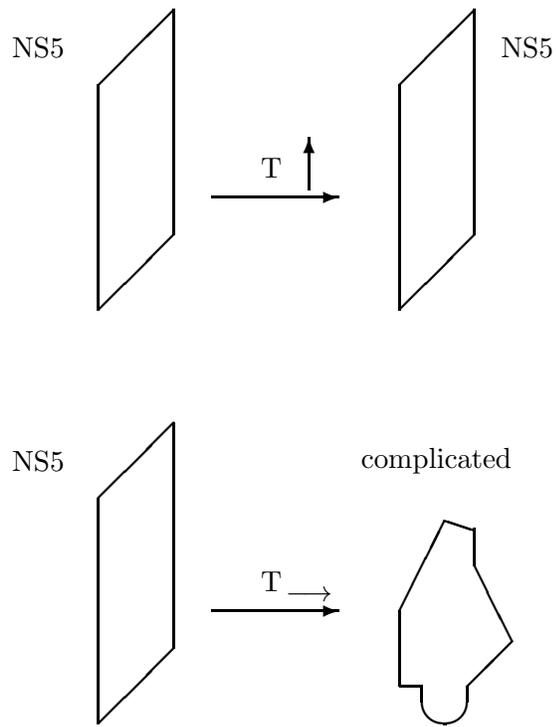
\begin{figure}[htp]
\begin{center}
\begin{picture}(130,250)
\setlength{\unitlength}{1mm}
\thicklines

\put(-20,55){\line(0,1){30}}
\put(-10,65){\line(0,1){30}}
\put(-20,55){\line(1,1){10}}
\put(-20,85){\line(1,1){10}}
\put(-28,90){\makebox(0,0){NS5}}

\put(-5,70){\vector(1,0){17}}
\put(3,74){\makebox(0,0){T}}
\put(8,71){\vector(0,1){7}}

\put(20,55){\line(0,1){30}}
\put(30,65){\line(0,1){30}}
\put(20,55){\line(1,1){10}}
\put(20,85){\line(1,1){10}}
\put(37,90){\makebox(0,0){NS5}}

\put(-20,0){\line(0,1){30}}
\put(-10,10){\line(0,1){30}}
\put(-20,0){\line(1,1){10}}
\put(-20,30){\line(1,1){10}}
\put(-28,35){\makebox(0,0){NS5}}

\put(-5,15){\vector(1,0){17}}
\put(3,19){\makebox(0,0){T}}
\put(8,17){\makebox(0,0){$\longrightarrow$}}

\put(25,35){\makebox(0,0){complicated}}

\put(20,5){\line(0,1){10}}
\put(20,15){\line(1,2){6}}
\put(26,27){\line(3,-1){4}}
\put(30,21){\line(0,1){5}}
\put(30,21){\line(1,-2){5}}
\put(29,5){\line(1,1){6}}
\put(26,5){\oval(6,10)[b]}
\put(20,5){\line(1,0){3}}

\end{picture}
\end{center}
\caption{T-duality acting on a NS5 brane. The detailed action of the transverse
duality is not indicated.}
\label{complicated}
\end{figure}

T-duality is a symmetry which has aspects peculiar to string theory
\cite{GPR}. In particular,
a closed bosonic string theory with one compact dimension whose radius in string units
is $R$, is identical to another bosonic string theory whose compact dimension in string
units is of radius $1/R$. It is the extended nature of the string which leads to this 
result. The mass $M$ of the particles depends on the compactification radius through the
formula:
\beq
M^2= {n^2 \over R^2} + m^2 R^2. \label{tdualmass}
\eeq
$n/R$ denotes the quantized momentum of the center of mass of the string. The term
${n^2\over R^2}$ is not particular to string theory, it describes also a point particle in a
Kaluza-Klein compactification. The second term $m^2 R^2$ reflects the extended nature
of the string. It describes those excitations in which the closed bosonic string
extends and winds around the compact dimensions $m$ times. For a small radius $R$ these
are very low energy excitations. All in all, an interchange of $n$ and $m$
simultaneously with an interchange of $R$ and $1/R$ in eq. (\ref{tdualmass}) gives an
indication of how T-duality works. T-duality can be generalized to an infinite discrete
symmetry and can be shown to actually be a gauge symmetry in the bosonic case. This
indicates that it persists non-perturbatively.
For supersymmetric string theories T-duality has some different manifestations. In 
particular the transformation $R \rightarrow 1/R$ maps a type IIA string theory
background with radius $R$ to a type IIB background with radius $1/R$ and vice-versa.
In the presence of D-branes one naturally distinguishes between two types of compact
dimensions: `longitudinal' dimensions, which are part of the world-volume of the brane,
and `transverse' dimensions, those dimensions which are not part of the  world-volume of the
brane. A  T-duality  involving a longitudinal  dimension will transform a D$p$-brane
into a D$(p-1)$-brane and will leave an NS5-brane intact. T-duality involving a transverse
direction transforms a D$p$-brane into a D$(p+1)$-brane. Its effect on a NS5-brane is more
complicated and we will not need it in this lecture 
(Figs.~\ref{TDualityDBrane},~\ref{complicated}).

\begin{figure}[htp]
\begin{center}
\begin{picture}(150,120)
\setlength{\unitlength}{1mm}
\thicklines

\put(-30,0)
{\begin{picture}(120,100)
\put(-10,10){\line(0,10){40}}
\multiput(-10,30)(4,0){6}{\line(10,0){3}}
\put(10,35){\makebox(0,0){F1}}
\put(-15,30){\makebox(0,0){D3}}
\put(20,30){\makebox(5,0){$\longrightarrow$}}
\put(23,25){\makebox(0,0){S}}
\put(40,10){\line(0,10){40}}
\put(40,30){\line(10,0){25}}
\put(60,35){\makebox(0,0){D1}}
\put(35,30){\makebox(0,0){D3}}
\put(60,25){\makebox(0,0){06}}
\put(40,5){\makebox(0,0){0789}}
\put(0,0){\makebox(0,0){IIB}}
\put(50,0){\makebox(0,0){IIB}}
\end{picture}}

\put(70,0)
{\begin{picture}(70,100)
\put(-30,30){\makebox(5,0){$\longrightarrow$}}
\put(-27,25){\makebox(0,0){T(1,2,3)}}
\put(-10,10){\line(0,10){40}}
\put(-10,30){\line(10,0){25}}
\put(10,35){\makebox(0,0){D4}}
\put(10,25){\makebox(0,0){01236}}
\put(-15,30){\makebox(0,0){D6}}
\put(-10,5){\makebox(0,0){0123789}}
\put(0,0){\makebox(0,0){IIA}}
\end{picture}}

\end{picture}
\end{center}
\caption{A combination of S and T duality transformations, establishing that
a D4 brane can indeed end on a D6 brane.}
\label{D4D6Proof}
\end{figure}
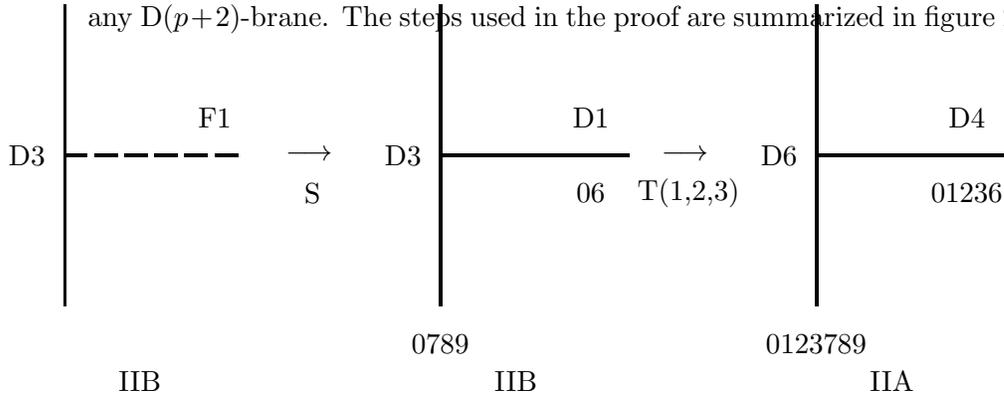

Equipped with this information we can continue the proof of the existence of a D4
configuration ending on a D6-brane. By performing a T-duality along three 
directions transverse to both the D3 and the D1-branes, for example the directions
$x_1, x_2$ and $x_3$, we obtain a D4-brane ending on a D6-brane.
Due to the odd number of T-duality transformations, one passes from a IIB background
to a type IIA background.
The proof thus rests on the validity of both S- and T-duality. There are many
indications that the former is correct, and there is firmer evidence of the validity
of T-duality. The construction sketched in this proof shows that any D$p$-brane can
end on any D$(p+2)$-brane. The steps used in the proof are summarized in 
figure~\ref{D4D6Proof}.

\begin{figure}[htp]
\begin{center}
\begin{picture}(150,100)
\setlength{\unitlength}{1mm}
\thicklines
\put (10,0){\line(0,10){40}}
\put (10,20){\line(10,0){30}}
\put(30,25){\makebox(0,0){D4}}
\put(5,20){\makebox(0,0){NS}}
\end{picture}
\end{center}
\caption{A D4 brane ends on a NS5 brane.}
\label{D4NS5}
\end{figure}
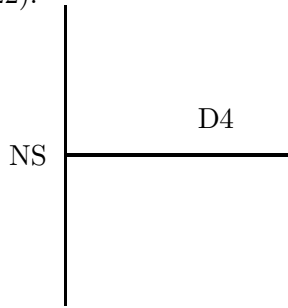

In a somewhat similar manner one can show that a D4-brane can end on a 
NS5-brane (Fig.~\ref{D4NS5}).

\begin{figure}[htp]
\begin{center}
\begin{picture}(150,300)
\setlength{\unitlength}{1mm}
\thicklines

\put(0,60){
\begin{picture}(130,110)
\put(-10,10){\line(0,10){40}}
\put(-10,30){\line(10,0){25}}
\put(5,35){\makebox(0,0){D1}}
\put(5,25){\makebox(0,0){06}}
\put(-15,30){\makebox(0,0){D3}}
\put(-10,5){\makebox(0,0){0345}}
\put(20,30){\makebox(5,0){$\longrightarrow$}}
\put(23,25){\makebox(0,0){T(1,2)}}

\put(40,10){\line(0,10){40}}
\put(40,30){\line(10,0){25}}
\put(55,35){\makebox(0,0){D3}}
\put(55,25){\makebox(0,0){0612}}
\put(35,30){\makebox(0,0){D5}}
\put(40,5){\makebox(0,0){034512}}
\put(70,30){\makebox(5,0){$\longrightarrow$}}
\put(73,25){\makebox(0,0){S}}
\put(0,0){\makebox(0,0){IIB}}
\put(50,0){\makebox(0,0){IIB}}
\end{picture}}

\put(0,0){
\begin{picture}(130,110)
\put(-10,10){\line(0,10){40}}
\put(-10,30){\line(10,0){25}}
\put(5,35){\makebox(0,0){0126}}
\put(5,25){\makebox(0,0){D3}}
\put(-10,5){\makebox(0,0){NS5}}
\put(-10,1){\makebox(0,0){012345}}
\put(20,30){\makebox(5,0){$\longrightarrow$}}
\put(23,25){\makebox(0,0){T(3)}}
\put(40,10){\line(0,10){40}}
\put(40,30){\line(10,0){25}}
\put(55,35){\makebox(0,0){D4}}
\put(55,25){\makebox(0,0){0612}}
\put(40,5){\makebox(0,0){NS5}}
\put(40,1){\makebox(0,0){012345}}
\put(0,-3){\makebox(0,0){IIB}}
\put(50,-3){\makebox(0,0){IIA}}
\end{picture}}

\end{picture}
\end{center}
\caption{S and T duality transformations establish that a D4 brane may end
on a NS5 brane.}
\label{D4NS5Proof}
\end{figure}
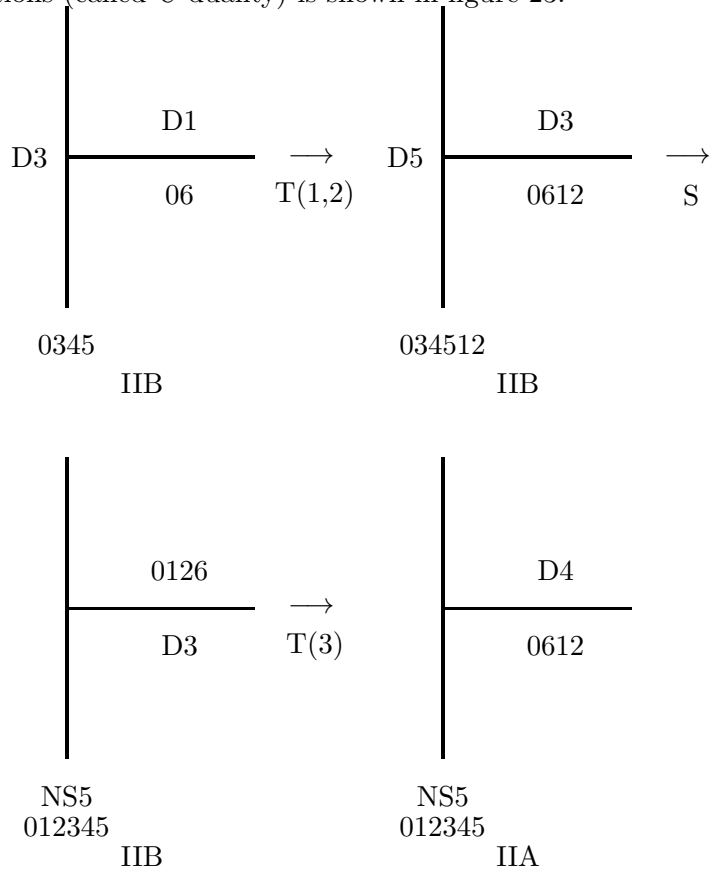

Starting from the by now established configuration of a D1-brane ending on  a D3-brane in type IIB
string theory, one performs T-duality along two transverse direction, $x_1$ and $x_2$
to obtain a D3-brane ending on a D5-brane. An S-duality transformation brings us to a D3-brane
ending on a NS5-brane. A T-duality along $x_3$, which is transverse to the newly formed
D3-brane but longitudinal to the NS5-brane, leads to the desired configuration in which 
a D4-brane ends on a NS5-brane. 
This series of S- and T-duality transformations (called U-duality)
is shown in figure~\ref{D4NS5Proof}.

\begin{figure}[htp]
\begin{center}
\begin{picture}(150,100)
\setlength{\unitlength}{1mm}
\thicklines
\put (10,5){\line(0,10){40}}
\put (60,5){\line(0,10){40}}
\put (10,25){\line(10,0){50}}
\put (35,30){\makebox(0,0){D4}}
\put(10,0){\makebox(0,0){NS5}}
\put(60,0){\makebox(0,0){D6}}
\end{picture}
\end{center}
\caption{A D4 brane suspended between and NS5 and D6 branes.
The effective field theory contains no massless particles and
is thus, at best, some topological theory.}
\label{NS5D4D6}
\end{figure}
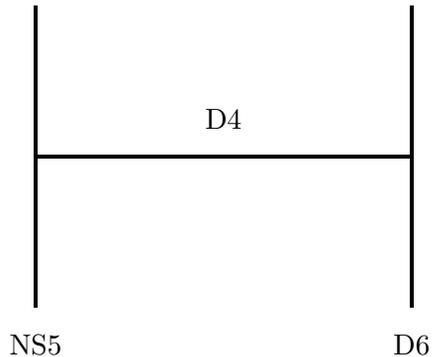

Let us consider first the effective low energy field theory related to 
the configuration in which the D4-brane is suspended
between a NS5-brane and a D6-brane (Fig.~\ref{NS5D4D6}).

The spatial extension of the various branes is summarized in the table below,
where $+$ and $-$ denote respectively directions longitudinal and transverse
to the relevant brane, and $=$ indicates that the brane is of finite extent in that
direction.
\beq
\begin{array}{ccccccccc}
 & x_{0123} & x_4 & x_5 & x_6 & x_7 & x_8 & x_9 \\
D4 & +      &  -  &  -  &  =  &  -  &  -  &  -   \\
D6 &  +     &  -  &  -  &  -  &  +  &  +  &  +  \\
NS5  &  +  &   +  &  +  &  -  &  -  &  -  &  -  
\end{array} \label{d4d6ns5tab}
\eeq

The effective field theory should contain the massless degrees of freedom of the system.
Massless degrees of freedom can also be identified in a geometrical manner.
Each possibility to displace the D4-brane along the D6 and the NS5 branes maintaining
the shape of the configuration corresponds to a massless particle. The D4-brane
left on its own could be displaced along the directions $x_4, x_5, x_6, x_7, x_8$ and $x_9$.
On the D6 side, the D4-brane is locked in the $x_4, x_5$ and $x_6$ directions.
On the NS5 side the D4 is locked along $x_6, x_7, x_8$ and $x_9$.
All in all, the D4-brane is frozen. It cannot be displaced in a parallel fashion
and therefore the effective theory contains no massless particles whatsoever.
Although the D4-brane would have allowed the propagation of 5-dimensional
photons had it been left on its own, suspended between the D6 and the NS5 it allows
no massless degrees of freedom to propagate on it. It is at best a topological field
theory. 

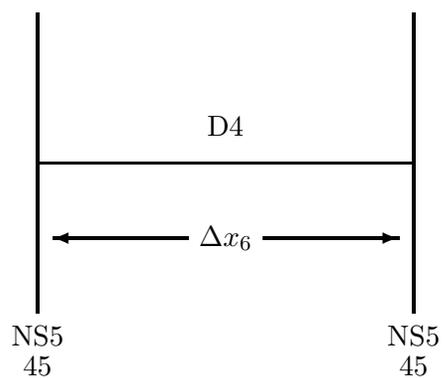
\begin{figure}[htp]
\begin{center}
\begin{picture}(150,110)
\setlength{\unitlength}{1mm}
\thicklines
\put (10,0){\line(0,10){40}}
\put (60,0){\line(0,10){40}}
\put (10,20){\line(10,0){50}}
\put (35,25){\makebox(0,0){D4}}
\put (30,10){\vector(-1,0){18}}
\put (40,10){\vector(1,0){18}}
\put (35,10){\makebox(0,0){$\Delta x_6$}}
\put(10,-3){\makebox(0,0){NS5}}
\put(10,-7){\makebox(0,0){45}}
\put(60,-3){\makebox(0,0){NS5}}
\put(60,-7){\makebox(0,0){45}}
\end{picture}
\end{center}
\caption{A D4 brane suspended between two NS5 branes.
The effective field theory is N=2 SUSY U(1) gauge theory.}
\label{NS5D4NS5}
\end{figure}

We thus turn to another attempt to build an effective four dimensional
field theory along the suspended D4-brane.
We now suspend it between two NS5-branes which are extended in the same directions
as the NS5 of the previous example \cite{HW} (Fig.~\ref{NS5D4NS5}).

\begin{figure}[htp]
\begin{center}
\begin{picture}(150,110)
\setlength{\unitlength}{1mm}
\thicklines
\put (10,0){\line(0,10){40}}
\put (61,0){\line(0,10){40}}
\put (10,20){\line(10,0){51}}
\multiput(10,30)(4,0){13}{\line(10,0){3}}
\multiput(13,21)(7,0){7}{\vector(0,1){7}}
\put(65,18){\vector(0,1){7}}
\put(73,20){\makebox(0,0){$x_4,x_5$}}
\end{picture}
\end{center}
\caption{The D4 brane can be parallely displaced along the $x_4$ and $x_5$
directions. This corresponds to two massless spin-0 particles appearing in the 
effective field theory. N=2 supersymmetry implies the existence of massless
spin-$\frac{1}{2}$ and spin-1 particles as well.}
\label{D4displaced}
\end{figure}
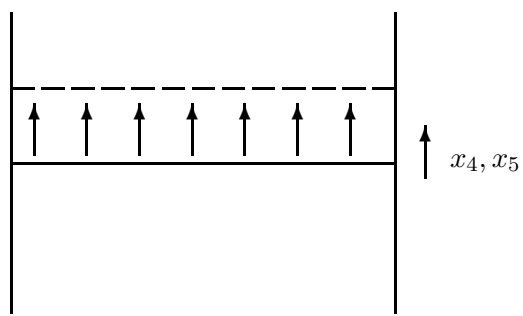

At both ends, the D4-branes is locked in the $x_6, x_7, x_8$ and $x_9$ directions.
Therefore now it can be displaced in the $x_4$ and $x_5$ directions, as shown in 
figure~\ref{D4displaced}.

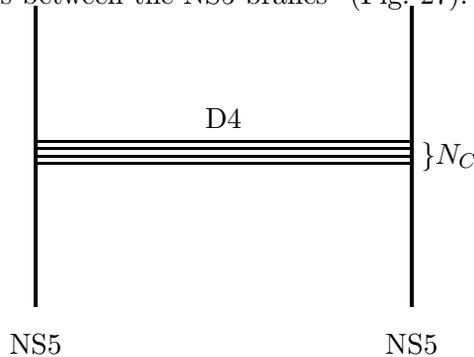
\begin{figure}[htp]
\begin{center}
\begin{picture}(150,100)
\setlength{\unitlength}{1mm}
\thicklines
\put (10,0){\line(0,10){40}}
\put (60,0){\line(0,10){40}}
\put (10,19){\line(10,0){50}}
\put (10,20){\line(10,0){50}}
\put (10,21){\line(10,0){50}}
\put (10,22){\line(10,0){50}}
\put (35,25){\makebox(0,0){D4}}
\put(10,-5){\makebox(0,0){NS5}}
\put(60,-5){\makebox(0,0){NS5}}
\put(65,20){\makebox(0,0){$\}N_C$}}
\end{picture}
\end{center}
\caption{The effective field theory is a D=4, N=2 $U(N_C)$ SUSY gauge theory.}
\label{UNC}
\end{figure}

Thus the effective field theory contains at least two massless spin 0 particles.
An analysis using (Eqs.~\ref{survivingcharge}), (\ref{survivingns}) and (Eq.~\ref{survivingd})
shows that in this configuration one half of the supersymmetry generators of the
single brane configuration survive. That is, 8 supercharges survive as symmetries,
implementing an N=2 supersymmetry in the effective four dimensional theory. The two
scalar particle identified geometrically form part of the N=2 vector multiplet.
That multiplet consists of spin 1, spin 1/2 and spin 0 particles. Thus the configuration
above describes an effective D=4, N=2, $U(1)$ supersymmetric gauge theory.
The effective four dimensional gauge coupling constant is related to the effective 
five dimensional gauge coupling constant in the usual Kaluza-Klein manner:
\beq g^2_{YM,4}={g^2_{YM,5} \over \Delta x_6}. \label{couplingkk}
\eeq
Changing the value of the separation $\Delta x_6$ amounts to rescaling $g^2_{YM,4}$.
As in the single brane case, the gauge symmetry can be enhanced to $U(N_C)$ by suspending 
$N_C$ D4-branes between the NS5-branes\footnote{The actual symmetry turns out to be
$SU(N_C)$. This is discussed in \cite{wittenM,GK}.} (Fig.~\ref{UNC}).

\begin{figure}[htp]
\begin{center}
\begin{picture}(150,100)
\setlength{\unitlength}{1mm}
\thicklines
\put (10,0){\line(0,10){40}}
\put (60,0){\line(0,10){40}}
\put (10,20){\line(10,0){50}}
\put (35,25){\makebox(0,0){D4}}
\put(10,-3){\makebox(0,0){D6}}
\put(60,-3){\makebox(0,0){D6}}
\end{picture}
\end{center}
\caption{The effective field theory on the D4 brane is a D=4, N=2 SUSY
field theory, containg an N=2 matter hypermultiplet.
It contains no massless gauge particles.}
\label{D6D4D6}
\end{figure}
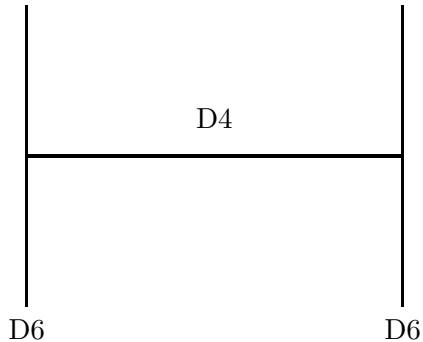

Rotating one of the NS5-branes from the $x_4, x_5$ directions to the $x_8, x_9$ directions
will lead to the desired D=4, N=1, $U(N_C)$ effective gauge theory.
Before pursuing this, it will be useful to study the effective field theory on a D4-brane 
suspended between two D6-branes (Fig.~\ref{D6D4D6}).

The spatial extension of the branes is:
\beq
\begin{array}{ccccccccc}
 & x_{0123} & x_4 & x_5 & x_6 & x_7 & x_8 & x_9 \\
D4 & +      &  -  &  -  &  =  &  -  &  -  &  -   \\
D6 &  +     &  -  &  -  &  -  &  +  &  +  &  +  
\end{array} \label{d4d6tab}
\eeq
This configuration also has 8 surviving supercharges and thus the effective field theory 
has N=2 supersymmetry in D=4. Similar considerations to the ones used above show that
the effective theory contains at least three massless spin 0 particles. These correspond
to the allowed translations in the directions $x_7, x_8$ and $x_9$. An N=2 supersymmetric
multiplet would require either two or four spin 0 particles. At the case at hand,
the multiplet is actually the N=2 hypermultiplet which contains four spin 0 and
four spin 1/2 degrees of freedom. It is important to note that the system contains
no massless spin 1 degrees of freedom, that is the low energy effective theory is not an unbroken 
gauge theory. Actually, later we will show that this configuration will be part of the description of the Higgs
phase of the supersymmetric gauge theory. The fact that the geometrical considerations were not
sufficient to identify all four spin 0 particles shows us one of the limitations of the
simple geometrical analysis. The fourth spin 0 particle can be identified in this case
with the component of a compactified gauge field, namely $A_6$.
At this stage of understanding of the gauge theory--brane correspondence, one finds
parameters in the brane picture with no clearly known field theoretical interpretation
and vice-versa.

Returning to the N=2 $U(N_C)$ gauge configuration, one notes that a separation of the
D4-branes along the directions $x_4, x_5$ leads to all rank preserving possible breakings of the
gauge symmetry. This is similar to what was described before in the case of the separations
of $N_C$ infinitely extended parallel D$p$-branes. The mass of the $W$ particles is again
proportional to $\Delta D4(x_4, x_5)$, which denotes the separation of two D4-branes
in the $x_4, x_5$ directions. The complex number of moduli is immediately read out of
the geometrical picture to be $N_C$. This coincides with algebraic analysis determining
the complex number of massless spin 0 particles surviving the breaking of the gauge group.

In figure~ref{Summary} we summarize the complex number of massless spin 0 particles appearing
in any of the four configurations discussed until now.

\begin{figure}[htp]
\begin{center}
\begin{picture}(150,360)

\put (-50,260){\begin{picture}(150,140)
\setlength{\unitlength}{1mm}
\thicklines
\put (10,0){\line(0,10){40}}
\put (60,0){\line(0,10){40}}
\put (10,20){\line(10,0){50}}
\put (35,25){\makebox(0,0){D4}}
\put (35,15){\makebox(0,0){2}}
\put(65,20){\makebox(0,0){D6}}
\put(5,20){\makebox(0,0){D6}}
\put(95,20){\makebox(0,0){N=2 Hyper-multiplet}}
\end{picture}}

\put (-50,130){\begin{picture}(150,140)
\setlength{\unitlength}{1mm}
\thicklines
\put (10,0){\line(0,10){40}}
\put (60,0){\line(0,10){40}}
\put (10,20){\line(10,0){50}}
\put (35,25){\makebox(0,0){D4}}
\put (35,15){\makebox(0,0){1}}
\put(65,20){\makebox(0,0){NS5}}
\put(5,20){\makebox(0,0){NS5}}
\put(95,20){\makebox(0,0){Vector SUSY U(1)}}
\end{picture}}

\put (-50,0){\begin{picture}(150,140)
\setlength{\unitlength}{1mm}
\thicklines
\put (10,0){\line(0,10){40}}
\put (60,0){\line(0,10){40}}
\put (10,20){\line(10,0){50}}
\put (35,25){\makebox(0,0){D4}}
\put (35,15){\makebox(0,0){0}}
\put(65,20){\makebox(0,0){D6}}
\put(5,20){\makebox(0,0){NS5}}
\put(95,20){\makebox(0,0){`Topological'}}
\end{picture}}

\end{picture}
\end{center}
\caption{A summary of the particle content of the low-energy effective theories
described in subsection~\ref{subseclabel}.}
\label{Summary}
\end{figure}
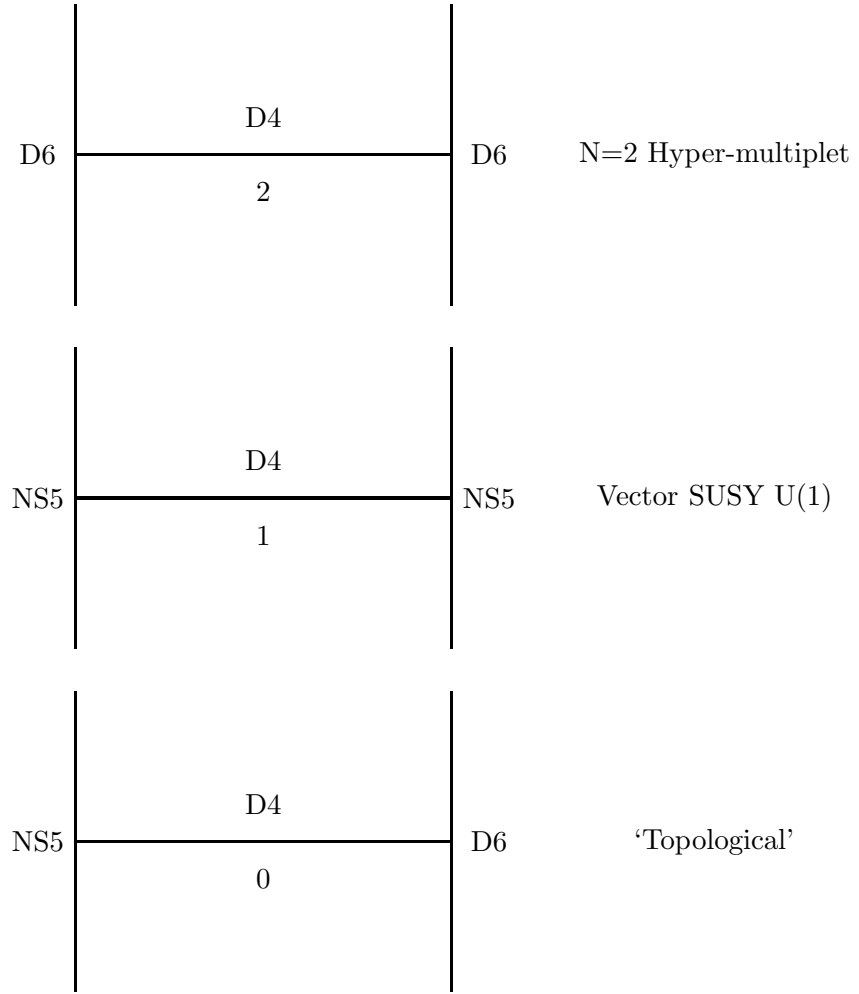

\subsection{An effective D=4, N=1, $U(N_C)$ gauge theory with matter}
\begin{figure}[htp]
\begin{center}
\begin{picture}(150,100)
\setlength{\unitlength}{1mm}
\thicklines
\put (5,0){\line(1,4){10}}
\put (60,0){\line(0,10){40}}
\put (10,20){\line(10,0){50}}
\put (35,25){\makebox(0,0){D4}}
\put(5,-3){\makebox(0,0){NS5}}
\put(60,-3){\makebox(0,0){NS5'}}
\put(5,-7){\makebox(0,0){45}}
\put(60,-7){\makebox(0,0){89}}
\end{picture}
\end{center}
\caption{The configuration leading to a D=4, N=1 U(1) SUSY gauge theory.
The NS5 extending also in the directions $x_8, x_9$ is labeled NS5'.}
\label{N=1SUSY}
\end{figure}
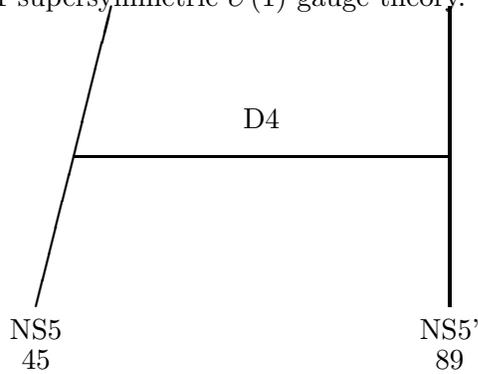

We construct now the configuration leading to N=1 
supersymmetry. The rotation of one of the NS5 from $x_4, x_5$ to the $x_8, x_9$ directions
corresponds to adding an infinite mass term to the scalar fields in the adjoint
representation, a rotation by different angles would have given rise in field theoretical 
language to a finite  mass term. The rotation
leads to a configuration with 4 surviving supercharges (Fig.~\ref{N=1SUSY}). 
The effective four dimensional
theory is a $U(1)$ gauge symmetry and has no moduli, as can be seen from the by now familiar
geometrical considerations. This agrees with a description by an effective D=4, N=1
supersymmetric $U(1)$ gauge theory.

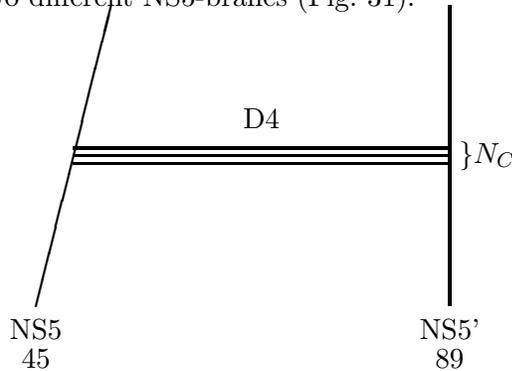
\begin{figure}[htp]
\begin{center}
\begin{picture}(150,100)
\setlength{\unitlength}{1mm}
\thicklines
\put (5,0){\line(1,4){10}}
\put (60,0){\line(0,10){40}}
\put (10,19){\line(10,0){50}}
\put (10,20){\line(10,0){50}}
\put (10,21){\line(10,0){50}}
\put (35,25){\makebox(0,0){D4}}
\put(5,-3){\makebox(0,0){NS5}}
\put(60,-3){\makebox(0,0){NS5'}}
\put(5,-7){\makebox(0,0){45}}
\put(60,-7){\makebox(0,0){89}}
\put(65,20){\makebox(0,0){$\}N_C$}}
\end{picture}
\end{center}
\caption{The configuration leading to a D=4, N=1 $U(N_C)$ SUSY gauge theory.}
\label{N=1UNC}
\end{figure}

The gauge symmetry can be enhanced to $U(N_C)$ by suspending $N_C$ D4-branes between the
two different NS5-branes (Fig.~\ref{N=1UNC}). 

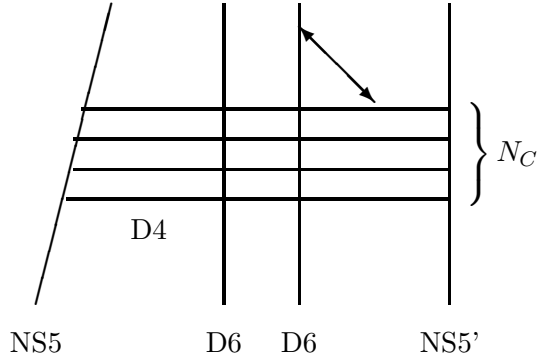
\begin{figure}[htp]
\begin{center}
\begin{picture}(150,100)
\setlength{\unitlength}{1mm}
\thicklines
\put (5,0){\line(1,4){10}}
\put (60,0){\line(0,10){40}}
\put (9,14){\line(10,0){51}}
\put (10,18){\line(10,0){50}}
\put (10,22){\line(10,0){50}}
\put (11,26){\line(10,0){49}}
\put(65,20){\makebox(0,0)
{$ \left.\begin{array}{c} \\ \\ \\ \end{array} \right\}N_C$}}
\put(5,-5){\makebox(0,0){NS5}}
\put(60,-5){\makebox(0,0){NS5'}}
\put (30,0){\line(0,10){40}}
\put (40,0){\line(0,10){40}}
\put(30,-5){\makebox(0,0){D6}}
\put(40,-5){\makebox(0,0){D6}}
\put (45,32){\vector(1,-1){5}}
\put (45,32){\vector(-1,1){5}}
\put(20,10){\makebox(0,0){D4}}
\end{picture}
\end{center}
\caption{Adding D6 branes allows the existence of matter in the effective
field theory. Open strings ending on D4 and D6 branes carry both color and flavor.}
\label{N1Flavor}
\end{figure}

The final ingredient needed is to add some flavor to the effective field theory. This is
done by distributing $N_F$ D6-branes along the $x_6$ direction (Fig.~\ref{N1Flavor}). 

The spatial extension of the various branes is:
\beq
\begin{array}{ccccccccc}
 & x_{0123} & x_4 & x_5 & x_6 & x_7 & x_8 & x_9 \\
D4 & +      &  -  &  -  &  =  &  -  &  -  &  -   \\
D6 &  +     &  -  &  -  &  -  &  +  &  +  &  +  \\
NS5  &  +  &   +  &  +  &  -  &  -  &  -  &  -  \\
NS5'&  +     &  -  &  -  &  -  &  -  &  +  &  + 
\end{array} \label{d4d6ns5ns5tab}
\eeq
The matter fields appear as  representations of diagonal vectorial subgroup of the flavor group 
$SU(N_F) \times SU(N_F) \times U(1)$. Geometrically they are associated with strings
connecting the D6-branes with the D4-branes. Touching the D6-branes endows the
open strings with flavor, and touching the D4-branes endows them with color. Their
directionality is responsible for the appearance of both the $N_F$ and $\bar{N}_F$.
In \cite{GK} the reader will find references for attempts to construct chiral 
configurations and to identify the full flavor group.
The masses of the squarks have a geometrical interpretation: they are proportional to the
distance between the D4 and D6-branes along the $x_4, x_5$ directions. The effective field
theory is thus a four dimensional N=1 $U(N_C)$ supersymmetric gauge theory which
has matter in the fundamental representation of the vector subgroup of the flavor 
symmetry. This is as close to our goal as we will reach in this lecture.

\begin{figure}[h]
\begin{center}
\begin{picture}(150,140)
\setlength{\unitlength}{1mm}
\thicklines

\put(15,3){
\begin{picture}(145,135)
\put (20,0){\line(1,4){10}}
\put (60,0){\line(0,10){40}}
\put (60,4){\line(-1,0){39}}
\put (60,8){\line(-1,0){38}}
\put(69,6){\makebox(0,0)
{$ \left.\begin{array}{c} \\ \end{array} \right\}\tilde{N}_C=2$}}
\put(20,-5){\makebox(0,0){NS5'}}
\put(60,-5){\makebox(0,0){NS5}}

\put (-50,0){\line(1,4){10}}
\put(-50,-5){\makebox(0,0){D6}}
\put (-40,0){\line(1,4){10}}
\put(-40,-5){\makebox(0,0){D6}}
\put (-30,0){\line(1,4){10}}
\put(-30,-5){\makebox(0,0){D6}}
\put (-20,0){\line(1,4){10}}
\put(-20,-5){\makebox(0,0){D6}}
\put (-10,0){\line(1,4){10}}
\put(-10,-5){\makebox(0,0){D6}}

\put(-20,43){\vector(1,0){20}}
\put(-20,43){\vector(-1,0){20}}
\put(-20,46){\makebox(0,0){$N_F$}}

\put(17,43){\makebox(0,0){$(N_F,\bar{N}_F)$}}
\put(7,33){\oval(10,16)[t]}
\put(12,36){\vector(0,-1){3}}
\put(2,36){\vector(0,-1){3}}

\put(20,31){\vector(-3,1){5}}
\put(20,31){\vector(3,-1){5}}

\put (28,33){\line(-1,0){30}}
\put (27,29){\line(-1,0){28}}
\put(-3,29){\oval(4,5)[t]}
\put (26,25){\line(-1,0){28}}
\put(-4,25){\oval(4,5)[t]}
\put (25,21){\line(-1,0){28}}
\put(-5,21){\oval(4,5)[t]}
\put (24,17){\line(-1,0){28}}
\put(-6,17){\oval(4,5)[t]}
\put (-5,29){\line(-1,0){8}}
\put (-6,25){\line(-1,0){6}}
\put(-14,25){\oval(4,5)[t]}
\put (-7,21){\line(-1,0){6}}
\put(-15,21){\oval(4,5)[t]}
\put (-8,17){\line(-1,0){6}}
\put(-16,17){\oval(4,5)[t]}

\put (-16,25){\line(-1,0){8}}
\put (-17,21){\line(-1,0){6}}
\put(-25,21){\oval(4,5)[t]}
\put (-18,17){\line(-1,0){6}}
\put(-26,17){\oval(4,5)[t]}

\put (-27,21){\line(-1,0){8}}
\put (-28,17){\line(-1,0){6}}
\put(-36,17){\oval(4,5)[t]}

\put (-38,17){\line(-1,0){8}}

\put(37,25){\makebox(0,0)
{$ \left.\begin{array}{c} \\ \\ \\ \\ \end{array} \right\}N_F=5$}}

\put(21,12){\vector(1,-1){4}}
\put(21,12){\vector(-1,1){5}}
\put(12,10){\makebox(0,0){$(N_F,\tilde{N}_C)$}}
\end{picture}}

\end{picture}
\end{center}
\caption{This is the dual configuration to that of a N=1 SUSY gauge theory
with $N_C$=3 and $N_F$=5 (Fig.~\ref{N1NC3}).
The configuration shown has $N_C$=5-3=2, $N_F$=5 and, in addition, $N_F^2$ color
singlet massless particles.}
\label{sophisticated}
\end{figure}
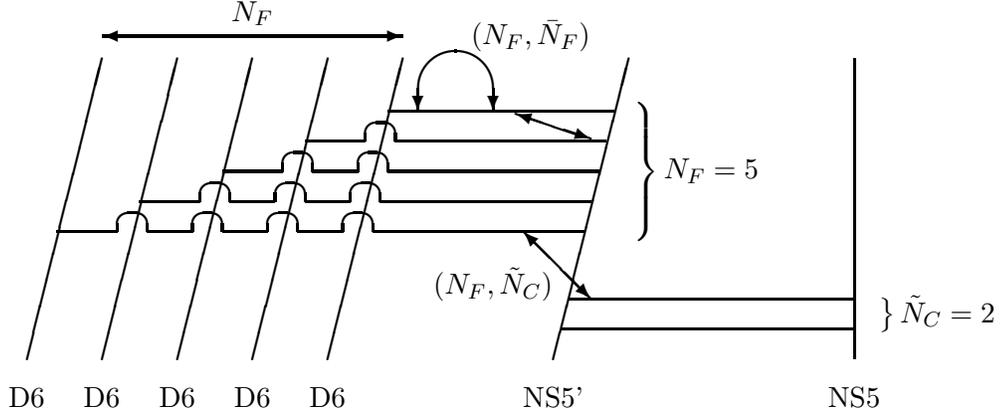

The way to obtain the Seiberg dual configuration is essentially to move the position
of the NS5-brane residing on the left hand part of the configuration all the way to the
right hand part. The resulting new configuration will have $SU(N_F-N_C)$ gauge symmetry
and $N_F$ and $\bar{N}_F$ colored matter as well as $N_F^2$, flavor adjoint color singlets,
which is exactly the result of Seiberg (Fig.~\ref{sophisticated}).
The $N_F^2$ fields are essentially the color-singlet particles appearing in the effective field
theory corresponding to the brane configuration (Fig.~\ref{D6D4D6}).

\subsection{More pieces of information}
In order to be able to perform this displacement of the NS5-brane, one needs three more
pieces of `information' as to the behavior of branes.

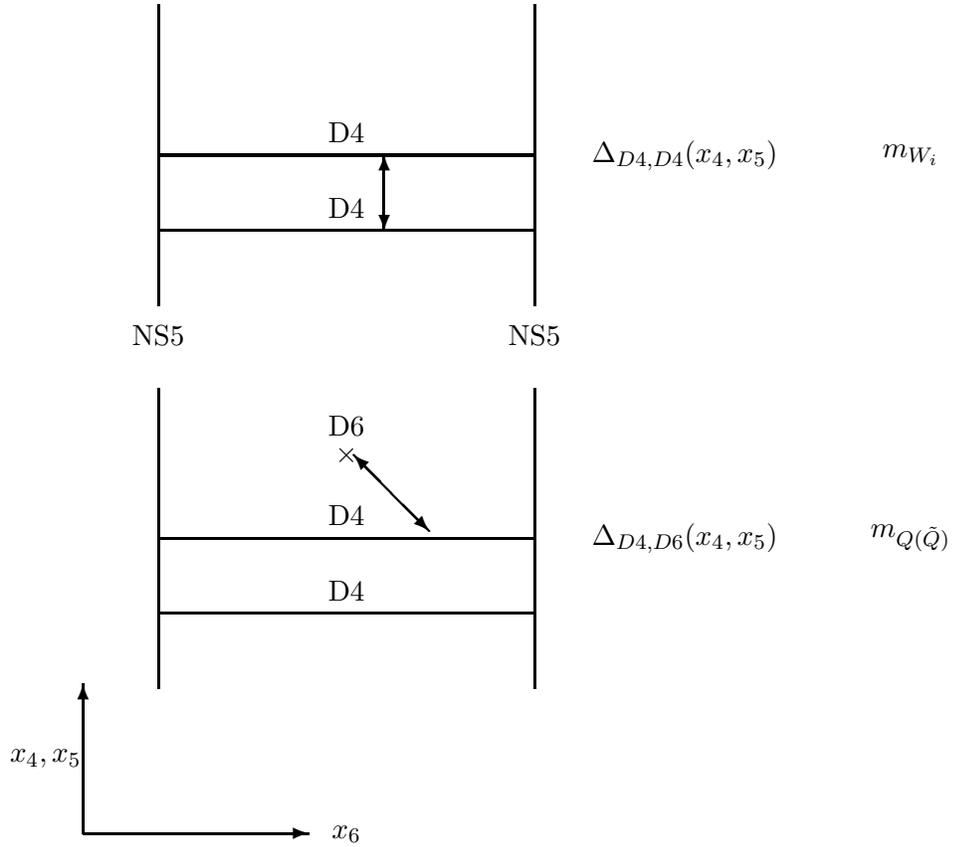
\begin{figure}[htp]
\begin{center}
\begin{picture}(150,300)

\put (-50,200){\begin{picture}(150,140)
\setlength{\unitlength}{1mm}
\thicklines
\put (10,0){\line(0,10){40}}
\put (60,0){\line(0,10){40}}
\put (10,20){\line(10,0){50}}
\put (35,23){\makebox(0,0){D4}}
\put (10,10){\line(10,0){50}}
\put (35,13){\makebox(0,0){D4}}
\put (40,15){\vector(0,-1){5}}
\put (40,15){\vector(0,1){5}}
\put(80,20){\makebox(0,0){$\Delta_{D4,D4}(x_4,x_5)$}}
\put(110,20){\makebox(0,0){$m_{W_i}$}}
\end{picture}}

\put (-50,55){\begin{picture}(150,140)
\setlength{\unitlength}{1mm}
\thicklines
\put(10,47){\makebox(0,0){NS5}}
\put(60,47){\makebox(0,0){NS5}}
\put (10,0){\line(0,10){40}}
\put (60,0){\line(0,10){40}}
\put (35,31){\makebox(0,0){$\times$}}
\put (41,26){\vector(1,-1){5}}
\put (41,26){\vector(-1,1){5}}
\put (35,35){\makebox(0,0){D6}}
\put (10,20){\line(10,0){50}}
\put (35,23){\makebox(0,0){D4}}
\put (10,10){\line(10,0){50}}
\put (35,13){\makebox(0,0){D4}}
\put(80,20){\makebox(0,0){$\Delta_{D4,D6}(x_4,x_5)$}}
\put(110,20){\makebox(0,0){$m_{Q(\tilde{Q})}$}}
\end{picture}}

\put (-50,0){\begin{picture}(150,140)
\setlength{\unitlength}{1mm}
\thicklines
\put (0,0){\vector(0,1){20}}
\put (0,0){\vector(1,0){30}}
\put(-5,10){\makebox(0,0){$x_4,x_5$}}
\put(35,0){\makebox(0,0){$x_6$}}
\end{picture}}
\end{picture}
\end{center}
\caption{The values of the masses of the W and squark particles is encoded in
this figure. They are proportional to the distances $\Delta_{D4,D4}(x_4,x_5)$
and $\Delta_{D4,D6}(x_4,x_5)$, respectively.
Both distances are indicated in the figure. Varying these distances smoothly changes
the parameters of the field theory.}
\label{Parameters1}
\end{figure}

1) {\em `Motions of branes'} \newline
Changes of the positions of branes correspond in some cases to smooth changes in 
field theoretical parameters. For example, a change in the relative displacement
in the $x_4, x_5$ directions of the D4-branes in the N=4 and N=2 supersymmetric configurations
led to a smooth change of the mass of the $W$'s (or equivalently to smooth changes
in the Higgs expectation values). A change in the relative position in the D4 and
D6-branes in the $x_4, x_5$ directions led to a change in the masses of the squarks
(it is actually equivalent to a smooth change in the expectation value of the
scalar coupled to fermions by a Yukawa coupling) (Fig.~\ref{Parameters1}).

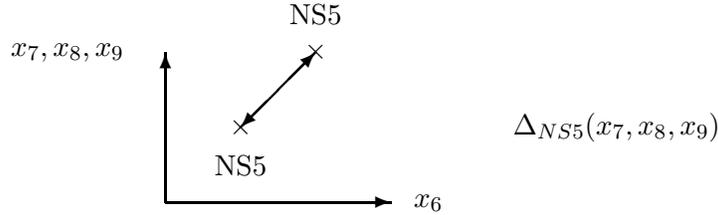
\begin{figure}[htb]
\begin{center}
\begin{picture}(150,90)
\setlength{\unitlength}{1mm}
\thicklines
\put (0,0){\vector(0,1){20}}
\put (0,0){\vector(1,0){30}}
\put (10,10){\makebox(0,0){$\times$}}
\put (10,5){\makebox(0,0){NS5}}
\put (20,20){\makebox(0,0){$\times$}}
\put (20,25){\makebox(0,0){NS5}}
\put (15,15){\vector(1,1){5}}
\put (15,15){\vector(-1,-1){5}}
\put(-13,20){\makebox(0,0){$x_7,x_8,x_9$}}
\put(35,0){\makebox(0,0){$x_6$}}
\put(60,10){\makebox(0,0){$\Delta_{NS5}(x_7,x_8,x_9)$}}
\end{picture}
\end{center}
\caption{The NS5 branes can be separated in the $x_7,x_8,x_9$ directions.
Each such separation corresponds to appropriate components of the Fayet-Iliopoulos
term in N=2 SUSY $U(N_C)$ gauge theory.}
\label{FI}
\end{figure}

A change in the relative distance between the NS5-branes corresponded to a change in the
coupling of the gauge theory on the D4-branes suspended between them.
A change in the relative position along the $x_7, x_8$ and $x_9$ directions of the
positions of two NS5-branes corresponds to adding a Fayet-Iliopoulos term in an N=2
field theoretical interpretation (Fig.~\ref{FI}).
A similar interpretation can be given for the separation along the $x_7$ direction 
between the NS5 and NS5$'$ in the N=1 case supersymmetric case.

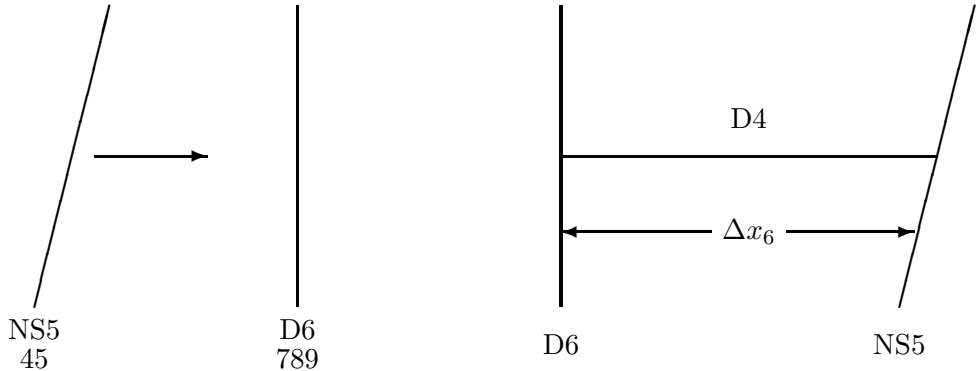
\begin{figure}[htp]
\begin{center}
\begin{picture}(170,120)
\setlength{\unitlength}{1mm}
\thicklines

\put(25,5){
\begin{picture}(80,100)
\put (10,0){\line(0,10){40}}
\put (55,0){\line(1,4){10}}
\put (10,20){\line(10,0){50}}
\put (35,25){\makebox(0,0){D4}}
\put(30,10){\vector(-1,0){20}}
\put(40,10){\vector(1,0){17}}
\put (35,10){\makebox(0,0){$\Delta x_6$}}
\put(10,-5){\makebox(0,0){D6}}
\put(55,-5){\makebox(0,0){NS5}}
\end{picture}}

\put(-40,5){
\begin{picture}(80,100)
\put (5,0){\line(1,4){10}}
\put (40,0){\line(0,10){40}}
\put(5,-3){\makebox(0,0){NS5}}
\put(40,-3){\makebox(0,0){D6}}
\put(5,-7){\makebox(0,0){45}}
\put(40,-7){\makebox(0,0){789}}
\put(13,20){\vector(1,0){15}}
\end{picture}}

\end{picture}
\end{center}
\caption{A D4 brane is created when the NS5 brane crosses the D6 brane, from left
to right in this figure.
The D4 brane carries no massless degrees of freedom.}
\label{D4Creation}
\end{figure}

 In other cases changes in the position of branes could have
more abrupt consequences. Consider a configuration consisting only of a NS5-brane 
and a D6-brane separated in the $x_6$ direction (Fig.~\ref{D4Creation}). 
Displacing the NS5-brane for example
in the $x_6$ direction is a smooth motion as long as the two branes do not intersect (this
is an example where a brane parameter, the relative separation in the $x_6$ direction,
has no clear impact on the field theory description). Eventually, the two branes have
to intersect.
However once they do, it is clear that something more singular may result.
What actually can be shown to occur \cite{HW,GK} is that as the NS5-brane `crosses'
the D6-brane a D4-brane is created and suspended between the NS5 and D6-brane.
Note that there is no particle content to the effective theory on that D4-brane and
thus no new degrees of freedom are created as the D4-brane is formed.

\begin{figure}[h]
\begin{center}
\begin{picture}(150,260)

\put(-10,153){\begin{picture}(130,100)
\setlength{\unitlength}{1mm}
\thicklines
\put (10,0){\line(0,1){40}}
\put (55,0){\line(1,4){10}}
\put (10,20){\line(10,0){50}}
\put (35,25){\makebox(0,0){D4}}
\put(10,-5){\makebox(0,0){NS5}}
\put(55,-5){\makebox(0,0){D6}}
\put(35,-5){\makebox(0,0){a)}}
\end{picture}}

\put(-10,3){\begin{picture}(130,100)
\setlength{\unitlength}{1mm}
\thicklines
\put (10,0){\line(0,1){40}}
\put (55,0){\line(1,4){10}}
\put (10,23){\line(10,0){51}}
\put (35,26){\makebox(0,0){D4}}
\put (10,17){\line(10,0){49}}
\put (35,12){\makebox(0,0){D4}}
\put(10,-5){\makebox(0,0){NS5}}
\put(55,-5){\makebox(0,0){D6}}
\put(35,-5){\makebox(0,0){b)}}
\end{picture}}

\end{picture}
\end{center}
\caption{SUSY allows only the existence of the configuration a).
Configuration b) is disallowed. However, if the NS5 is exchanged
with an NS5', configuration b) is allowed.}
\label{SRule}
\end{figure}
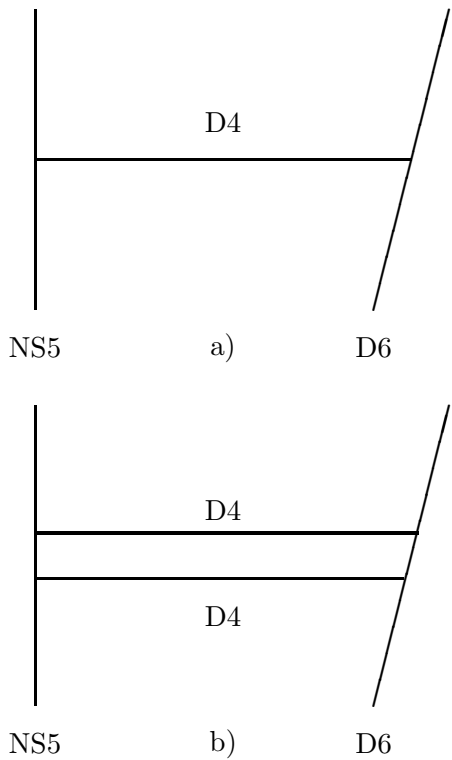

2) {\em Supersymmetry restrictions} \newline
Supersymmetry allows that only a single D4-brane can be suspended between a NS5-brane
and a D6-brane \cite{HW,GK} (Fig.~\ref{SRule}).

\begin{figure}[hbp]
\begin{center}
\begin{picture}(150,250)
\setlength{\unitlength}{1mm}
\thicklines

\put(-10,50){
\begin{picture}(130,90)
\put (10,0){\line(0,1){40}}
\put (35,0){\line(0,1){40}}
\put (60,0){\line(0,10){40}}
\put (10,20){\line(10,0){50}}
\put (53,23){\makebox(0,0){D4}}
\put(10,-5){\makebox(0,0){NS5}}
\put(60,-5){\makebox(0,0){NS5}}
\put(35,-5){\makebox(0,0){D6}}
\put(35,42){\makebox(0,0){$N_F=1$}}
\put(68,20){\makebox(0,0)
{$ \left.\begin{array}{c} \\ \end{array} \right\}N_C=1$}}
\end{picture}}

\put(-10,0){
\begin{picture}(130,90)
\put (10,0){\line(0,1){40}}
\put (35,0){\line(0,1){40}}
\put (60,0){\line(0,10){40}}
\put (10,15){\line(10,0){25}}
\put (35,25){\line(10,0){25}}
\put (53,28){\makebox(0,0){D4}}
\put (15,18){\makebox(0,0){D4}}
\put(10,-5){\makebox(0,0){NS5}}
\put(60,-5){\makebox(0,0){NS5}}
\put(35,-5){\makebox(0,0){D6}}
\put(37,20){\vector(0,1){5}}
\put(37,20){\vector(0,-1){5}}
\put(48,20){\makebox(0,0){$<Q>$}}
\end{picture}}

\end{picture}
\end{center}
\caption{The Higgs phase of field theory is realized by the D4 branes connecting
the two NS5 branes, breaking and separating on the D6 brane.
This indicated distance of separation corresponds to the expectation value of the
Higgs matter field.
The effective field theory in this example carries no massless degrees of freedom
at all. In the presence of additional D6 branes, the effective theory would contain massless
spin-0 and spin-$\frac{1}{2}$ degrees of freedom, but no massless spin-1 degrees of
freedom, as appropriate for a Higgs phase (Fig.~\ref{D6D4D6}).}
\label{HiggsPhase}
\end{figure}
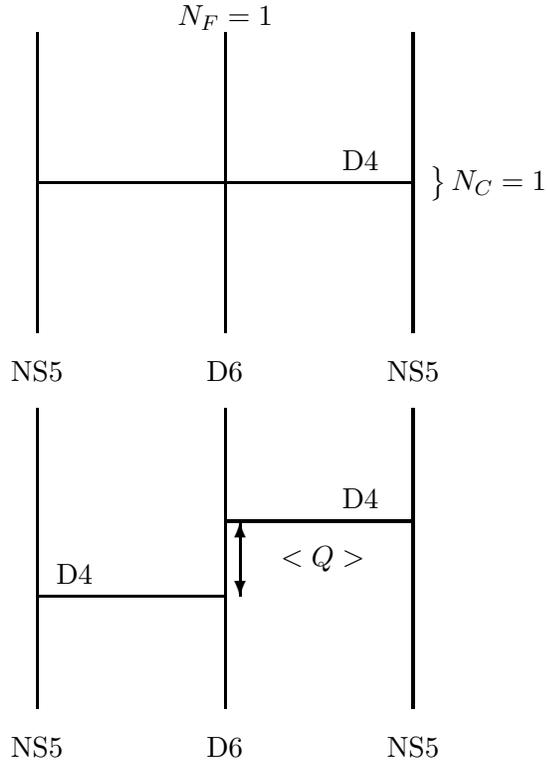

3) {\em The brane realization of the Higgs phase} \newline
We have seen that the Coulomb phase of N=4 gauge theories is spanned by the separation
among the D4-brane. The same is true the N=2 case. The N=1 case has no Coulomb phase.
Also the Higgs phase has a simple geometrical realization in terms of branes. Consider
an N=2 configuration in which in addition of having a D4-brane suspended between two
NS5-branes a D6-brane is inserted between the NS5-branes (Fig.~\ref{HiggsPhase}). 
The separation of the D4-branes
on the D6-brane along the $x_4, x_5$ directions maintains the supersymmetry. However
the effective field theory on each of the D4-branes contains no spin 1 massless particles.
One can show that this breaking indeed represents the Higgs phenomenon, the separation
among the two D4-branes along the   $x_4, x_5$ directions being proportional
to the expectation value of the scalar field responsible for the Higgsing.
This result is true also for the N=1 configurations. For the N=1 case one needs also
to realize that the effective theory on a D4-brane suspended between a D6-brane
and a NS5$'$-brane contains one massless complex spin 0 field,  one massless spin 1/2 field and no
massless spin 1 fields.

The brane geometrical picture has a classical weakly coupled taste. In order to be able to
trust the pictures as one moves the branes around it is advisable to be in a weak coupling
situation as long as possible. To enforce that, one starts the journey by setting the
D=4 N=1 $U(N_C)$ supersymmetric gauge theory with $N_F$ flavors in the Higgs phase.
Retaining the Higgs phase, one is able  not only to reproduce the particle content
required by Seiberg's duality, but also his analysis of comparing the dimensions of the moduli
space at both ends of the duality transformation, as well as comparing the numbers of
the relevant operators. One is also able to show in the brane picture as Seiberg has
done in field theory that masses as one end of the duality pair correspond to expectation
values on the other side. 

\subsection{Obtaining the dual field theory}
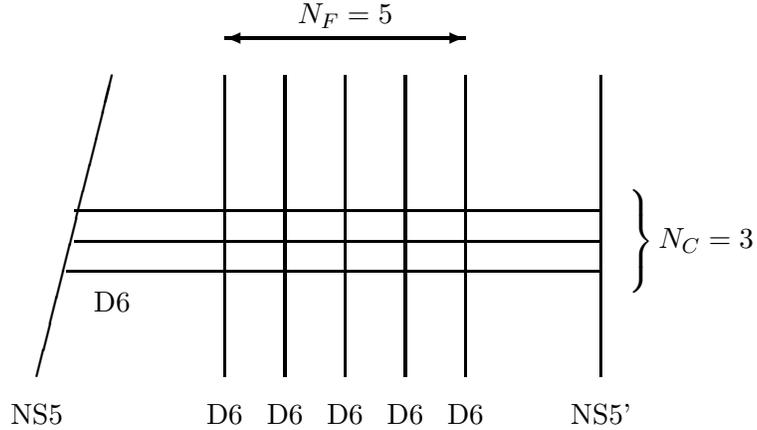
\begin{figure}[htp]
\begin{center}
\begin{picture}(150,140)
\setlength{\unitlength}{1mm}
\thicklines
\put (-5,5){\line(1,4){10}}
\put (70,5){\line(0,10){40}}
\put (-1,19){\line(10,0){71}}
\put (0,23){\line(10,0){70}}
\put (0,27){\line(10,0){70}}
\put(5,15){\makebox(0,0){D6}}
\put(80,23){\makebox(0,0)
{$ \left.\begin{array}{c} \\ \\ \\ \end{array} \right\}N_C=3$}}
\put(-5,0){\makebox(0,0){NS5}}
\put(70,0){\makebox(0,0){NS5'}}

\put (20,5){\line(0,10){40}}
\put(20,0){\makebox(0,0){D6}}
\put (28,5){\line(0,10){40}}
\put(28,0){\makebox(0,0){D6}}
\put (36,5){\line(0,10){40}}
\put(36,0){\makebox(0,0){D6}}
\put (44,5){\line(0,10){40}}
\put(44,0){\makebox(0,0){D6}}
\put (52,5){\line(0,10){40}}
\put(52,0){\makebox(0,0){D6}}

\put (36,50){\vector(1,0){16}}
\put (36,50){\vector(-1,0){16}}
\put(36,53){\makebox(0,0){$N_F=5$}}

\end{picture}
\end{center}
\caption{A D=4, N=1 SUSY, $N_C$=3, $N_F$=5 configuration.}
\label{N1NC3}
\end{figure}

At this stage in the lecture a movie\footnote{An attempt is made to digitalize the movie.} 
composed out of the comic strips was shown. This enabled
an easy visualization of the continuous features of the duality transformation. The reader is
actually equipped by now to embark on this journey on her/his own (or by reading
\cite{EGK,EGKRS,GK}). As a navigational aid, I will briefly mention the different stages of the
journey.
The starting point is a configuration depicted in figure~\ref{N1NC3}, which has $N_F$=5
and $N_C$=3.

{\em First stage:} the NS5-brane is displaced across the D6-branes respecting the rule mentioned
in 1) above concerning D4-brane creation.

{\em Second stage:} the system is driven into the Higgs phase according to rule 3) above.

{\em Third stage:} it is realized that if the NS5-brane will directly collide with the NS5$'$
brane the gauge coupling will diverge, leaving the protected weak coupling regime.
In order to avoid that, the branes are separated along the $x_7$ direction before they
are made to coincide in the $x_6$ direction. This separation, as stated in 1) above,
corresponds in some sense \cite{EGK,EGKRS,GK} to the turning on of a Fayet-Iliopoulos coupling in field
theory. At this stage, the appropriate reconnections of the branes shows that the
configuration represents the Higgs phase of a $U(N_F-N_C)$ gauge theory.

{\em Fourth stage:} rejoining the branes in the $x_7$ direction requires some quantum 
adjustments which we will indicate below.
The final outcome is shown in figure~\ref{sophisticated}, the gauge theory has
$N_F$=5 flavors and $N_C$=5-3=2 colors. In addition, the theory has $5^2$ singlet
complex spin-0 and spin-$\frac{1}{2}$ massless particles. 
This is the realization of the dual theory obtained by Seiberg.
The result here is simply the outcome of the NS5 moving according to the brane
rules from the left of the NS5' to its right.
The simple procedure allows one to obtain the infra-red dual of many other brane
configurations. 

\subsection{Concluding remarks}

\noindent {\em Infrared duality} 

What has actually been shown is that in a continuous motion starting from one N=1
supersymmetric gauge theory one reaches another. That does not really demonstrate that
the two are identical in the infrared or at all. Taking a bus ride from Chamonix to
Les Houches does not constitute a proof that the two cities are the same. What has emerged
naturally is one gauge group out of another as well as the presence of $N_F^2$ color singlets.
A more detailed analysis of the resulting massless particle spectrum (chiral ring)
and its properties is needed to conclude infrared duality  \cite{EGK,EGKRS,GK}.
\vskip .6cm

\noindent {\em Other groups} 

It can be shown that $Sp(N)$,  $O(N)$ and product gauge groups can be constructed as well
\cite{GK,EGK,EGKRS}.
\vskip .6cm

\noindent {\em Generalized infrared dualities}

By allowing configurations with $k$ NS5-branes and $k'$ NS5$'$-branes one can test new and
old generalizations of Seiberg's duality in the presence of a richer matter content
and various Landau-Ginzburg-like interactions \cite{EGK,EGKRS,GK}.
\vskip .6cm

\noindent {\em M-theory context}

The brane--field theory correspondence obtains even more geometrical features
once embedded in M-theory \cite{wittenM,GK,kings}.
\vskip .6cm

\noindent {\em Quantum corrections}

The classical brane picture needs to be amended by quantum considerations \cite{HW,wittenM}.
For example, recall that two D4-branes were not allowed to be suspended between an NS5
and a D6-brane. One could ascribe this to a quantum repulsive force between two D4-branes on the
same side of a NS5-brane in a non-BPS configuration. The agreement between the dimensions
of the moduli spaces of the two dual models was obtained in the presence of a quantum attractive
interaction between two D4-branes on the opposite sides of a NS5-brane.
At this stage, these are additional postulates they have however immediate consequences in 
allowing a unified description of D=4 and D=3 gauge theories with 4 supercharges. For example,
consider the D=4 N=1 $U(N_C)$ gauge theory with no flavors. Such a system has a ground
state \cite{Amati,SI}. However the same system in D=3 has no ground state \cite{AHISS}. 
This is indicated in the brane picture in the following manner: compactify one dimension, say $x_3$, of the
world-volume of the D4-brane, $0\leq x_3 \leq 2\pi R$. It is more convenient to perform
a longitudinal T-duality along the $x_3$ direction. The resulting system is a type IIB
string theory whose third direction has radius $1/R$. Thus a four dimensional IIA
system corresponds to a type IIB system with vanishing radius. For any finite radius,
the type IIB effective theory corresponds to a field theory with more than three dimensions
in type IIA. Only an infinite type IIB radius will correspond to an effective three dimensional
gauge theory. Due to the repulsive force between the D4-branes, and due to the compact nature
of the $x_3$ coordinate, a stable configuration will form for any finite value of
$x_3$, that is, such a vacuum state exists for any effective supersymmetric gauge 
theory with no flavors in more than three dimensions. 
Once the IIB radius is infinite, the equilibrium state exists no more,
the three dimensional gauge theory has no vacuum.

\section{Final Remarks}
In this set of lectures we have had a panoramic vista of the rich structure of supersymmetric gauge theories. The classification of possible phases of gauge theories were understood well before concepts like confinement could have been analysed analytically in four dimensional continuum theories. Perhaps a similar course can be followed for theories whose symmetry includes general coordinate invariance in analogy to figures 4,5 and 8. The qualitative understanding of the phase structure of such systems will proceed their complete quantitative analysis. As a prototype for a phase diagram of gravity we propose the following figure \ref{gravfig}, \cite{Barbon}.

\begin{figure}[h] 
\centering
\includegraphics[width=7.truecm]{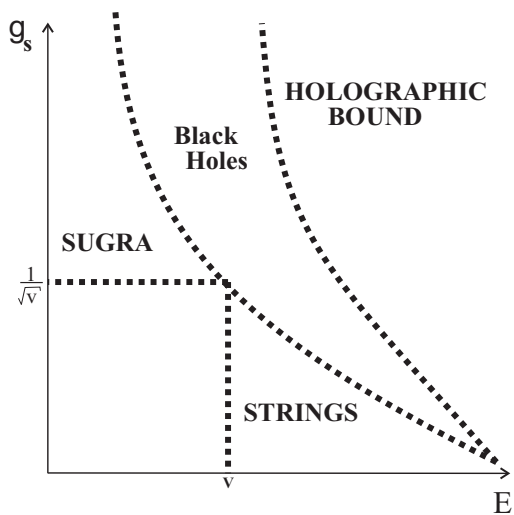} 
\caption[]{A prototype of a phase diagram for a gravitational theory}
\label{gravfig} \end{figure}

{\section{Acknowledgements}}

This work is supported by BSF- the American-Israeli Bi-National
Science foundation, the Israel academy of sciences and
humanities-Centers of excellence program, the German-Israeli
Bi-National science foundation, the european RTN network
HPRN-CT-2000-00122 and DOE grant DE-FG02-90ER40560. E.~R. acknowledges
S.~ Elitzur, A.~ Forge, A.~ Giveon, D.~ Kutasov and A.~ Schwimmer for
collobartions and many very valuable discussions. D.~B. acknowledges
the inspirational atmosphere created by the groups of the Hebrew
University and the Weizmann institute. We are also grateful to
A.~Sever with help preparing the manuscript.

\newpage

The references given in this article though detailed are not complete.

\end{document}